\documentclass[preprint2]{pp7}
\input{pp7.h}

\usepackage{color}
\usepackage{xspace}
\usepackage{units}
\usepackage{cancel}
\usepackage{float}

\usepackage[version=3]{mhchem}

\newcommand{\CeightO}{\ce{C^{18}O}\xspace}
\newcommand{\ammo}{\ce{NH3}\xspace}
\newcommand{\ntwohp}{\ce{N2H+}\xspace}
\newcommand{\hone}{H{\sc i}\xspace} 
\newcommand{\hion}{H{\sc ii}\xspace} 

\newcommand{\htwo}{\ce{H2}\xspace}
\newcommand{\etal}{{\it et al.}}

\newcommand{\msun}{\ensuremath{\rm M_\odot}\xspace}
\newcommand{\msunpc}{\ensuremath{\rm M_\odot\,pc^{-1}}\xspace}
\newcommand{\msunyr}{\ensuremath{\rm M_\odot\,yr^{-1}}\xspace}
\newcommand{\cc}{\ensuremath{\rm cm^{-3}}\xspace}
\newcommand{\kms}{\ensuremath{\rm km\,s^{-1}}\xspace}
\newcommand{\kkms}{\ensuremath{\rm K\,km\,s^{-1}}\xspace}

\def\ltsima{$\; \buildrel < \over \sim \;$}
\def\simlt{\lower.5ex\hbox{\ltsima}}
\def\gtsima{$\; \buildrel > \over \sim \;$}
\def\simgt{\lower.5ex\hbox{\gtsima}}

\begin{document}
\title{\textbf{\LARGE From Bubbles and Filaments to Cores and Disks: 
Gas Gathering and Growth of Structure Leading to the Formation of Stellar Systems}}

\author {\textbf{\large Jaime E. Pineda}$^{1}$,
         \textbf{\large Doris Arzoumanian}$^{2,3}$,
         \textbf{\large Philippe Andr\'e}$^{4}$,
         \textbf{\large Rachel K. Friesen}$^{5}$,
         \textbf{\large Annie Zavagno}$^{3}$,
         \textbf{\large Seamus D. Clarke}$^{6}$,
         \textbf{\large Tsuyoshi Inoue}$^{7}$,
         \textbf{\large Che-Yu Chen}$^{8}$,
         \textbf{\large Yueh-Ning Lee}$^{9}$,
         \textbf{\large Juan D. Soler}$^{10}$,
         \textbf{\large Michael Kuffmeier}$^{11}$}
\affil{$^{1}$\small\it Max-Planck-Institut f\"ur extraterrestrische Physik,%
$^{2}$\small\it National Astronomical Observatory of Japan, %
$^{3}$\small\it Aix-Marseille Universit\'e \& IUF, %
$^{4}$\small\it CEA Paris-Saclay, %
$^{5}$\small\it University of Toronto, %
$^{6}$\small\it Academia Sinica Institute of Astronomy and Astrophysics, %
$^{7}$\small\it Konan University, %
$^{8}$\small\it Lawrence Livermore National Laboratory, %
$^{9}$\small\it National Taiwan Normal University, %
$^{10}$\small\it Max Planck Institute for Astronomy, %
$^{11}$\small\it University of Virginia}

\begin{abstract}
\baselineskip = 11pt
\leftskip = 1.5cm 
\rightskip = 1.5cm
\parindent=1pc
{\small Abstract
{
The study of the development of structures on multiple scales in the cold interstellar medium has experienced rapid expansion in the past decade, 
on both the observational and the theoretical front. 
Spectral line studies at (sub-)millimeter wavelengths over a wide range of physical scales have provided unique probes 
of the kinematics of dense gas in star-forming regions, and have been complemented by extensive, high dynamic range dust continuum surveys of the column density structure 
of molecular cloud complexes, while dust polarization maps 
have highlighted the role of magnetic fields. 
This has been accompanied by increasingly sophisticated numerical simulations including new physics (e.g., supernova driving, cosmic rays, non-ideal magneto-hydrodynamics, radiation pressure) and new techniques such as zoom-in simulations allowing multi-scale studies.
Taken together, these new data have emphasized the anisotropic growth of dense structures on all scales, from giant ISM bubbles driven by stellar feedback 
on $\sim$50--100 pc scales through parsec-scale molecular filaments down to $<$0.1pc dense cores and $<$1000 au protostellar disks. 
Combining observations and theory, we present a coherent picture for the formation and evolution of these structures and synthesize 
a comprehensive physical scenario for the initial conditions and early stages of star and disk formation.
}
\\~\\~\\~}
\end{abstract}

\section{INTRODUCTION}

Studying the growth of structures on multiple scales in the cold interstellar medium (ISM) is crucial for improving our understanding 
of the general inefficiency of the star formation process, the origin of the stellar initial mass function (IMF), 
and ultimately the birth of protoplanetary disks.
This field of research has  thrived in the past decade, thanks to both high dynamic 
range observations and numerical simulations.
In particular, spectral line studies at (sub-)millimeter wavelengths over a wide range of physical 
scales with, e.g., the Robert C. Byrd Green Bank Telescope (GBT),  
the Atacama Large Millimeter/submillimeter Array (ALMA), and the Northern Extended Millimeter Array (NOEMA) 
have provided unique probes of the kinematics of dense gas in star-forming regions, 
complementing deep, extensive dust continuum surveys of the column density structure of molecular cloud complexes with, e.g.,~{\it Herschel}. 
Dust polarization maps from {\it Planck} and the James Clerk Maxwell Telescope (JCMT) have also highlighted the role of magnetic fields. 
We review how these newly-acquired observations of gas structure 
and dynamics impact our understanding of the physics of star formation by
comparing them with state-of-the-art advanced hydrodynamical (HD) and magnetohydrodynamical (MHD) models. 
Taken together, the new data enabled the community
to establish clear physical connections between a broad hierarchy of 
cold interstellar structures ranging from giant ISM bubbles driven by stellar feedback on $\sim$50--100 pc scales (Sect.~\ref{sec:bubble}) 
through parsec-scale molecular filaments (Sect.~\ref{sec:filament}) 
down to  $< 0.1$ pc dense cores (Sect.~\ref{sec:core}) and $< 1000$~au protostellar disks (Sect.~\ref{sec:disk}).
The observational results emphasize the anisotropic growth of dense structures on all scales, 
with shell-like accretion of gas 
from bubbles to filaments (Sects.~\ref{sec:bubble} and \ref{sec:filament}), 
axial contraction from filaments to cores (Sects.~\ref{sec:filament} and \ref{sec:core}), 
followed by non-axisymmetric accretion through streamers from cores to disks (Sect.~\ref{sec:disk}). 
We follow the flow of material throughout the different levels of the ISM hierarchy, with each level accreting from its parent level and funneling material to its sub-level. 
While each level in the hierarchy is interconnected in this way, different geometries and physical processes dominate 
on each size scale, meaning that there exists a partial decoupling between consecutive levels. 
It is therefore productive to discuss each level separately as a set of discrete structural entities. 
The chapter sections follow the flow of dense gas from large to small scales, or from bubbles to disks. 
To conclude the chapter, we synthesize a comprehensive updated paradigm  
for the initial conditions and early stages of star and disk formation.

\section{\uppercase{Interstellar Bubbles and Their\\ Connections to the Formation of\\ Filamentary Structures}\label{sec:bubble}}
\index{Bubble|(}\index{Filament|(}

\begin{figure*}[ht]
\centering
\includegraphics[width=0.9\textwidth]{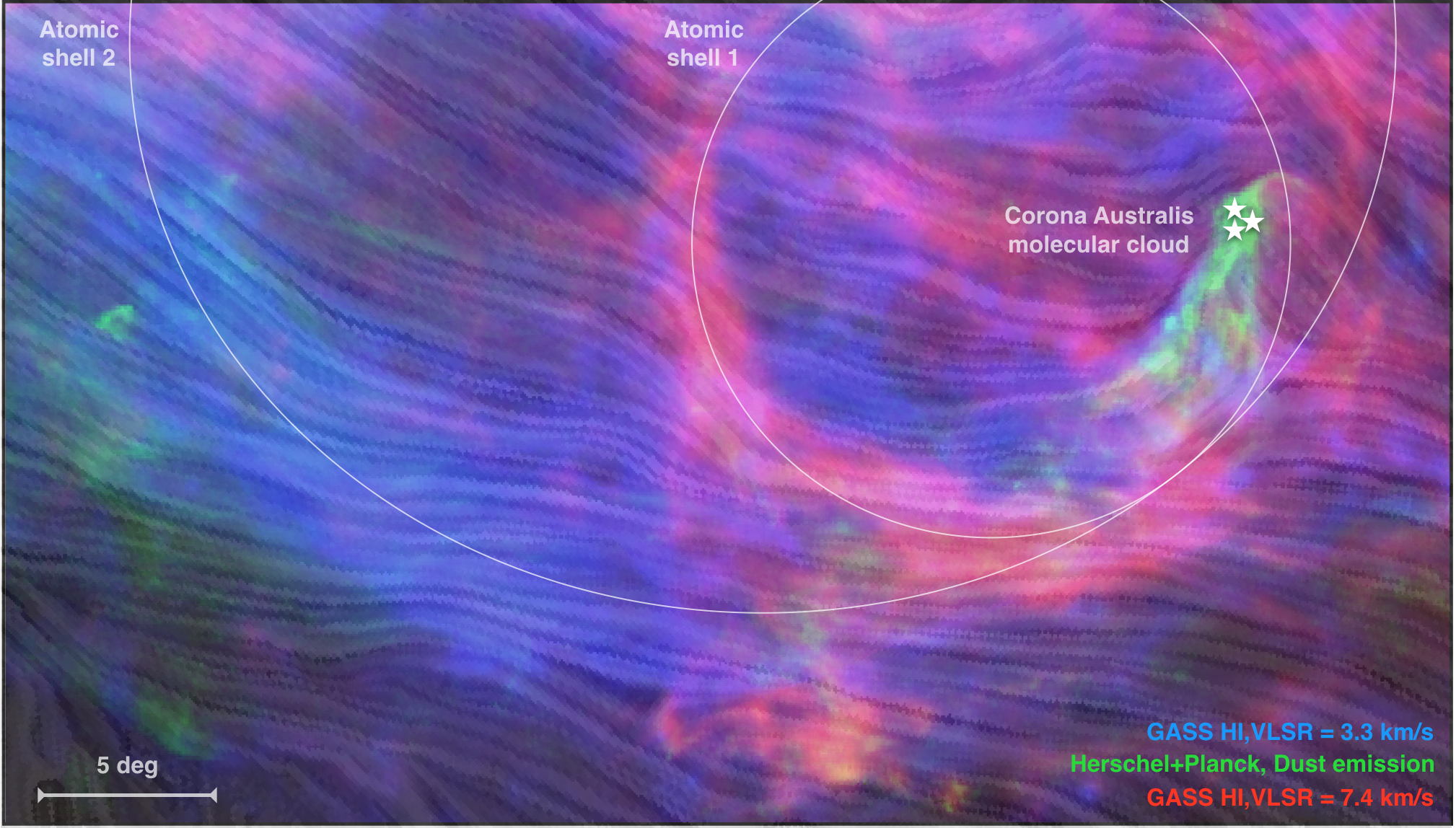}
 \caption{\small
Color composite  image showing the \hone emission at 7.4 \kms (red) and at 3.3 \kms (blue) and the column density map derived from combined {\it Herschel}+{\it Planck} data (green). 
The white circles trace the two \hone shells\index{\hone shells} associated with the Corona Australis (CrA) molecular cloud\index[obj]{Corona Australis Molecular Cloud} traced in green. 
The magnetic field structure derived from {\it Planck} polarization observations is shown as a drapery pattern 
\citep[from][]{Bra20}. The 5\,deg scale corresponds to 13\,pc at the 150\,pc distance of the CrA  cloud.
}
 \label{fig_HiShell_fil}
\end{figure*}

The cold interstellar medium is observed to be organized in bubbles and filamentary structures.
Interstellar bubbles of both neutral and ionized gas are associated with high-mass stars at different phases of their evolution. %
The expanding nature of these bubbles shapes the surrounding medium and possibly plays a role in the formation and evolution of  interstellar filaments.
In this section we present recent observations of both neutral and ionized bubbles %
and their relationship with filamentary structures  identified in atomic and molecular clouds.
We then review existing theories of filament formation and discuss the important role of 
expanding bubbles in the formation process of molecular filaments in shock-compressed layers\index{Shock!Shock-compressed layers}.

\subsection{Observations of Bubbles and  Filamentary\\ Structures in the Interstellar Medium \label{sec:2.1}}
\subsubsection{Observations of Bubbles and their Relation with the Local Interstellar Medium \label{sec:2.1.1}}

In our Galaxy, bubbles of both neutral and ionized gas are ubiquitously observed.
The expanding nature of bubbles, with velocities $\sim$5--20\,\kms \citep{Spitzer1978}, 
means that bubbles may sweep up and gather diffuse gas and  participate in the formation of new dense filamentary structures, 
but may also compress or disrupt existing structures. 
It is important to note that the  neutral and ionized bubbles have markedly different origins and also affect different size scales. 
We present recent observations and description of the two types of bubbles below.

\hone shells\index{\hone Shell}, observed in 21\,cm \hone  atomic emission \citep{Dai07,Ehl13}, 
are generated mainly by supernovae or multiple \hion regions\index{\hion Region} and result from the expansion, recombination, and cooling of 
the hot and ionized expanding gas \citep{Tomisaka1981}.
\hone  supershells  interact with the ISM at  large-scales ($\sim$100\,pc) and can drive strong large-scale compressive flows, 
which has been observationally suggested  to help the formation of molecular gas \citep{Ehl16,rob18}. 
In the early 90's, \hone filaments and arcs delineating the Orion-Eridanus bubble were observed, which
suggested a relation between the wind-blown bubble and the form and structures of the surrounding medium \citep{bro95}. 
The spatial distribution of CO clumps, correlated with and observed in excess toward the walls of \hone shells, 
reinforces the idea of the role of \hone shells in the formation (and shaping) of molecular clouds \citep{Daw08}.  
Recent observations support a scenario of filamentary molecular cloud formation triggered by supersonic compression 
of cold magnetized \hone gas from at least two expanding bubbles \citep[][see Fig.\,\ref{fig_HiShell_fil}]{Bra20}.
The formation of many nearby molecular clouds may have been driven by the expansion of the Local Bubble 
resulting from multiple supernova explosions \citep{Zucker+22}.

\begin{figure*}[ht!]
\centering
\includegraphics[width=\textwidth]{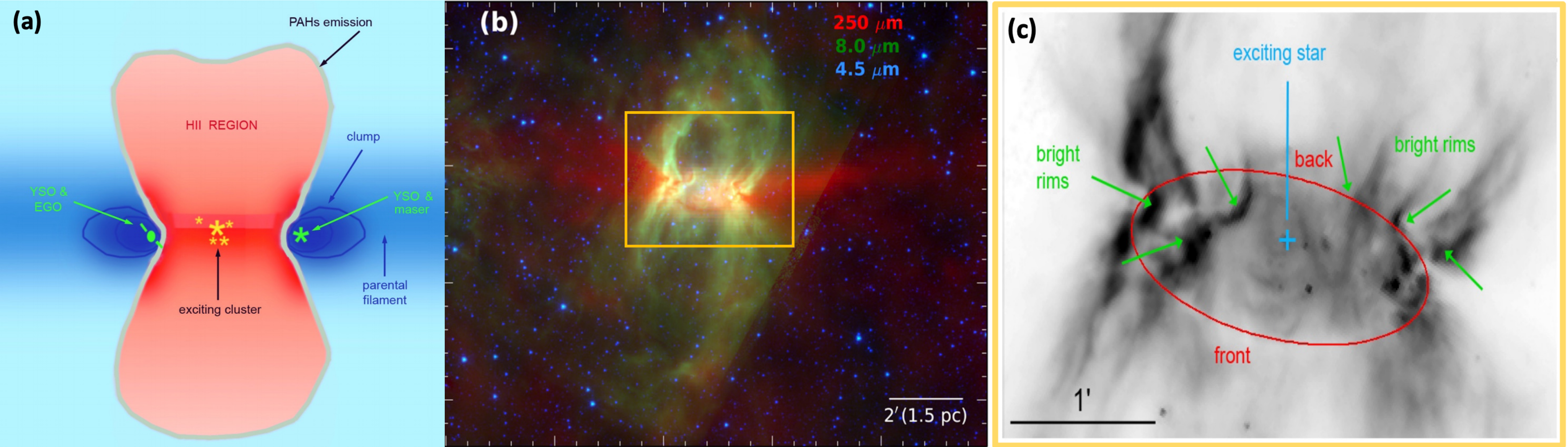}
\caption{\small 
{\bf a)} Illustration of a generic bipolar \hion region\index{\hion Region} and its environment.
The ionized gas, the molecular gas, and  the PAH emission of the PDR are shown in red, blue, and grey, respectively. 
{\bf b)} Composite image of  G$319.88+00.79$\index[obj]{G319.88+00.79}, which is an observed example of a bipolar \hion  region.
{\bf c)}  Center of G$319.88+00.79$ at 8.0\,$\mu$m showing the dense waist of the nebula (detected in absorption) %
the bright rims, and the position of the exciting star.
Panels a, b, and c are adapted from \citet{deh15} and \citet{Man18}.
 }
\label{fig_BipolarHII}
\end{figure*}

\begin{figure}[ht!]
\begin{center}
\includegraphics[width=\columnwidth]{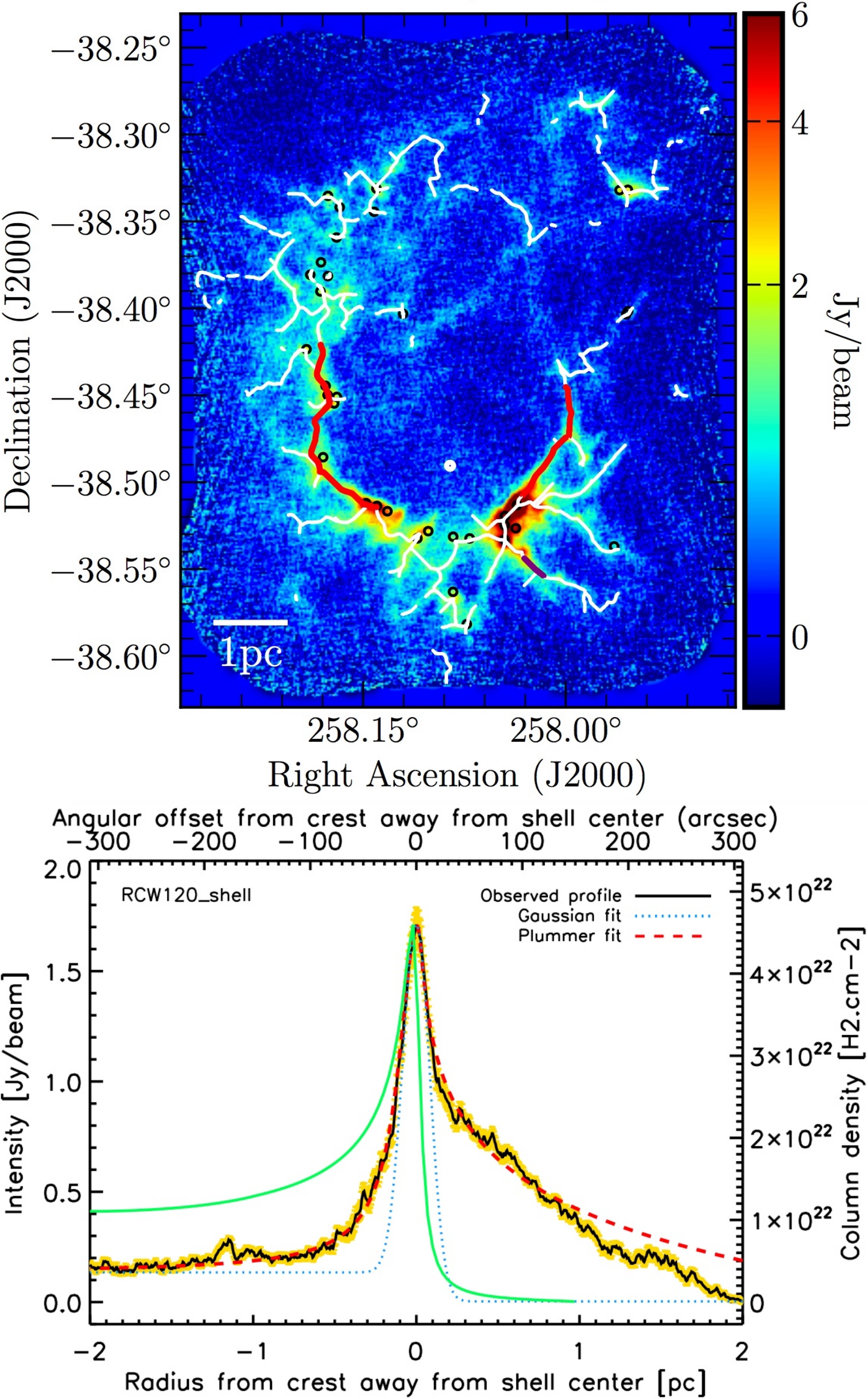}
\caption{\small 
    {Top:} Combined ArT\'eMiS+{\it Herschel}-SPIRE 350\,$\mu$m map of the \hion region\index{\hion Region} RCW 120\index[obj]{RCW 120}. 
    The crests of the filaments are overlaid in white. The black circles show the {\it Herschel} young sources. 
    The white circle shows the location of the ionizing star. 
    {Bottom:} The observed mean radial column density profile (black curve) of the RCW 120 dense shell 
    (measured across and averaged along the crest shown in red in the top panel), with positive offsets going away from the shell center. 
    A Gaussian fit (blue dotted curve) and a Plummer fit (red dashed curve) are shown. 
    The green full line shows the profile expected for a spherical shell, 
    which does not reproduce well the observations. 
    This suggests that RCW 120 might be a 3D ring, rather than a spherical structure. \citep[From][]{Zav20}.
}
\label{fig_HII}
\end{center}
\end{figure}

\hion regions\index{\hion Region}, detected in H${\alpha}$, infrared, 
and/or radio emission,  interact with the nearby matter of the parental cloud  ($\sim$1--5\,pc scales) and trace the 
present-time stellar feedback due to  stellar winds, ionizing radiation, and outflows of massive stars.
The \textit{Herschel} far-infrared survey of the Galactic Plane, Hi-GAL \citep{Mol10}, clearly reveals the importance 
of ionized bubbles and their key role in shaping their surrounding material and influencing subsequent 
star formation \citep{pal17} and especially the formation of new generations of high-mass stars 
\citep[as shown by][]{zha20,zha21}. 
\hion regions are observed to be surrounded by dense molecular gas. 
The interaction between possibly pre-existing dense molecular gas and the expansion of \hion regions is revealed by 
the variation of the abundance ratio of CO isotopologues (\ce{^{13}CO}/\ce{C^{18}O}) that results from the selective 
photodissociation by the FUV radiation from 
embedded OB stars, since this variation cannot be explained solely by the  effect of  the interstellar 
radiation field \citep[][]{Shimajiri2014,Are19}.

The impact of ionized regions on their surrounding medium strongly depends on the evolutionary stage 
of the ionized region, the ionized gas geometry, 
and the original configuration of the molecular gas that gave birth to the ionizing stars (see Fig.\,\ref{fig_BipolarHII}). 
Classical ionized regions (Str\"omgren spheres) collect their surrounding material between their 
expanding ionization front and the shock\index{Shock} front 
that precedes it into the surrounding neutral medium. 
Compression of the surrounding medium by the ionized region is accompanied by a 
local density increase 
and by the formation of filamentary structures (see Fig.\,\ref{fig_HII}).\index{Filament!Formation}  
Ionized regions might also expand in a pre-existing filamentary medium and 
play a role in compressing the already formed filaments, dispersing them  or preventing their dispersion \citep{xu18}. 
There are examples of 
bipolar \hion regions\index{\hion Region} (see Fig.\,\ref{fig_BipolarHII}) where the ionizing stars 
form in and impact the parental sheet-like\index{Sheet} filamentary molecular cloud \citep{Xu17}, 
which itself have been formed at the edge of an expanding \hone shell.\index{\hone Shell}
Hub-filament systems, which are centrally converging networks of filaments into high density hubs, 
have been proposed to be the main sites of star-clusters and high-mass star formation 
\citep{Myers_2009,bau18,Wil18,Kumar_HFS_2020,dew21,liu21}. 
The hub-filament configuration may allow the ionizing pressure and radiation from the high-mass stars 
formed in the hub to escape through the inter-filamentary gaps, while dense material continues 
flowing along the filaments onto the hub.  
These flows replenish the hub with matter allowing the formation of massive 
clusters and high-mass stars \citep{Kumar_HFS_2020,Kumar2021}. 
At an evolved stage, the parental filamentary cloud is always observed to be associated 
with the particular morphology of a bipolar ionized region. 
The waist of these bipolar regions hosts young and active high-mass star formation 
(see Fig.\,\ref{fig_BipolarHII}), possibly as a direct consequence 
of the ionized region expansion in the parental filaments \citep{deh15,Man18}. 
In other cases, \hion regions are observed to expand into filamentary molecular clouds but the ionizing stars are 
not observed to be directly associated with the presently observed dense molecular material. 
This is the case, for example, of the RCW 120\index[obj]{RCW 120} \hion  region \citep[see Fig.\,\ref{fig_HII} and][]{Zavagno2010,Zav20}. 
These different configurations of the matter structure and  of the local density of the material associated with the
 ionizing star may be linked to the evolutionary stage of the expanding \hion region and of the star formation event \citep{Kumar_HFS_2020}.

These observations pose the question of the \textit{original} density distribution of the surrounding material prior to 
both \hone and \hion bubbles' expansion.\index{\hion Region}\index{\hone Shell} 
If the material is organized into filaments from its origin, the bubble's %
expansion may mostly have a shaping role and be probably in turn shaped by the 
filamentary structures (e.g., in hub-filament systems as explained above); 
whereas if the material is not filamentary at its origin, then the bubble's
expansion can have a direct role in the formation of filaments. \index{Filament!Formation}

\subsubsection{Observations of the Filamentary Interstellar Medium\label{sec:2.1.2}}
\label{sec:FilaCat}

Both the atomic and molecular phases of the Galactic ISM are observed to be filamentary 
\citep[e.g.,][]{McClure-Griffiths2006,Clark2014,Ragan2014,Mattern2018}.
In the diffuse atomic ISM, some of these filamentary structures correspond to elongated structures 
``crawling away from the Galactic plane'' \citep{heiles_1984} or parallel to the Galactic plane \citep{soler_2020}. 
Others are hair-like  extended structures aligned with the interstellar magnetic field \citep{Clark2014,kalberla2016}. 
Some of these structures are probably embedded in a warm ionized medium, as suggested by their correlation 
with radiopolarimetry observations \citep{kalberla2017,Bracco2020LOFAR}.
Descriptions of current filament identification and characterization methods are presented in Sect.\,\ref{sec:CoreIdn}.

As for the molecular ISM, it is comprised of complex filamentary networks \citep[see review in][]{Andre2014-PPVI}  
from large to small scales in a diversity of environments. 
On average, $\sim$15\% of the total mass of  molecular clouds is observed to be in the form of molecular filaments, 
while up to 
\hbox{$\sim$60--90\%} of the dense gas mass (defined as the mass of gas with 
column density $N_{\ce{H2}}>7\times10^{21}$\,cm$^{-2}$) 
is in the form of filaments \citep{Arzoumanian2019-Filament_properties,Roy+19,Kumar2021}.
In the literature, various names have been employed to describe the multitude of filaments 
observed at different scales using  different tracers. 
These different types of filamentary structures are discussed in detail by 
\citet[][in this volume]{Hacar+2022}.
%
At scales $\gtrsim 10$\,pc and distances $\gtrsim 1$\,kpc, filamentary structures have been referred 
to as  giant molecular filaments \citep[e.g.,][]{Ragan2014} 
or Galactic bones  \citep[e.g.,][]{Goodman_bones_2014,Zucker2015}, including (collections of) 
extraordinarily elongated infrared dark clouds (IRDCs).  
At scales of $\sim$1--10\,pc and at distances of $\lesssim$1\,kpc, the term of 
interstellar or molecular filaments has mostly been used \citep[e.g.,][]{Andre2014-PPVI,Arzoumanian2019-Filament_properties}. 
Molecular filaments as traced in the dust continuum may further break down into filamentary  substructures in 
position-position-velocity space when observed in emission from molecular lines. 
These velocity-coherent substructures are sometimes termed fibers when they are filamentary and overlapping 
in projection \citep[][]{Hacar_fiber_2013,Hacar2018-ALMA_Orion}.

Filaments are often not isolated objects, but are observed to form systems 
with multiple junctions and intersections.
These systems may be identified as: 
1) ridge--filament systems when side-filaments are connected from the side to a (usually denser and 
more massive) star-forming main-filament sometimes referred to as a ``ridge'' \citep[e.g.,][]{Hennemann_ridge_2012}, 
or 2) hub--filament systems when multiple filaments join from various directions
into a hub \citep[e.g.,][see also Sect.\,\ref{sec:2.1.1}]{Myers_2009,Kirk_SerpensS_2013,Peretto2014}. 
Some side-filaments are star-forming \citep[as in, e.g., the DR21 filament system,][]{Hennemann_ridge_2012}\index[obj]{DR21}, while others, 
similar to \hone filaments, are magnetically-aligned thin, hair-like linear structures, referred to as ``striations'' 
\citep[][]{Goldsmith2008,Palmeirim2013,Cox_Musca_2016,Malinen_striation_2016}.  
When  observations of the velocity field are available, side-filaments appear to be channels of gas flows feeding 
a main-filament (see reviews in \citealt{Andre2014-PPVI,Motte_HMreview_2018}, or see e.g., 
\citealt{Wil18,Trevino_Morales_MonR2_2019,Chen20,liu21} for more recent observations). 
Indeed, the physical intersections between filaments usually exhibit multiple velocity components, 
which trace the individual velocities of each of the merging filaments.

Most of these different filament names are based on the apparent morphology of the systems, and are not well defined 
in terms of measurable physical properties such as mass per unit length (or line mass, denoted by $M/L$ or $M_{\rm line}$).\index{Filament!Line Mass} 
Nevertheless, filaments with line masses below and above the thermal value of the critical mass per unit length 
$M_{\rm line, crit} = 2\, c_s^2/G $ \citep[e.g.,][]{Ost64} are usually called thermally subcritical and thermally supercritical, respectively 
\citep[e.g.,][see also \citealt{Arzoumanian2019-Filament_properties} and Sect.~\ref{sec:width} below for a refined classification based on $M_{\rm line}$]{Andre2014-PPVI}. 
The term  `striations' is typically used to refer to fainter subcritical structures aligned with the magnetic field, while filaments, 
filamentary substructures, (molecular) fibers, and side-filaments are generally indistinguishable in terms 
of their line mass and other properties.

\subsection{Theoretical Models of Filament Formation}
\index{Filament!Formation|(}
The expansion of bubbles compresses the ISM, which drives not only the evolution of the diffuse ISM into cold \hone and 
molecular clouds, but also induces filament formation in the compressed molecular cloud \citep{IIIH15}.
\cite{NBFH11} demonstrated that collisions of super-bubbles triggers the formation of highly 
structured filamentary molecular gas  ($\sim$100\,cm$^{-3}$).
Recent kpc-scale galactic simulations allowed for bubble formation by including 
supernova explosions, radiation, and winds from massive stars,
and concluded that bubbles are critical in controlling the evolution of the 
ISM \citep[e.g.,][]{Padoan2017,KO18, Rathjen+21}.
Although large-scale simulations can capture the global ISM evolution, detailed processes of filament formation 
and evolution are often outside the resolvable scales. 
In the following, we focus on the formation of parsec-scale  atomic and molecular filaments, which have more 
direct relations to star formation.
In particular, the parsec scale studies of filamentary cloud formation 
by shock\index{Shock} compression 
reported below describe the impact of expanding bubbles on the ISM.

\subsubsection{Overview of Recent Models for Filament Formation}

While several formation mechanisms have been proposed for different types of filaments,
the fact that filamentary structures appear in both 2D and 3D \mbox{simulations of} clouds with or without turbulence, 
magnetic field, and/or self-gravity
(e.g.,\,\citealt{IF13,Gomez_VS_14,Smith+14,VanLoo_etal_14,Kirk_etal_15,Federrath16}, also see review in \citealt{Andre2014-PPVI}
renders the dominant mechanism for filament formation uncertain.
In addition, filaments in simulations are often dynamic structures that are continuously evolving \citep{Smith+14,Li_Klein_2019}. 
This makes the characterization of  filament properties in simulations challenging.

In the \hone  phase of the interstellar medium, it is generally recognized that filamentary structures 
are created in shock compressed layers\index{Shock!Shock-compressed layers} \citep{II09, HSH09}.
In such layers, \cite{II16} showed that \hone  clouds created via shock induced thermal instability 
are stretched by turbulent shear flows along the magnetic field 
\citep[see also][]{Hennebelle13,solerANDhennebelle2017}.
The resulting filamentary \hone  clouds resemble the observed \hone filaments \citep{Clark2014}.

In the denser regions of the ISM, molecular filaments have been considered to be the products of 
either direct compression by interstellar turbulence or gravitational fragmentation at the cloud scale
\citep[see review in][]{Andre2014-PPVI}.
The earliest models of filament formation often considered the semi-analytic fragmentation of 
sheet-like\index{Sheet} clouds due to gravitational instability (see references for  the `type-G' mechanism described in 
Sect.~\ref{sec:2.2.3} below).
It has also been shown that thermal instability in quiescent clouds can
lead to the formation of clumpy and filamentary structures \citep{Wareing_2016,Wareing_2019}.

With the advances in computational capability over the past decade, fully-3D simulations with prescribed 
turbulence have become the common approach to studying the formation and evolution of star-forming clouds. 
Simulations considering the evolution of individual clouds usually start from uniform gas in a cubic simulation 
box or from a spherical clump with scale-dependent velocity perturbation, using 
either periodic or open boundary conditions \citep[e.g.,][]{Smith+14,Kirk_etal_15,LiPS_2015,Federrath16}. 
These simulations successfully produce filaments and  show that filamentary structures are the inevitable 
outcome of cloud evolution. 

In the scenario of global hierarchical collapse \citep[GHC;][]{VS_GHC_2019}, 
filaments are described as dynamic structures that continuously accrete from the ambient gas while 
feeding dense cores within them, with gravity being the main driving force of filament formation 
\citep[see also][]{Gomez_VS_14,NaranjoRomero_GHC_2015,VS_2017}. 
Without invoking global gravitational collapse, the combination of large-scale compressive flows 
and local self-gravity can lead to similarly dynamic filaments
continuously accreting and feeding their cores (see more details in Sect.\,\ref{sec:2.2.3} below).

As for the formation of the faint, thermally subcritical  and periodically spaced linear 
striations often observed in the immediate surroundings of star-forming filaments 
\citep[e.g.,][see also a review in \citealt{Andre2014-PPVI}]{Cox_Musca_2016,Malinen_striation_2016}, 
recent theoretical studies have proposed various scenarios. 
The models include anisotropy in magnetized turbulence \citep{Vestuto2003,Heyer_striation_2008, XSL19}, 
density oscillations induced by MHD\index{MHD} waves \mbox{\citep{Heyer_striation_2016,TT_striation_16,heyer_2020},} 
and the corrugation of ultra-thin shock-compressed layers\index{Shock!Shock-compressed layers} \citep{Chen_striation_17}. 
Numerical simulations also suggest that striations could be more commonly present in 
star-forming clouds than what has been observed, because these faint structures 
are easily washed out due to projection overlapping \citep{Chen_striation_17,Li_Klein_2019}.

\subsubsection{Anisotropic Filament Formation in Shock-\\Compressed Layers}
\label{sec:2.2.3}\index{Sheet|(}\index{Shock!Shock-compressed layers|(}

With increasing observational evidences of the lack of symmetry between the plane-of-sky extent 
and the line-of-sight depth of molecular clouds \citep[e.g.,][]{CLASSy_Storm_2014,KLee_2014_CLASSy,Arzoumanian2018-Filament_Nobeyama,Shimajiri_Taurus_2019}, 
\mbox{it has been suggested that filaments form} in sheet-like\index{Sheet} gas structures rather than from spherical gas clouds 
 \citep{Heitsch13,CO15,Auddy_etal_16}. 
Hence, simulations with convergent flows or colliding clumps that generate shock-compressed 
dense gas layers became an alternative way to numerically investigate the star formation process 
in the ISM \citep[e.g.,][]{IF13,Gomez_VS_14,CO14,CO15,GO15,Wu_GMC_2017,Inoue+18}.
In this view, filament formation is still directly connected to shock compression 
but as a two-step process. Firstly, a large-scale shock wave (e.g. an expanding \hone  bubble) or 
a supersonic turbulent flow in the molecular cloud compresses the gas to form a dense layer (a 2D structure). 
String-like 1D filaments then form within this locally flat region via gas accretion in a preferred direction parallel 
to the shock-compressed layer \citep{Inoue+18,Chen20}.

\begin{figure}[t]
\centering
\includegraphics[width=\columnwidth]{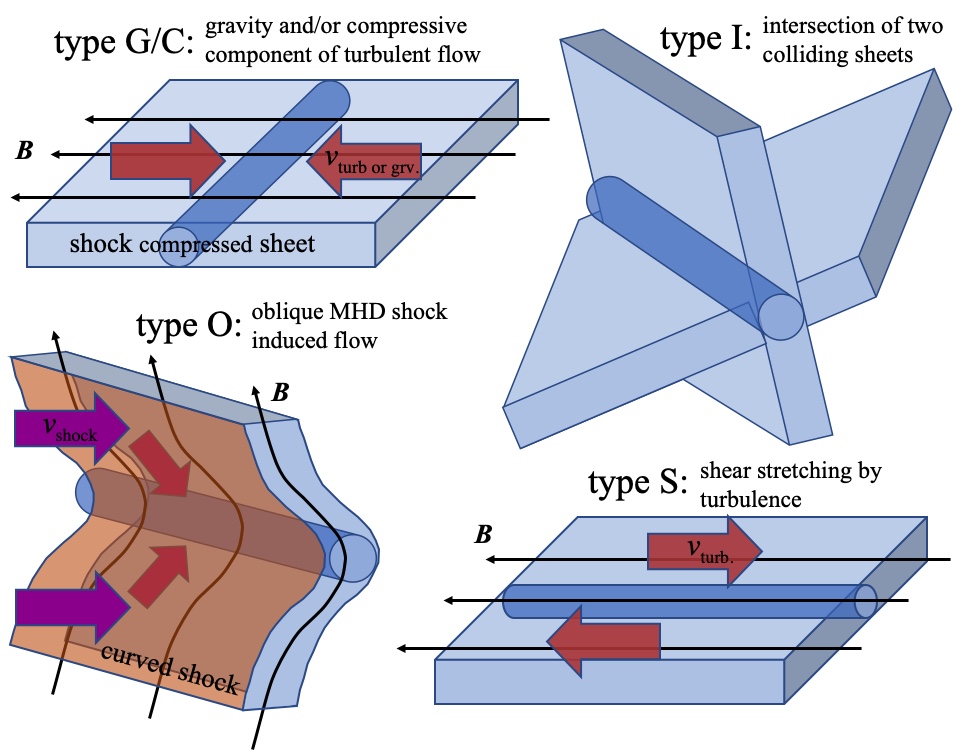}
\caption{\small
    Schematics of the filament formation mechanisms, where the blue sheets are shock compressed layers in which 
    filaments are formed, the black thin arrows represent the magnetic fields, and the thick arrows show the gas flow orientations.
  }
\label{fig_class}
\end{figure}

In \cite{AIIM21},  existing filament formation models
associated with shock-compressed layers were clas\-sified
into five  types,  as follows  (see Fig.~\ref{fig_class}). 
\mbox{{\bf Type-C}:}  \mbox{filament formation}
induced by\, local\, velocity perturba\-tions within shock-compressed layers
\citep{PN99, CO14,GO15}.
{\bf Type-O}: Filaments formed at the convergent point of material flows within bent oblique MHD\index{MHD} shock fronts induced by the clumpiness of the medium \citep{IF13, VHF13}.
{\bf Type-G}: Gravity-induced fragmentation of sheet-like clouds \citep{TI83, NIM98, KW07}.
{\bf Type-S}: Shear flows associated with turbulence stretching existing
clumps, which  become elongated structures \citep{Hennebelle13,II16}. 
Filaments formed in this way generally have small line-masses, like  \hone   filaments.
{\bf Type-I}: Filaments formed at the intersection of two shock-compressed sheets \citep{PN99, PK13, MDS15}.
This mode works only in highly super-Alfv\'enic or unmagnetized turbulence.
Note that type-G filaments could also grow alongside type-C and type-O filaments within  shock-compressed layers after a 
few free-fall times of the sheet formation, and gravity-induced accretion is also seen in type-C and type-O filaments, as described in
the two-step, gravity-induced preferred-direction accretion picture by \cite{Chen20}.

\begin{figure}[ht!]
\centering
\includegraphics[width=0.975\columnwidth]{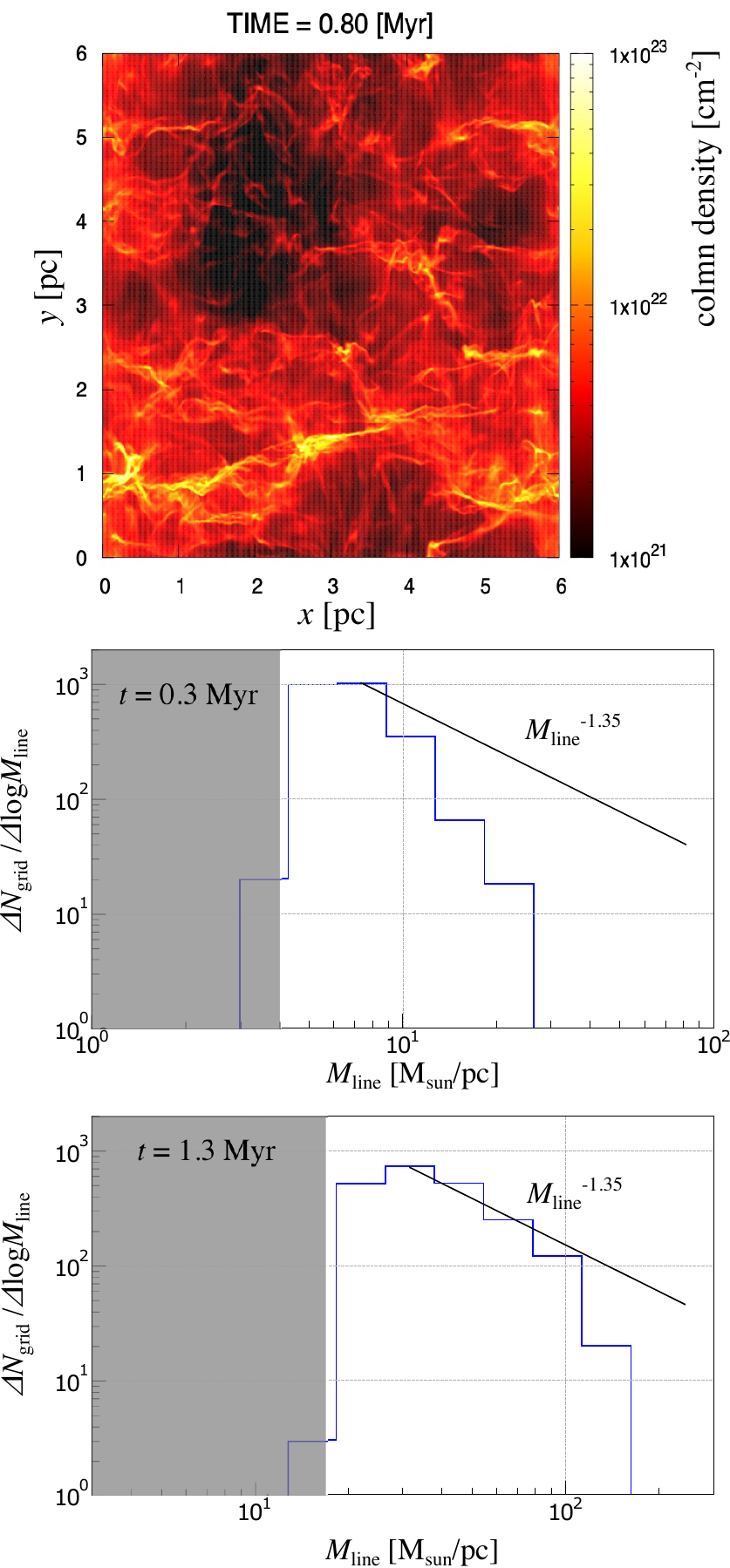}
 \caption{\small 
    Top: Face-on view of a simulated shock compressed layer by \citet{AIIM21}, where
    most major filaments are created by the type-O mechanism.
    Middle (bottom): Filament line mass function (FLMF)\index{Filament!Line Mass Function}  %
    at $t=0.3$ ($t=1.3$) Myr in the shocked layer.
    The grey regions show the range below the filament detection threshold.
}
\label{fig_simulaiton}
\end{figure}

By performing MHD\index{MHD} simulations of shock-compressed molecular clouds under various conditions (see e.g., 
top panel of Fig.~\ref{fig_simulaiton}), \cite{AIIM21} investigated which mechanism is primarily responsible 
for filament formation in shock compressed layers.
They found that the type-O mechanism accounts for the majority of filament formation, when the velocity of 
the shock that compresses the molecular cloud is higher than $v_{\rm cr}\sim$5\,\kms (dependent on the 
density and magnetization level). 
At shock velocities lower than $v_{\rm cr}$, type-C becomes the dominant  mechanism of filament formation.
Because of this dependence on the shock velocity, the type-O mechanism creates more massive filaments than the type-C. 
Pure type-G filament formation can only be clearly seen when the initial structure of the molecular cloud 
is featureless (no initial density fluctuations and weak turbulence).
It is important to note that while type-G formation is rarely dominant with realistic initial conditions, 
gravity-assisted accretion helps the growth of filaments formed via type-C/O and may collect 
separately-formed filaments into bundles.
Type-S filaments appear everywhere due to turbulence, but filaments formed by shear motions 
do not become dense enough to form stars,
and type-I is not clearly observed in a medium with a  realistic magnetization level.

\index{Filament!Line Mass Function|(}
Remarkably, the line-mass distribution of filaments (or FLMF)  created by both type-O and type-C 
mechanisms (cf. Fig.~\ref{fig_simulaiton})
resembles the observed, Salpeter-like FLMF \citep[][see Fig.~\ref{flmf_CMF}a  and discussion in Sect.~\ref{sec:FLMF} below]{Andre2019}.
The middle and bottom panels of Fig.~\ref{fig_simulaiton} show the FLMF obtained at two time steps 
in a simulation of the dominant (type-O) filament formation mechanism \citep{AIIM21}.
Throughout the evolution, smaller line-mass filaments always dominate in mass. 
Once the shock wave attenuates, many subcritical filaments stop evolving due to the finite type-O accretion duration 
and eventually disappear due to shear out or expansion. 
The type-O accretion duration depends on the scale of the pre-shock molecular clumps, 
which have a broad size distribution in turbulent molecular clouds. 
However, once a filament evolves into the supercritical regime, it is long-lived
and can continue to accumulate mass due to gravitational accretion (cf.  Sect.~\ref{sec:accretion}).
Since the type-O formation mechanism is induced by a high-velocity shock, the resulting filaments can become as massive 
as $\gtrsim 100$ M$_{\sun}$ pc$^{-1}$ on a timescale of a fraction of a Myr, 
which is reminiscent of recent works connecting high-velocity gas flows and massive star/cluster formation \citep{FHI21}.
\index{Filament!Line Mass Function|)}
\index{Sheet|)}\index{Shock!Shock-compressed layers|)}

\subsection{Synergy of Observations and Theory\label{sec:2.3}}

Numerical simulations of  molecular cloud formation out of the magnetized atomic medium stress 
the necessity of multiple compressions for the formation of the cold molecular 
medium \citep[e.g.,][and see above]{II09, II12, IIIH15}. 
Such multiple \mbox{compressions} from expanding \hone shells (see Fig.\,\ref{fig_HiShell_fil}) are suggested by  
observations  showing sheet-like extended structures connected in velocity (as seen in PV diagrams) 
to cylindrical filaments  \citep[][]{Arzoumanian2018-Filament_Nobeyama,Arzoumanian2022,Shimajiri_Taurus_2019}.
The sheet-like geometry of molecular clouds is also suggested by the column density structure of 
compressed shells  observed around  \hion  regions\index{\hion Region}. These shells  are better described by a 3D ring geometry 
rather than a spherical geometry \citep[see Fig.\,\ref{fig_HII} and][]{Zav20}.  
The 3D cylindrical structure of {\it Herschel} (star-forming) filaments 
is supported by the successful detection of dense  molecular gas tracers
\citep[such as \ntwohp, \ce{H^{13}CO+}, \ce{HC3N}, and \ammo, cf.,][]{LiGoldsmith2012}.
This favors a truly high density in these filaments, rather than  high column density sheet-like structures seen edge-on. 
In addition,  \citet[][]{Bonne2020} used radiative transfer modelling of two transitions of the \ce{^{13}CO} emission (3--2) and (2--1) 
observed with APEX and showed that  the Musca filament \citep[][]{Cox_Musca_2016} is a cold ($\sim$10\,K), 
dense ($n_{\ce{H2}} \sim10^4$\,\cc) velocity-coherent \citep[][]{Hacar_2016_Musca} 10\,pc long structure, 
which is best described as a $\sim$0.1\,pc-wide  cylindrical structure.

The theoretical classification of filament formation mechanisms helps in understanding some observations.
In the early filament formation stage of the type-O mechanism, a characteristic V-shaped structure can be seen in the 
position-velocity (PV) map across the filament, and a curved magnetic field around the filament can be observed.
Such signatures have been reported recently in observations of velocity structure 
\citep{Arzoumanian2018-Filament_Nobeyama,KTS20,Bonne2020}  and magnetic field structure \citep{Tahani2018,Tahani2019}. 
The different curvature of the velocity pattern in the PV maps (V-shaped or $\Lambda$-shaped) toward filaments in the 
same region may suggest different episodes of compressions \citep{Arzoumanian2022}. 
In this scenario of filament formation, one would expect to observe a population of subcritical, low-column-density filaments 
oriented perpendicular to the magnetic field (B-field).
This seems in contradiction with previous results derived from polarization data
\citep[e.g.,][]{planck2016-XXXII,Planck2016-XXXV,Palmeirim2013,Cox_Musca_2016,Soler2016} that
subcritical filaments almost always align parallel to the surrounding magnetic field lines on the plane-of-the-sky. 
This discrepancy may be due to the rapid transition of a newly formed, young filament from the thermally subcritical to the thermally 
transcritical/supercritical regime resulting from the fast  accretion of surrounding matter onto the evolving filament
\citep[$\sim$10\,M$_{\odot}$\,pc$^{-1}$ in $\sim$0.2\,Myr, cf.,][]{Arzoumanian2018-Filament_Nobeyama}. 
For statistical reasons, such short timescales may be difficult to observe.

In the type-O/C/G models, velocity gradients across the filaments are generated  
as a consequence of anisotropic gas accumulation from a flattened layer. 
This preferred-direction of gas accretion is a distinguishing kinematic feature of such a 
filament formation scenario, which has been reported in 
observations of both nearby star-forming clouds and distant IRDCs (see Fig.~\ref{velo_grad} and Sect.~\ref{sec:accretion} below).
Overall, the available observations seem to be consistent with a scenario in which dense molecular filaments form initially through the type-O 
or type-C mechanisms and subsequently grow in mass due to gravity-induced accretion. 
\cite{Chen20} proposed that, by comparing the ratio between the kinetic energy of the 
flow transverse to the filament and the gravitational potential energy of the filament gas, 
one could distinguish between filaments formed purely due to direct turbulent compression 
and those formed due to gravity-induced accretion (type-O/C/G). 
Following this theoretical argument, \cite{Dhabal_2019} proposed that the large velocity gradient 
observed across the south-east filament of NGC\,1333\index[obj]{NGC 1333} is due to the collision between a 
large-scale turbulent cell and the cloud, and that this filament is likely at the front-end of an expanding bubble.

The proposed type-C/O filament formation models  
are also compatible with statistical results on the relative orientation between  filament axis and  magnetic field  lines, 
as derived from observations of dust polarized emission with {\it Planck} and starlight polarization \citep{planck2016-XXXII,planck2016-XXXIII,Planck2016-XXXV,Palmeirim2013,Cox_Musca_2016,Soler2016,Wang2020ic5146}.
These studies show that  the plane-of-the-sky (POS) B-field orientation is mostly perpendicular 
to high column density supercritical filaments. 
Recently, higher sensitivity and angular resolution observations with SOFIA/HAWC+ and JCMT/POL2 
have revealed the complex but organized small-scale structure of the B-field within filament networks. 
While the relative orientation of the B-field may differ from one filament to another, 
the B-field often shows an ordered structure along a given filament \citep[e.g.,][]{Doi2020,PSLi2022}.
Toward hub-filament systems, the B-fields\index{Magnetic field} are observed 
mostly perpendicular to the filaments in their outer parts, away from the hubs. 
In the inner parts, where  filaments merge with the hubs, 
the POS B-field becomes mostly parallel  to the filaments 
\citep{Wang2020,Pillai2020,Arzoumanian2021}. 
These changes of the relative B-field orientation from perpendicular to 
parallel in the interior of dense filaments connected to hubs 
suggest a coupled evolution of the B-field and the filaments. 
Such a re-organization of the B-field along the filaments, due to local velocity flows of matter in-falling onto the hubs, 
is also suggested by MHD\index{MHD} simulations  \citep[e.g.,][]{Gomez2018}.

As mentioned above, the \hone  medium is itself filamentary, as seen in observations \citep{Clark2014,Clark2019} and simulations \citep{II16}. 
The conditions under which \hone  filaments, formed by thermal instability, may evolve into molecular star-forming filaments are still poorly known. 
However, since these \hone filaments are thermally stable and long-lived, they might be gathered and merged by type-O/C/G flows induced by 
other \hone shocks, and potentially evolve into molecular filaments as observations suggest (see Sect.\,\ref{sec:FilaCat}). 
Similarly, the shock wave associated with an \hion region\index{\hion Region} expanding in an already filamentary atomic and molecular
medium may induce the formation of new filaments, as well as reshape, compress, and gather  pre-existing filaments 
playing an important role in the evolution, star formation history, and properties of the cores formed from filament 
fragmentation (see Sect.\,\ref{sec:filament} below).
Future observational and theoretical  studies \citep{Haid2019}  are needed to refine our understanding of the 
possible evolutionary link between atomic and molecular filaments and the role of \hone and \hion  regions 
(i.e., formation and/or feedback) on the surrounding atomic and molecular filamentary ISM. 
\index{Bubble|)}\index{Filament|)}\index{Filament!Formation|)}

\section{EVOLUTION AND FRAGMENTATION OF DENSE MOLECULAR FILAMENTS\label{sec:filament}}
\index{Filament|(} 
\subsection{Inflow from Cloud to Dense Filaments\label{sec:accretion}}
\subsubsection{Evidence of Non-isotropic Inflow of Ambient Gas in Sheet-like or Bubble-like Parent Cloud Structures}

\begin{figure*}[th!]
\centering
\includegraphics[width=\textwidth]{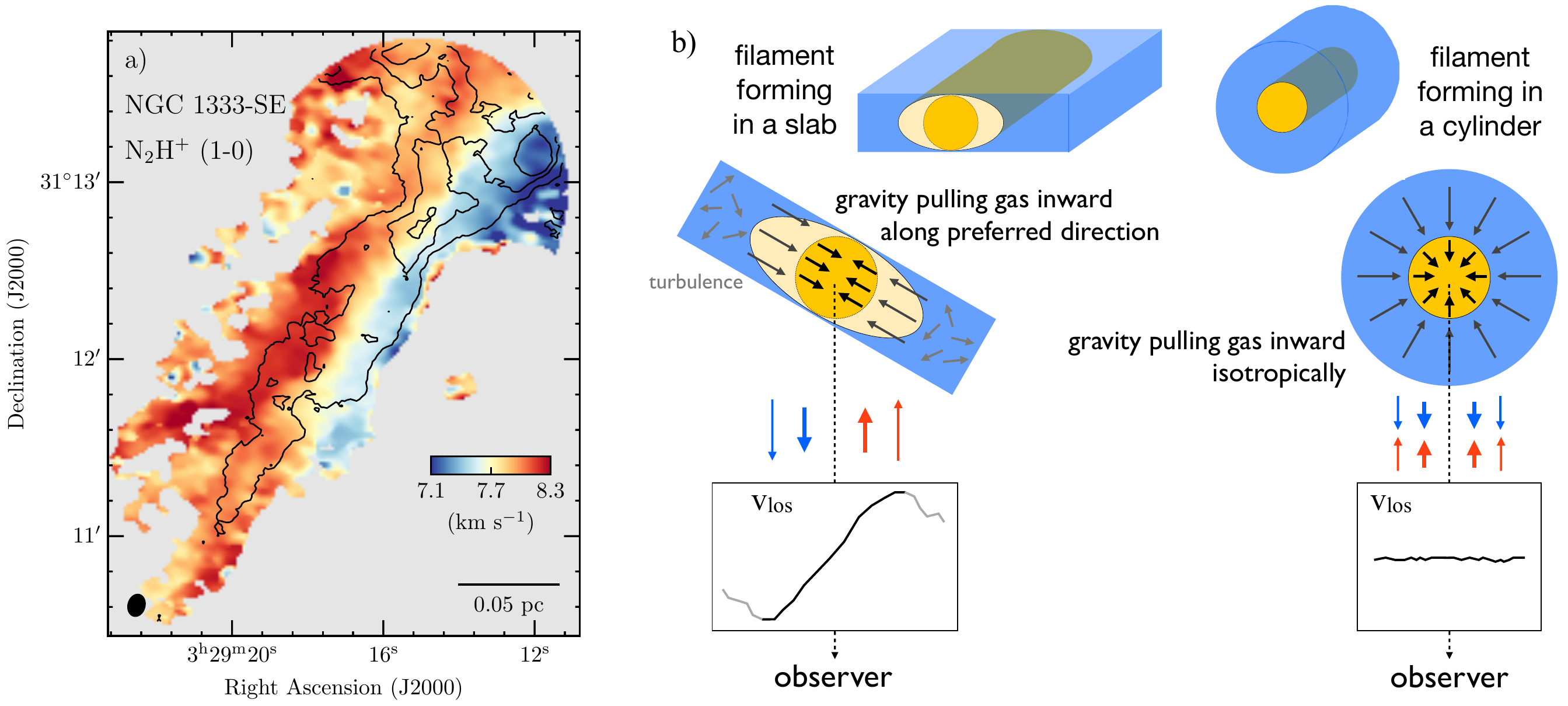}
 \caption{\small 
    (a) Example of a  consistent transverse velocity gradient observed in \ntwohp (1--0) across a filament in the 
    NGC~1333-SE\index[obj]{NGC 1333} region at $d \approx$300\,pc \citep[adapted from][]{Dhabal_2018}, which can be interpreted 
    as accretion in a sheet-like cloud as illustrated in (b) \citep[adapted from][]{Chen20}.
 }
 \label{velo_grad}     
\end{figure*}

In agreement with the scenario of filament formation favored 
in the previous section, molecular line studies of the projected velocity 
field observed within nearby filament systems support the view that dense star-forming filaments continuously grow 
through accretion\index{Filament!Accretion}  or inflow from flattened cloud layers.
Velocity gradients both along and perpendicular to the major axis 
of molecular filaments have been detected \citep[e.g.,][]{Peretto+06, Kirk_SerpensS_2013, Friesen+13, Shimajiri_Taurus_2019, M_Chen+20}, 
but the gradients across filaments 
often dominate in amplitude \citep[e.g.,][]{Fernandez-Lopez2014-CLASSY,Dhabal_2018}. 
Such transverse velocity gradients, sometimes seen 
consistently along the entire length of the filamentary structure \citep[][]{Beuther2015}, 
would not be observed if the filaments were embedded in (and isotropically accreting from) a spherical or cylindrical ambient cloud (see Fig.~\ref{velo_grad}b). 
Transverse velocity gradients which switch directions several times over the length of a filament 
are nevertheless consistent with \textit{anisotropic} cylindrical  inflow
\citep{Clarke+18}.

These transverse gradients may a priori arise from large-scale inflow, 
rotation, shearing motions, or a combination 
of these types of motions. However, large-scale solid-body rotation of filaments around their main axis 
at the level implied by the observed gradients is unlikely, as it would severely distort the filaments and 
their radial density structure \citep[][]{Recchi+14,GZhang+20}. 
Shearing motions can produce filamentary structures with transverse velocity gradients, but the resulting 
filaments are expected to be non-self-gravitating or subcritical \citep[][]{Hennebelle13}. 
Therefore, transverse velocity gradients across self-gravitating filaments 
are most readily explained if these filaments are forming and growing inside sheet-like or shell-like structures (see Fig.~\ref{velo_grad}b). 

Numerical MHD\index{MHD} simulations of filament formation within shock-compressed layers\index{Shock!Shock-compressed layers} generated by large-scale supersonic flows 
\citep[e.g.,][]{Inoue+18,Chen20} 
can reproduce this kinematic pattern (cf. model types C/O in Sect.~\ref{sec:bubble}). 
The observed transverse gradients appear to be roughly aligned with the large-scale magnetic field\index{Magnetic field} \citep[e.g.,][]{Palmeirim2013,Bonne2020}
and in some cases, such as Orion~A\index[obj]{Orion A Molecular Cloud} and Perseus\index[obj]{Perseus Molecular Cloud}, 
there are hints that the magnetic field has an  arc-shaped morphology in 3D 
\citep[e.g.,][]{Tahani2019,Tahani2022-Perseus_3D_Bfield,Tahani2022-OrionA_3D_Bfield}. 
This is consistent with a picture of magnetically-aligned inflow of matter
from a shell-like parent cloud.

\subsubsection{Models and Simulations of Accreting Filaments}
\index{Filament!Accretion|(} 
Accretion from inflowing gas in the surrounding medium is highly important to filaments as it provides an external 
pressure via ram pressure, as well as delivering mass and energy into the filament. 
The presence of an external pressure acts to confine filaments to finite radii and to produce shallower 
density profiles, $\rho(r)$, which are consistent with the observed logarithmic slopes $p \equiv -d\ln{\rho}/d\ln{r} \sim2$ 
\citep{Fis12}; without such a pressure filaments  extend to infinite radii and possess a steep $p=4$ density profile \citep{Ost64}.

Additionally, the effect that accretion has on filaments, due to their delivery of additional mass as well as kinetic energy, 
is profound in controlling filament evolution \citep{Heitsch13,Hei13b,Clarke_2016_filament,Smi16,Clarke_2017_filament,Heigl18,Clarke_2020_filament,Heigl20}. 
The key effect is that of accretion driven turbulence, 
where a fraction of the kinetic energy of the accreted material is converted into turbulent energy. 
Thus an accreting filament gains mass while simultaneously gaining in turbulent support, 
which may explain the observations that dense filaments are close to virial equilibrium \citep[cf.][]{Arzoumanian+13} 
and have a common half-power width $\sim$0.1\,pc (cf. Fig.~\ref{fil_width} and \S~\ref{sec:width} below). 
Moreover, this continuous input of energy and mass leads to the conclusion that hydrostatic equilibrium may 
never be achieved in filaments, unless the accretion timescale is considerably longer 
than the filament crossing time. Accretion also has a considerable impact on filament fragmentation 
into cores as discussed later.
\index{Filament!Accretion|)}

\subsection{Filamentary Substructures}\index{Filament!Substructure|(}
\subsubsection{Observational Evidence}

A frequent feature of many dense molecular filaments is the presence of significant substructures 
observed in the form of velocity-coherent features, called {\it fibers}. 
The presence of fibers was first reported by \citet{Hacar_fiber_2013}  
in the Taurus B211/B213\index[obj]{B211/B213} filament ($d\sim$140 pc), for which a friends-of-friends algorithm in velocity (FIVE)  
was used to identify at least 20 velocity-coherent components in \ntwohp and \CeightO.
Subsequently, similar velocity-coherent components were 
detected in \ntwohp in other regions, 
including the IRDC G035.39-00.33\index[obj]{G035.39-00.33} \citep{Henshaw+14}, 
the NGC~1333\index[obj]{NGC 1333} protocluster \citep{Hacar+17}, 
IRDC G034.43+00.24\index[obj]{G034.32+00.24} \citep{Barnes+18}, 
the Orion A\index[obj]{Orion A Molecular Cloud} integral-shaped filament \citep{Hacar2018-ALMA_Orion}, 
and the NGC~6334\index[obj]{NGC 6334} main filament \citep[][see Fig.~\ref{ngc6334}b below]{Shimajiri+19b}. 
The velocity-coherent substructures identified in NGC~1333 and Orion~A 
are well separated in the plane of sky, however, and may differ in nature 
from those observed in  Taurus and  NGC~6334 
which are intertwined. 
Moreover, not all molecular filaments consist of multiple fiber-like substructures. 
The Musca filament, for instance, is a 6-pc-long velocity-coherent filament with much less 
velocity substructure than the Taurus B211/3 filament, 
and may be interpreted as a single-fiber system
\citep{Hacar_2016_Musca,Cox_Musca_2016}.

Interestingly, most of the line-identified fibers can also be detected in {\it Herschel} dust continuum maps   
when large-scale emission is filtered out, enhancing the contrast of small-scale structures in the data 
\citep[cf.][]{Menshchikov_getfilaments_2013,Andre2014-PPVI}.
Moreover, the {\it Herschel} data suggest that these fibers are somehow 
linked to fainter, magnetically-aligned striations often observed around the main filaments,
almost perpendicular to their long axis. 
In the Taurus B211/B213\index[obj]{B211/B213} and Musca\index[obj]{Musca} filaments, for instance, hair-like strands or spur-like features, 
which appear to be the tips of larger-scale striations, are visible in the immediate vicinity of the filaments, 
attached to their main body \citep[][]{Palmeirim2013,Cox_Musca_2016,Bonne2020}. 
This is suggestive of a direct physical connection between striations and fibers, and 
is consistent with the observed striations tracing accretion flows onto the 
main filaments, possibly influencing the formation of  their fiber-like substructure.

\subsubsection{Physical Origin}

Numerical simulations have shown that the presence of fiber-like substructures 
is most likely linked to accretion flows onto the main filament. 
Both \citet{Smi16} and \citet{Clarke_2017_filament} show that accretion\index{Filament!Accretion}  from an inhomogeneous and 
turbulent medium leads to the formation of small substructures within larger parent filaments, 
albeit via two distinct mechanisms. 
\citet{Smi16} show that the turbulence within a larger cloud leads to numerous small 
filaments forming independently of each other which are subsequently swept together into a 
single filament or accreted onto already formed filaments. This mechanism is termed 
\textbf{fragment and gather}. 
Meanwhile, \citet{Clarke_2017_filament} show that the turbulence driven by accretion 
of clumpy material leads to numerous coherent shocks and a strong vorticity field in 
a filament. This driven turbulence leads to the formation in situ of substructures similar to the 
\textbf{fray and fragment} mechanism proposed by \citet{Tafalla_Hacar_2015}.  
It is currently unclear which formation mechanism for filamentary substructures is dominant 
and in which scenarios.

The substructures formed in simulations are reminiscent of the velocity-coherent 
fiber structures seen in observations \citep[e.g.,][]{Hacar_fiber_2013}; although it must be noted that synthetic observations \citep{Clarke+18} 
and purely numerical works \citep{Moe15,Zam17} show that projected structural and velocity coherence are insufficient 
conditions for a real filament and high levels of caution must be applied when dealing with fibers. 
\index{Filament!Substructure|)}

\subsection{The Common Width of Molecular Filaments}
\label{sec:width}\index{Filament!Width|(} 
\subsubsection{Observational Evidence}

The  filamentary structures detected with {\it Herschel} span broad ranges in length, 
central column density, and mass per unit length \citep[e.g.,][]{Schisano+14,Arzoumanian2019-Filament_properties}.
In contrast, detailed analysis of the radial column density profiles 
indicates that, at least in nearby ($d <$450\,pc) molecular clouds, 
 {\it Herschel} filaments are characterized 
by a narrow distribution of half-power widths with a typical 
value of $\sim$0.1\,pc and a dispersion of less than a factor of 2  
when the data are averaged over the filament crests  
\citep[][]{Arzoumanian+11,Arzoumanian2019-Filament_properties}. 
In particular, it is remarkable that molecular filaments appear to share approximately 
the same inner width in the {\it Herschel} data
regardless of their mass per unit length $M_{\rm line}$, 
whether they are thermally {\it subcritical} with $M_{\rm line}   \lesssim 0.5\, M_{\rm line, crit}$,  
{\it transcritical} with  $0.5\, M_{\rm line, crit} \simlt M_{\rm line} \simlt 2\, M_{\rm line, crit}$, 
or  {\it thermally supercritical} with $M_{\rm line}  \simgt 2\, M_{\rm line, crit}$, 
where $M_{\rm line, crit} = 2\, c_s^2/G $ is the thermal value of the critical mass per unit length \citep[e.g.,][]{Ost64}, 
i.e., $\sim$16\,$M_\odot \, {\rm pc}^{-1} $ for a sound speed $c_{\rm s} \sim$0.2\,\kms 
or a gas temperature \hbox{$T \approx$10\,K}.

Independent submillimeter continuum studies of filament widths in nearby clouds 
have generally confirmed this result 
\citep[e.g.,][]{AlvesdeOliveira+2014,Koch_FilFinder_2015,Salji+15, Rivera-Ingraham+16}.
Measurements of filament widths obtained in molecular line tracers 
\citep[e.g.,][]{Pineda+11, Fernandez-Lopez2014-CLASSY, Panopoulou+14, Hacar2018-ALMA_Orion, Monsch2018-NH3_OrionA_VLA, Schmiedeke2021-Filament_Properties}
have been less consistent with the {\it Herschel} dust continuum results, however.  
For instance, using \ce{^{13}CO} emission, \citet{Panopoulou+14} found a broad distribution of widths in Taurus, with a peak of 0.4~pc. 
\citet{Hacar2018-ALMA_Orion} found a median width of 0.035~pc for Orion ``fibers'' 
in the integral-shaped filament of Orion\index[obj]{Orion A Molecular Cloud}, 
combining \ntwohp ALMA and IRAM 30-m observations. 
Likewise, \citet{Schmiedeke2021-Filament_Properties} found a typical width of 0.03~pc along two filaments 
in Perseus-B5\index[obj]{Barnard 5} using combined VLA and GBT \ammo\ observations.
These differences can be attributed to the lower dynamic range of  densities 
sampled by observations in any given molecular line tracer compared to dust observations 
\citep[][]{Priestley+20,Shimajiri_2023_OrionB}. 
More specifically, \ce{^{12}CO} or \ce{^{13}CO} data only trace low-density gas 
and cannot reliably measure the whole (column) density 
profile (hence the width) of a dense molecular filament. 
Likewise, commonly observed transitions of \ntwohp or \ammo
only trace relatively high density gas 
(typically above the effective densities $n_{\rm efff}$ of the transitions --  \citealp[cf.][]{Shirley_2015}) 
and cannot reliably measure the whole profile of a filament either. 
In contrast, submillimeter dust continuum images obtained from space 
(with {\it Herschel}) achieve a significantly higher dynamic range and are sensitive to 
both the low density outer parts and the dense inner parts of filaments. 
While some changes in dust properties (emissivity and temperature) 
may occur in the interior of the densest filaments and affect the width measurements 
obtained in the dust continuum \citep[cf.][]{Schmiedeke2021-Filament_Properties}, 
these effects appear to be rather modest \citep[cf.][]{Schuller+21}.
For moderately dense filaments, the most reliable molecular gas tracer seems to be
the \CeightO (1--0) line which results in 
widths approximately around 0.1 pc, in agreement 
with the dust continuum findings in both simulations and observations, 
albeit with a somewhat broader spread of values 
\citep{Clarke+18,Suri+19,Orkisz+19}.

\begin{figure}[htb!]
\centering
\includegraphics[width=\columnwidth]{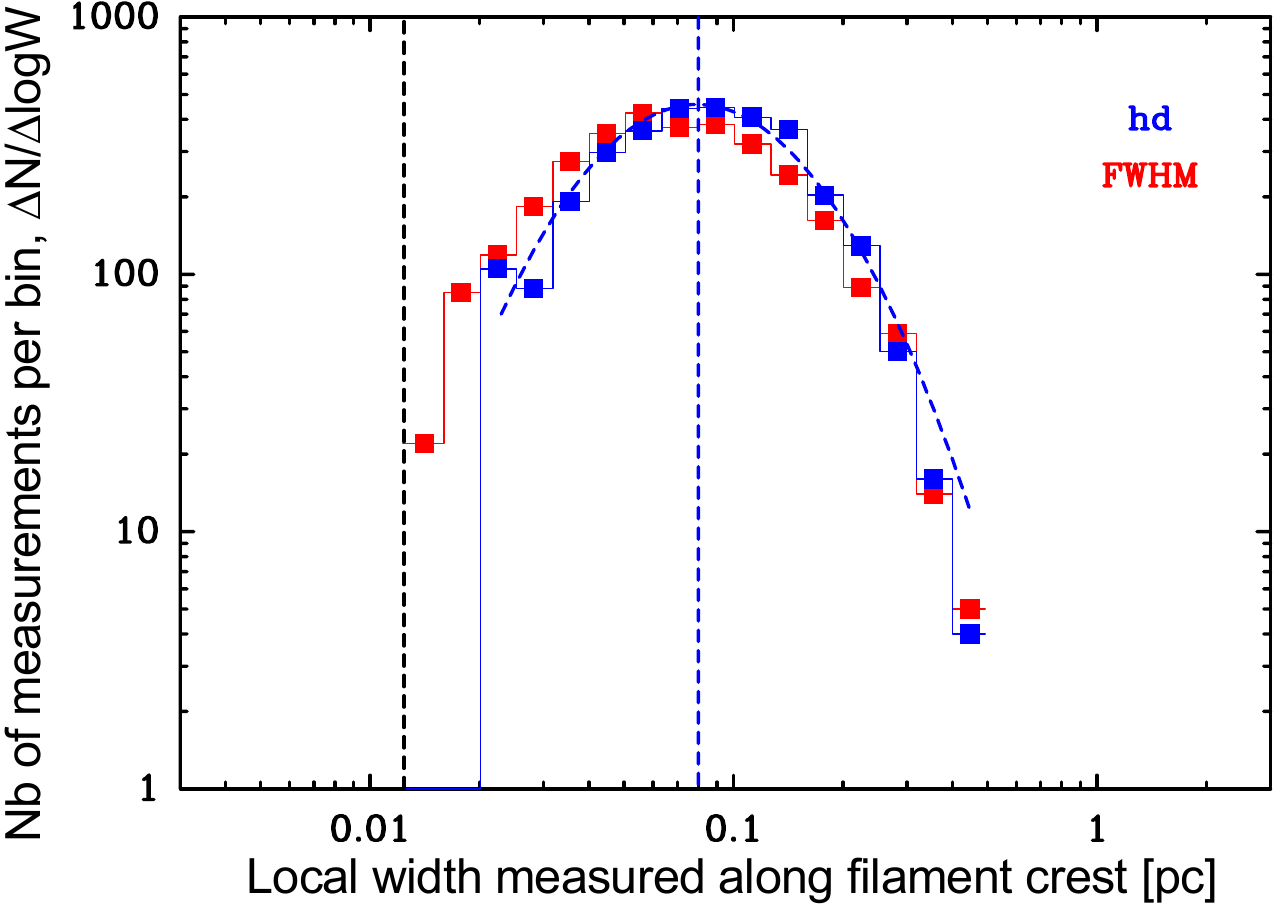}
 \caption{\small 
    Distribution of local FWHM widths derived from more than 3000 independent radial profile 
    measurements along the crests of 205 {\it Herschel} filaments  in three nearby molecular 
    clouds (Taurus, Ophiuchus, Pipe) at $d \sim 140\,$pc \citep[adapted from][]{Arzoumanian2019-Filament_properties}. 
    The blue and red histograms correspond to two methods of estimating the FWHM width.
    The dashed blue curve is a lognormal fit to the blue histogram, which peaks at the median 
    width $\sim$0.08\,pc (vertical blue dashed line). 
    The black dashed line marks the {\it Herschel} resolution $\sim$0.012\,pc in these regions.
}
 \label{fil_width}     
\end{figure}

Some concerns have been raised about the reliability of the filament widths 
found with {\it Herschel} \citep[][]{Smith+14,Panopoulou+17,Panopoulou+22}.
In particular, \citet{Panopoulou+17} pointed out an apparent contradiction between the existence of a characteristic filament width 
and the essentially scale-free nature of the power spectrum of interstellar cloud images 
(well described by a single power law from $\sim$0.01\,pc to $\sim$50\,pc; \citealp{Miville-Deschenes+10,Miville-Deschenes+16}). 
However, \citet{Roy+19} showed that there is no contradiction 
given the only modest area filling factors ($\simlt$10\%) 
and column density contrasts ($\leq$100\% in most cases) 
derived by \citet{Arzoumanian2019-Filament_properties} for the filaments seen in {\it Herschel} images. 
For realistic filament filling factors and column density contrasts, filamentary structures contribute  
only a negligible fraction of the image power spectra.

Another caveat pointed out by \citet{Panopoulou+17,Panopoulou+22} is the potential presence of systematic biases in filament width 
measurements, especially those based on Gaussian fits. Indeed, the radial 
density profiles of molecular filaments often feature pronounced power-law wings with logarithmic slopes in the range $1.5 \leq p \leq 2.5$ 
and tend to be better represented by Plummer-like profiles than by Gaussian distributions \citep[e.g.,][]{Palmeirim2013}. 
As it is well known that the FWHM size of a power-law (scale-free) distribution convolved with a Gaussian beam is always slightly 
larger than the beam size \citep[e.g.,][]{Ladd+1991}, the role of observational resolution must be considered with care. 
For cylindrical power-law density distributions with indices $1.5 \leq p \leq 2.5$, 
the apparent FWHM width is expected to be $\sim$5--90\% larger than the beam size. The Hi-GAL filaments 
analyzed by \citep{Schisano+14} have observed FWHM widths only $\sim$90\% broader than the {\it Herschel} 500$\,\mu$m beam size, 
consistent with power-law wings with $p \sim 1.5$ and either unresolved or non-existent flat inner density profiles. 
However, the {\it Herschel} 500$\,\mu$m beam size ($36.3\arcsec$) corresponds to $\sim$0.2--0.5\,pc at $d \sim$1--3\,kpc 
and is not sufficient to resolve an inner width of $\sim$0.1\,pc at the typical distances of Hi-GAL filaments. 
In contrast, the observed half-power widths of the nearby ($d <$450\,pc) filaments studied by \cite{Arzoumanian2019-Filament_properties} 
are a factor $\sim$2.5--8 broader than the {\it Herschel} 250$\,\mu$m beam size ($18.2\arcsec$), 
inconsistent with scale-free density profiles \citep{Andre2022-Filament_Width_Test}.
Moreover, 
performing numerous tests using synthetic data, \citet{Arzoumanian2019-Filament_properties} showed 
that their method of measuring filament profiles and widths was reliable and free of significant biases, 
at least when the contrast of the filaments over the local background 
exceeds $\sim$50\%, which is the case for $\sim$70\% of the {\it Herschel} filament population they measured 
and $>$80\% of star-forming filaments. 
These tests suggest
that the typical half-power width $\sim$0.1\,pc obtained through the analysis of the 
radial profiles of nearby {\it Herschel} filaments is robust \citep{Andre2022-Filament_Width_Test}.
Nevertheless, there are sometimes large local deviations, by a factor of $\simgt$2--4, 
from the typical width
\citep[e.g.,][]{Juvela+12, Ysard+13}. 

\begin{figure}[ht!]
\centering
\includegraphics[width=0.8\columnwidth]{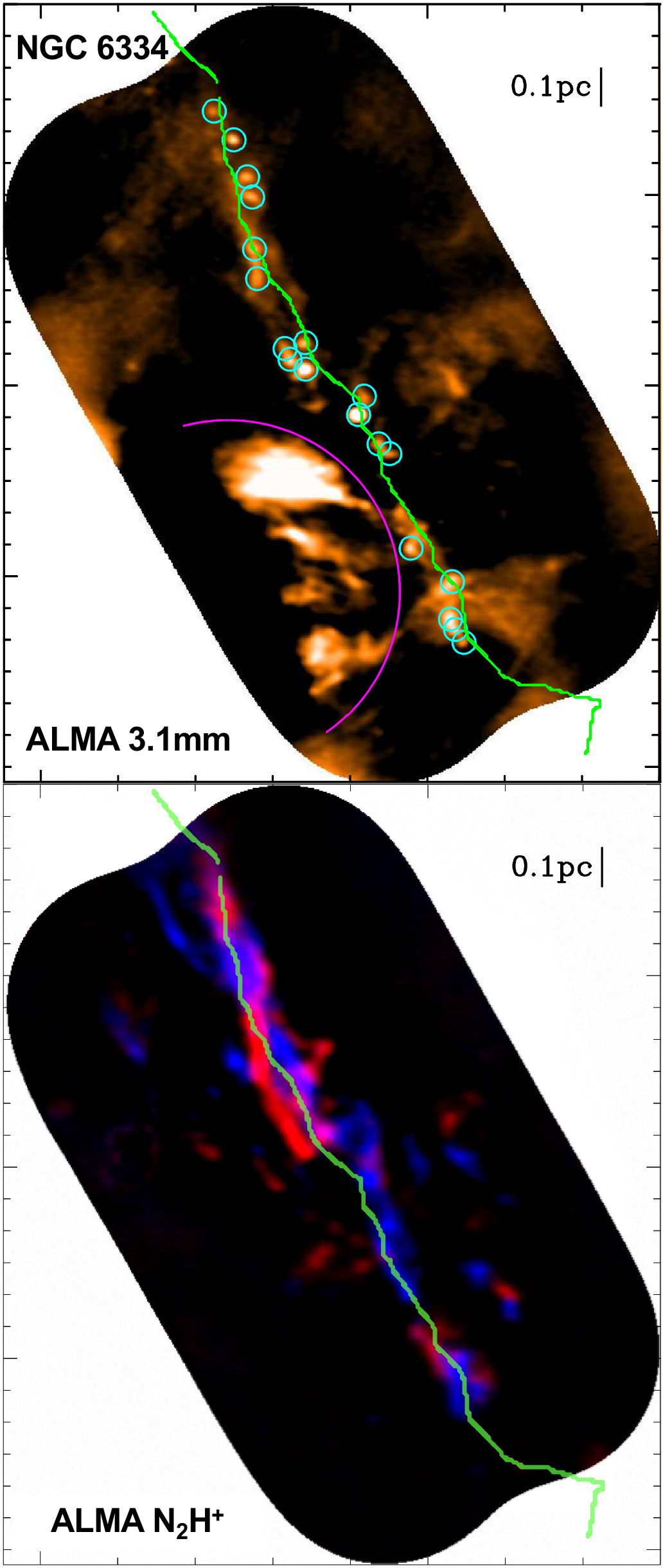}
 \caption{\small 
    Dense cores\index{Dense Core} ({\it top}) and intertwined fiber-like substructures  ({\it bottom}) 
    as viewed by ALMA  in the massive NGC~6334\index[obj]{NGC 6334} filament with 
    $M_{\rm line} \sim 500\, M_\odot \, {\rm pc}^{-1}$  \citep[][]{Shimajiri+19b}.
    The top panel is a 3.1~mm dust continuum map. 
    The bottom panel shows an overlay of 
     two \ntwohp (1--0) maps, integrated over two distinct velocity ranges separated by 1.1~\kms. 
     Green curves show the crest of the filament as traced by the DisPerSE algorithm in the APEX/ArT\'eMiS 
     350\,$\mu$m dust continuum map of the region \citep{Andre+16}. 
     The purple curve in the top panel marks the boundary of a compact \hion  region, 
     whose expansion may be compressing the filament from the side.
 }
 \label{ngc6334}
\end{figure}

The distribution of local widths found by \citet{Arzoumanian2019-Filament_properties} 
for 205 molecular filaments in Taurus\index[obj]{Taurus Molecular Cloud}, 
Ophiuchus\index[obj]{Ophiuchus Molecular Cloud}, Pipe\index[obj]{Pipe Molecular Cloud} (at $d \sim$140\,pc)
prior to averaging along the filament crests is well described by a lognormal 
function centered at $0.08\pm 0.01\,$pc 
with a standard deviation of $0.3\pm0.02$ dex, 
corresponding to a factor of $\sim$2 on either side of the median width 
(see Fig.~\ref{fil_width}).
A very similar distribution of individual half-power widths was recently reported 
by \citet{Schuller+21} for the Orion integral-shaped filament\index[obj]{Orion A Molecular Cloud}. 
\citet{Schuller+21} further show that both the choice of the filament-tracing algorithm 
and the effect of uncertainties in dust properties have only a small influence on 
the derived column density profiles, hence the estimated filament widths. 
The median half-power diameter $\sim$0.1\,pc measured with {\it Herschel} 
therefore appears to reflect the presence of a true common scale in the filamentary structure of 
molecular clouds. Further high-resolution submillimeter continuum studies are nevertheless required to confirm  
that the same common width also holds in clouds beyond a distance of 
$\sim$$500$\,pc \citep[see][]{Andre+16, Schuller+21}. 
Although no clear trend has been found concerning a possible time evolution of the filament width 
in a statistically significant sample of filamentary structures, tentative evidence of a weak correlation between local width 
and central density has been reported for a set of transverse cuts observed toward two Perseus filaments\index[obj]{Barnard 5} 
\citep{Schmiedeke2021-Filament_Properties}. This type of study also needs to be followed-up.

\subsubsection{Theoretical Interpretation}

Without extra support, isolated thermally supercritical filaments should 
undergo fast radial contraction on a free-fall timescale \citep{Inutsuka_Miyama_1992_filament}. 
However, there exists a population of `wide', supercritical filaments where $M_{\rm line}  \gg  M_{\rm line, crit}$, such as 
the filament shown in Fig.~\ref{ngc6334} which has $M_{\rm line} \sim 500\, M_\odot \, {\rm pc}^{-1}$ and 
a half-power width $\sim$0.15\,pc, suggesting that they cannot be so transient. 
There currently exists two categories of explanation for this: accretion driven turbulence and magnetic fields.\index{Filament!Accretion}

\index{Magnetic field|(}
\citet{Heitsch13} and \citet{HennebelleAndre13} propose that accretion driven turbulence may delay 
the moment of observational collapse and explain the observational decorrelation 
between filament width and peak column density \citep{Arzoumanian2019-Filament_properties}. 
However, it cannot prevent radial gravitational collapse.
This is supported by the simulations performed by \citet{Heigl20} which show that unmagnetized accretion 
driven turbulence cannot contribute to the stability of the filament by increasing the critical line-mass as 
the turbulent pressure does not possess a radial profile, i.e. it cannot provide a supporting force, 
but does act to widen a filament. 
\citet{Clarke_2017_filament} show that this behavior changes when the accretion flow is highly inhomogeneous; 
the anisotropic, clumpy nature of the accreted material produces both sub-filament sub-structure 
and local regions of rotation. They show that this allows a thermally supercritical filament (1.3$\times M_{\rm line, crit} $) 
to remain wide and have a diameter of 0.12\,pc. \citet{Clarke_2020_filament} show increasingly 
unstable filaments ($\sim$3$\times M_{\rm line, crit} $) with comparably wide diameters. 
However, these simulations did not include filaments with $M_{\rm line}  \gg  M_{\rm line, crit}$, 
and so it is currently unclear to what line-mass a filament thus supported could be maintained.

Support from magnetic fields has been a natural answer to the question of supercritical filament widths 
\citep{Nag87,Fie00,Tom14,Seifried_2015_filament,Inoue+18}. 
Observations indicate that some filaments are magnetically transcritical \citep[][in this volume]{PSLi2022,Pattle+2022}, 
suggesting the magnetic field may indeed play a significant role.
Due to the anisotropy of a filament 
the magnetic field direction is the dominant decider of its role in support. 
Magnetic field configurations which have the field parallel to the filament's axial axis are highly effective at 
providing support against gravity as radial collapse requires gas movement perpendicular to the field lines.
As noted above, this is rarely observed for the {\it ambient} magnetic field in the immediate vicinity of dense filaments. 
There is a hint, however, from recent high-resolution polarization observations with ALMA and SOFIA \citep{Dallolio+19,Pillai2020} 
that there may be a transition from perpendicular field to parallel field orientation in the {\it interior} of some supercritical filaments. 
More systematic high-resolution polarization studies will be key to assessing whether this is a generic trend.

Perpendicular magnetic field configurations have been shown semi-analytically to provide support and may increase 
the critical line-mass to an arbitrarily large value depending on the field strength \citep{Tom14,KasTom21}. 
This has not been seen in some simulations designed to investigate the impact of the magnetic field direction \citep{Seifried_2015_filament}, 
but simulations considering the oblique-shock formation mechanism for filaments \citep{Inoue+18} appear to 
be consistent with the result of \citet{Tom14}. 
The apparent disagreement may be due to the relatively weak field strength considered in the \citet{Seifried_2015_filament} simulations. 
Further numerical work focused on this question would be highly beneficial. 
It should also be noted that the high magnetic field strength necessary to provide radial support to a filament when the field is perpendicular, 
may also curb filament fragmentation\index{Filament!Fragmentation}  and core formation \citep{Han17}, which is in tension with
the numerous highly fragmented super-critical fila\-ments seen in observations (for example Fig. \ref{ngc6334}). 
When considering more complicated helical fields, one finds that the ratio of toroidal and poloidal components 
may increase or decrease the critical line-mass \citep[see][for more details]{Fie00}. 
It is thus of considerable interest to observationally determine the magnetic field geometry 
within the $\sim$0.1\,pc filament width to better constrain theoretical works invoking supporting magnetic fields.

Both explanations, accretion driven turbulence\index{Filament!Accretion} and magnetic fields, show promise in answering the question of 
additional support; however, neither explanation is currently conclusive. Further work focused on this question is 
necessary, with particular attention paid to the limits of these support mechanisms as well as the combination 
and interaction of the two mechanisms.
\index{Filament!Width|)} \index{Magnetic field|)}

\subsection{The Filament-Core Connection\label{sec:fil_core}}
\index{Dense Core|(} 

Thanks to the high surface-brightness sensitivity and spatial dynamic range achievable from space, 
a big step forward with {\it Herschel} imaging surveys compared to earlier submillimeter ground-based observations 
has been the ability to {\it simultaneously} probe compact structures such as dense cores {\it and} larger-scale structures 
within the parent clouds such as filaments. This has provided, for the first time, an unbiased view of both the spatial 
distribution of dense cores and the link between dense cores and the texture of molecular clouds. 
In particular, {\it Herschel} Gould Belt Survey (HGBS) observations have shown that most (75\%$_{-\,5\%}^{+15\%}$) prestellar cores 
are located within filamentary structures of typical column densities $N_{\htwo} \simgt 7 \times 10^{21}\, {\rm cm}^{-2}$,  
corresponding to visual extinctions $A_{\rm V} \simgt$7\,mag \citep[e.g.,][]{Andre+10,Konyves_2015,Marsh_2016_HerschelL1495cores}. 
Moreover, most prestellar cores lie very close to the crest of their parent filament 
\citep[e.g.,][]{Bresnahan_2018_HerschelCrA,Konyves_OrionB_2020, Ladjelate+20}, 
that is within the flat  inner $<$0.1\,pc portion of the filament radial profile \citep[cf.][]{Arzoumanian+11,Arzoumanian2019-Filament_properties}.

\begin{figure}[hb!]
\centering
\includegraphics[width=\columnwidth]{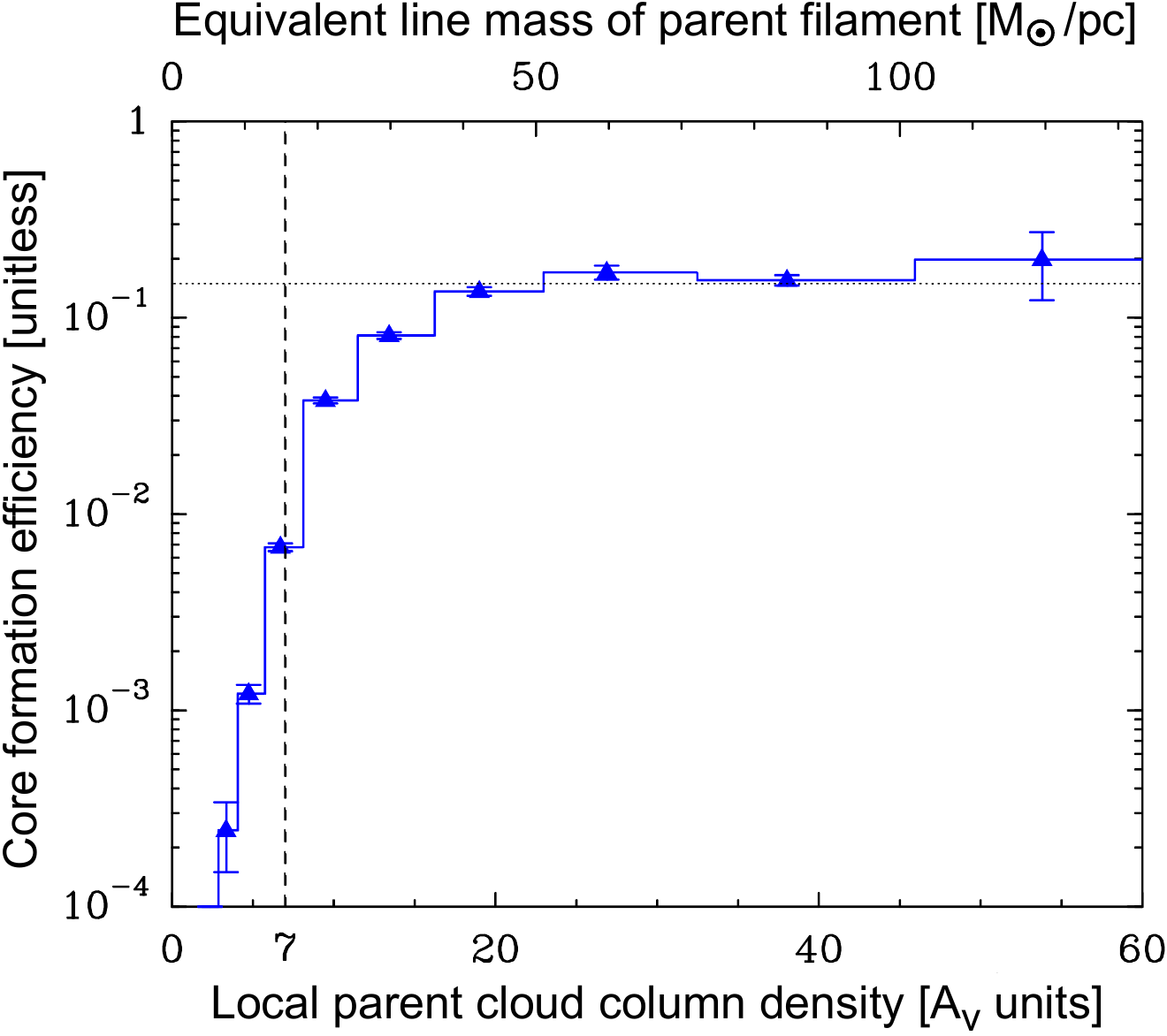}
 \caption{\small 
    Prestellar core formation efficiency 
    [$ {\rm CFE}(A_{\rm V}) = \Delta M_{\rm cores}(A_{\rm V})/\Delta M_{\rm cloud}(A_{\rm V}) $]
    in the Aquila molecular cloud (blue histogram with error bars)
    as a function of background column density expressed in $A_{\rm V}$ units
    using the simple conversion $N_{\htwo} ({\rm cm}^{-2}) = 0.94 \times 10^{21}\, A_{\rm V}$ \citep{Bohlin+78}. 
    The latter is dominated by the column density of the parent filaments 
    at high $A_{\rm V}$. (From \citealp{Konyves_2015}.)
}
 \label{aquila_cfe}     
\end{figure}

The column density transition above which prestellar cores are found in filaments is quite pronounced. 
Indeed, the observed core formation efficiency, defined as 
$ {\rm CFE_{obs}}(A_{\rm V}) = \Delta M_{\rm cores}(A_{\rm V})/\Delta M_{\rm cloud}(A_{\rm V}) $
where $\Delta M_{\rm cores}(A_{\rm V})$ is the mass in the form of prestellar cores 
in a given bin of background $A_{\rm V}$ values and $\Delta M_{\rm cloud}(A_{\rm V})$ is the total cloud 
mass in the same bin, 
resembles a smooth step function when plotted against  the ``background'' column density 
of the parent filaments
\citep[cf. Fig.~\ref{aquila_cfe} and][]{Konyves_2015}. 
There is a natural interpretation of this column density transition for prestellar core formation 
in terms of simple theoretical expectations for the gravitational instability of nearly isothermal gas cylinders. 
Adopting the typical inner width $W_{\rm fil} \sim$0.1\,pc measured  for nearby molecular filaments 
with {\it Herschel} \citep[][]{Arzoumanian+11,Arzoumanian2019-Filament_properties} 
and using the relation $M_{\rm line} \approx \Sigma_0 \times W_{\rm fil}$ between the central 
gas surface density $\Sigma_0$ and the mass per unit length $M_{\rm line}$ of a filament, 
there is a very good match between 
the transition at $A_{\rm V}^{\rm back} \sim$7\,mag or $\Sigma_{\rm gas}^{\rm back} \sim $150\,$M_\odot \, {\rm pc}^{-2} $ 
and the critical mass per unit length 
$M_{\rm line, crit} = 2\, c_{\rm s}^2/G \approx 16\, M_\odot \, {\rm pc}^{-1} $  
of isothermal long cylinders in hydrostatic equilibrium for a sound speed $c_{\rm s} \sim$0.2\,\kms, i.e., 
a typical gas temperature $T \approx$10\,K \citep{Ost64}. 
Therefore, the observed column density transition essentially corresponds to thermally transcritical filaments 
with line masses within a factor of $\sim$2 of $M_{\rm line, crit}$, which are 
prone to gravitational fragmentation along their length \citep{Inutsuka_Miyama_1992_filament, Inutsuka_Miyama_1997_filament, Fis12}.\index{Filament!Fragmentation}

It is important to stress that it is the {\it local} value of the mass per unit length which matters for the fragmentation 
of a particular filament segment, and that the above-mentioned transition does not correspond to a perfectly sharp threshold. 
While most prestellar cores appear to form in filaments that are locally transcritical or supercritical 
($M_{\rm line} \geq M_{\rm line, crit}/2 $), examples of prestellar cores in globally subcritical 
(but locally transcritical) 
filaments have been observed in regions such as Lupus, Cepheus, and Perseus \citep[][]{Benedettini+18, How19, DiFrancesco_HGBS_2020,Pezzuto_HGBS_2021}. 
This is supported by simulations which show that a locally supercritical line-mass is a sufficient 
condition for fragmentation \citep{Chira18}.

\subsection{Core Separations Along Molecular Filaments} 
\index{Dense Core!Separation|(}
\subsubsection{Observations of Core Spacings} 

\begin{figure}[htb!]
\centering
\includegraphics[width=\columnwidth]{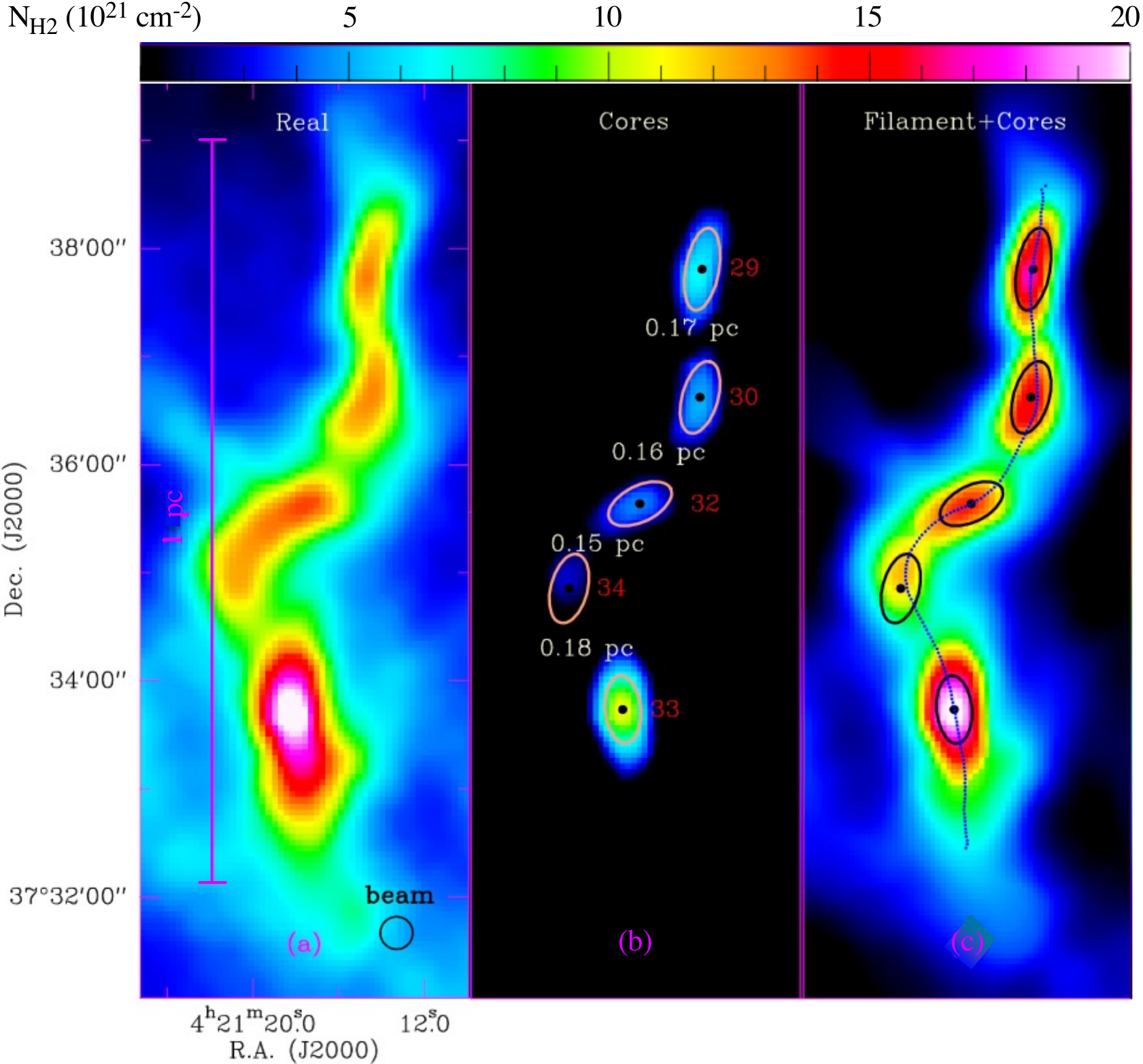}
 \caption{\small 
Quasi-periodic separations of dense cores along a transcritical filament in the 
California molecular cloud\index[obj]{California Molecular Cloud} \citep[from][]{GZhang+20}. 
From left to right, the panels display the original column density map of the filament as derived 
from {\it Herschel} data at $18.2\arcsec$ resolution, 
and two filtered versions of this map emphasizing 
the dense cores identified along the filament with \texttt{getsources} \citep[][]{mensh_2012_getsources}, 
and the cores overlaid on the filament (after subtraction of the non-filamentary background), respectively.
}
 \label{xshape_fil}
\end{figure}

The observed separations of dense cores along filaments are not consistent with the predictions of 
standard semi-analytic cylinder fragmentation models without turbulence or magnetic fields. 
Linear fragmentation models for infinitely long, isothermal equilibrium  cylinders 
indeed predict a characteristic core spacing of $\sim$\,4$\times$ the filament width \citep[e.g.,][]{Inutsuka_Miyama_1992_filament}. 
In contrast, the \mbox{spacing} observed between {\it Herschel} prestellar cores is generally not periodic 
and the median value of the projected core separation is found to be close to the typical $\sim$0.1\,pc 
half-power width of the parent filaments \citep[e.g.,][]{Andre2014-PPVI,Konyves_OrionB_2020}. 
A few good examples of quasi-periodic chains of dense cores have also been found (see Fig.~\ref{xshape_fil}, 
\citealp{Tafalla_Hacar_2015}, and \citealp{GZhang+20}), 
but again the corresponding characteristic spacing appears to be comparable to, rather than $\sim$\,4$\times$ larger than, 
the diameter of the parent filament.
Moreover, complementary high-resolution studies with interferometers 
\citep[][see Fig.~\ref{ngc6334}a]{Takahashi+13,Teixeira+16,Kainulainen+13,Kainulainen+17,Shimajiri+19b} 
have provided some evidence of two distinct fragmentation modes within at least some thermally supercritical filaments: 
a) a ``cylindrical'' fragmentation mode corresponding to clumps or groups of cores 
with a separation consistent with $\sim$\,4$\times$ the filament width taking projection effects into account; 
\mbox{and b)} a ``spherical'', Jeans-like fragmentation mode corresponding to a typical spacing $\simlt$0.1\,pc 
between cores (and within groups). 
This discrepancy between observations and simple theoretical predictions may be understood
by realizing that real molecular filaments are not isolated cloud structures in perfect hydrostatic equilibrium. 
Dynamic fragmentation models therefore appear much more appropriate (see below). 

\subsubsection{Interpretation of Observed Core Spacings} 
\index{Filament!Fragmentation|(}
Supplementing classical near-equilibrium models of filament fragmentation \citep{Sto63,Nag87,Inutsuka_Miyama_1992_filament,Fie00,Fis12,Han17,Han19},  
dynamic mod\-els using numerical simulations considering accretion, turbulence, and magnetic fields 
have recently been proposed to interpret the observed core spacings \citep{Seifried_2015_filament,Clarke_2016_filament,Clarke_2017_filament,Gritschneder_2017_filament,Clarke_2020_filament}. 

When filaments are accreting from a turbulent medium, \citet{Clarke_2017_filament} showed that fragmentation 
appears in a two-tier manner where large-scale fragmentation (separation length-scales) occurs first, followed 
by small-scale fragmentation at the effective Jeans length, which is consistent with observations. 
For filaments in which gravitational energy dominates over turbulent energy,
the large-scale fragmentation occurs at the cylindrical fragmentation 
length scale found by \citet{Clarke_2016_filament}; 
when the two are roughly equivalent, the large scale fragmentation 
is determined by the turbulence itself, i.e. those locations which happen to be predominately compressive, 
and is thus random. In both cases, 
two-tier fragmentation occurs as the large-scale gravity/turbulent fragmentation leads to local regions exceeding 
the critical line-mass, a necessary condition for fragmentation as shown by \citet{Chira18}, 
even while the filament may still be globally subcritical. 
Subsequently, as the filament continues 
to gain mass and the entire filament reaches the critical line-mass, smaller scale perturbations may collapse as greater sections of the filament 
becomes unstable and the local collapse time-scale decreases. Due to the complexity of this fragmentation it is difficult to statistically and robustly 
detect the presence of characteristic fragmentation length-scales with the low number of cores typically present in a filament \citep{Cla19}.

The formation of cores is also affected by the presence of sub-filaments \citep{Smi16,Clarke_2020_filament}. 
Sub-filaments may act in two distinct manners:  they may 
fragment independently of each other to form cores (termed \textit{isolated} cores by \citealt{Clarke_2020_filament}) or they may form small hub systems which form a 
grouping of cores (termed \textit{hub} cores by \citealt{Clarke_2020_filament}). These sub-filaments can then act as channels for mass into cores \citep{Smi16}, such that 
hub cores are typically more massive than isolated cores \citep{Clarke_2020_filament}. Moreover, the intermediate fragmentation step of forming sub-filaments leaves 
no strong evidence of quasi-periodic fragmentation with most core spacings being located at approximately the effective Jeans length \citep{Clarke_2020_filament}.

Magnetic fields\index{Magnetic field} have typically been found to suppress 
filament fragmentation but the degree to which fragmentation is suppressed is dependent on the magnetic field ori\-entation \citep{Nag87,Fie00,Seifried_2015_filament,Han17,Han19}. 
Axial magnetic field models suggest that there is still an expectation of quasi-periodic cores, but the spacing is larger than that expected from 
purely hydrodynamical\index{HD} models \citep{Nag87,Seifried_2015_filament}. Perpendicular fields may also lead to quasi-periodic cores with larger spacings but in cases where the 
magnetic field is fixed at large distances from the filament, even weak magnetic fields may completely suppress fragmentation \citep{Han17}. 
\citet{FiegePudritzII2000}  showed that, for highly toroidal fields, a fast magneto-instability may lead to fragmentation instead of gravity.

\begin{figure*}
\centering
\includegraphics[width=\textwidth]{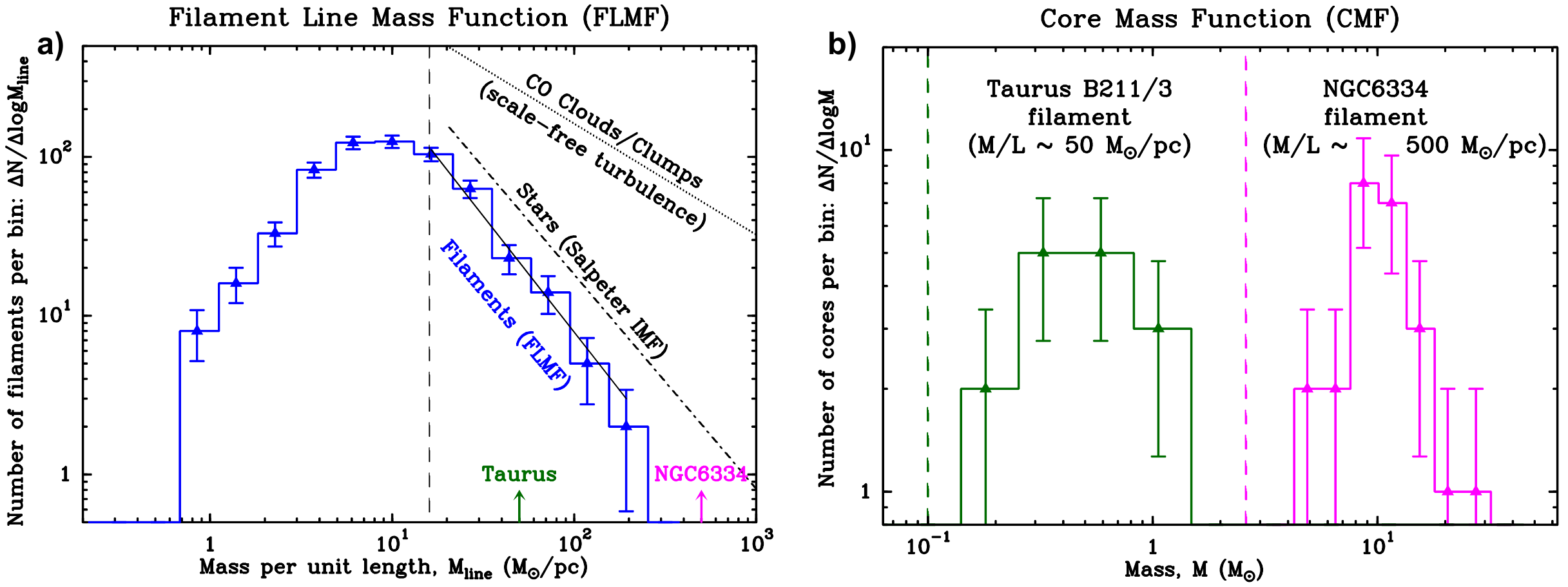}
 \caption{\small 
Potential role of filaments in the origin of the prestellar CMF and stellar IMF: As higher-mass cores form in higher-line mass 
filaments [panel (b), adapted from \citealp{Shimajiri+19b} and \citealp{Marsh_2016_HerschelL1495cores}], 
the Salpeter slope of the global CMF and IMF may be inherited from the filament line mass function (FLMF) [panel (a), adapted from \citealp{Andre2019}].
 }
 \label{flmf_CMF}     
\end{figure*}

It is worth noting that other forms of fragmentation ex\-ist beyond gravo/magneto-instabilities. 
One such scenario is geometric fragmentation, where \textit{spatial} rather than density perturbations may lead to fragmentation \citep{Gritschneder_2017_filament}. 
This process may occur on timescales comparable to other fragmentation processes and could lead to the formation 
of cores at any wavelength. Another process is that cores quickly form at the ends of filaments due to the end-dominated mode of 
global longitudinal collapse \citep{Bas83,Pon12,Cla15}. 
Core formation due to global longitudinal 
collapse ought to produce massive cores/clumps preferentially toward the ends of filaments due to gravitational focusing. There is some 
observational evidence for such fragmentation \citep{Zer13,Dew19,Yuan20,Bha20,Liu20} and \citet{Clarke_2020_filament} found tentative signs of such core properties 
in their simulations of turbulent, accreting filaments.
\index{Filament!Fragmentation|)} 
\index{Dense Core!Separation|)} 

\subsection{From the Filament Line Mass Function to the Prestellar Core Mass Function}
\label{sec:FLMF}\index{Filament!Line Mass Function|(}\index{Filament!Line Mass|(}\index{Dense Core!Mass Function|(} 

The mass function of molecular clouds (MCs)
and CO clumps within MCs is known to be rather shallow,  $\Delta N / \Delta \log M \propto M^{-0.6\pm0.2}$ \citep[e.g.,][]{Solomon+87,Blitz93, Kramer+98,Rice+16}, 
and significantly shallower than the Salpeter initial mass function \citep[IMF,][]{Salpeter55}. 
This implies that most of the molecular gas mass in the Galaxy resides in the {\it most massive} MCs 
and within the MCs  themselves in the most massive CO clumps. 
In contrast, the mass distribution of self-gravitating
prestellar cores or prestellar {core mass function} (CMF) 
broadly resembles the stellar IMF in both shape and mass scale, 
\citep[e.g.,][]{Motte+98,Johnstone+01,Nutter+07,Alves+07,Konyves_2015,Marsh_2016_HerschelL1495cores, DiFrancesco_HGBS_2020,Pezzuto_HGBS_2021}. 
The difference in shape between the observed mass distribution of MCs or CO clumps and that of prestellar cores 
may a priori arise from the use of different tracers, typically CO for clouds or clumps and dust continuum for prestellar cores.  
However, millimeter/submillimeter 
dust continuum studies have also reported mass functions shallower than the Salpeter IMF 
for both small MCs/large clumps \citep[e.g.,][]{Ellsworth15}
and unbound starless cores \citep[e.g.,][]{Marsh_2016_HerschelL1495cores}. The mass functions of the latter types of cloud structures 
therefore appear to genuinely differ from the IMF and prestellar CMF.

Recently, a good estimate of the filament mass function (FMF) and filament line mass function (FLMF) 
in nearby molecular clouds has been derived using a comprehensive study of filament properties from 
HGBS observations \citep{Arzoumanian2019-Filament_properties,Andre2019}. 
The FLMF is well fit by a power-law distribution in the supercritical mass per unit length regime 
(above 16 \msunpc), $\Delta N / \Delta \log M_{\rm line} \propto M_{\rm line}^{-1.6\pm0.1}$ 
(see Fig.~\ref{flmf_CMF}a). 
The FMF is very similar in shape to the FLMF and also follows a  
power-law distribution at the high-mass end (for $M_{\rm tot} > 15\, M_\odot$),  
$\Delta N / \Delta \log M_{\rm tot} \propto M_{\rm tot}^{-1.4 \pm0.1}$, 
which is significantly steeper than the MC mass function. 
Both the FLMF and the FMF are reminiscent of the form of the IMF at the high-mass end ($M_\star \geq 1\, M_\odot$),  
which scales as the Salpeter power law $dN / d\log M_\star \propto M_\star^{-1.35}$ in the same format. 
Thus, molecular filaments may represent the key evolutionary step in the hierarchy of cloud structures at which 
a steep Salpeter-like mass distribution is established. 
The filament mass function differs in a fundamental way from the MC 
mass function in that most of the filament mass lies in {\it low-mass} filaments. 
In particular, this result implies that most of the mass of star-forming filaments lies 
in thermally transcritical filaments with line masses within a factor 2 of the critical value $M_{\rm line, crit}$. 
Interestingly, the numerical study of \citet{AIIM21} shows that the FLMF 
resulting from the type-O filament formation mechanism introduced in \S~\ref{sec:2.2.3} 
quickly becomes Salpeter-like and similar to the observed FLMF (see Fig.~\ref{fig_simulaiton} above).
The same FLMF shape is also seen in type-C induced filaments, 
but for the pure type-G mechanism \cite{AIIM21} report a significantly narrower FLMF 
which peaks at the (thermal) critical line-mass.

The most massive prestellar cores identified with {\it Herschel} 
(with masses between $M \sim 2$ and $10\, \msun$)
\mbox{tend to} be spatially segregated in the highest column density parts/filaments 
of the clouds, suggesting that the prestellar CMF is not homogeneous {\it within} a given cloud
but depends on the local column density (or line mass) of the parent filaments  
(\citealp{Konyves_OrionB_2020}; see also \citealp{Shimajiri+19b}). 
In Orion~B\index[obj]{Orion B}, for instance, there is a marked trend for the prestellar CMF 
to broaden and shift to higher masses in higher density areas \citep{Konyves_OrionB_2020}. 
This supports the view that the global prestellar CMF results from the superposition 
of the CMFs produced by individual filaments \citep{YNLee_2017, Andre2019}. 

The close link between the FMF (or FLMF) and the prestellar CMF may be understood if we recall that 
the thermally supercritical filaments observed with {\it Herschel} in nearby clouds 
have a typical inner width \mbox{$W_{\rm fil}$$\sim$0.1\,pc} 
and are virialized with $ M_{\rm line} \sim  \Sigma_{\rm fil} \times W_{\rm fil}  \sim M_{\rm line, vir} \equiv 2\, \sigma^2_{\rm tot} / G$,   
where $\sigma_{\rm tot}$ is equivalent to the effective sound speed \citep{FiegePudritz00, Arzoumanian+13}.
This implies that the effective Bonnor-Ebert mass  $M_{\rm BE, eff}  \sim 1.3\, \sigma_{\rm tot}^4 /(G^2 \Sigma_{\rm fil} $) scales roughly 
as  $\Sigma_{\rm fil}$ or $ M_{\rm line} $ in supercritical filaments. 
Thus, higher-mass cores may form 
in higher $M_{\rm line} $ filaments, as indeed suggested by observations  \citep[][see Fig.~\ref{flmf_CMF}b]{Shimajiri+19b}. 
If the CMF produced by a single supercritical filament were a narrow $\delta$ function peaked at $M_{\rm BE, eff} $, 
then there would be a direct correspondence between the FLMF and the prestellar CMF \citep[cf.][]{Andre2014-PPVI}.
In reality, the prestellar CMF generated by a single filament is expected 
to be broader than a $\delta$ function \citep[][]{Inutsuka_2001_IMF}, 
and observationally it appears to broaden 
as $M_{\rm line} $ increases \citep[][]{Konyves_OrionB_2020}, 
although for statistical reasons, this is difficult to constrain very accurately. 
The global prestellar CMF therefore results from a ``convolution'' of the FLMF with the CMFs 
produced by individual filaments \citep{YNLee_2017}. 
It can be shown, however, that the high-mass end of the global CMF is primarily driven by the power-law shape 
of the FLMF in the supercritical regime 
and depends only weakly on the breadths of the individual CMFs \citep[cf. Appendix B of][]{Andre2019}.  

\index{Filament|)}\index{Filament!Line Mass Function|)}\index{Filament!Line Mass|)}\index{Dense Core!Mass Function|)}\index{Dense Core|)} 

\section{DENSE CORES IN MOLECULAR CLOUDS
\label{sec:core}}
\index{Dense Core|(}

Dense, gravitationally bound molecular cores are the immediate precursors of stars.
Generally speaking, ``dense cores'' can be used to refer to all overdense (relative to the background) structures at sub-pc scale, 
while ``clumps'' usually represent pc-scale complex structures that may further fragment into multiple cores.
Following previous authors \citep{diFrancesco2007-PPV,Andre2014-PPVI}, we define {\it starless} cores as those that show 
no evidence of the presence of an embedded protostar, 
via the observational detection of a compact infrared or millimeter emission source, 
or outflows in molecular line emission, in contrast to {\it protostellar} cores. Cores that appear unstable to gravitational collapse 
are categorized as {\it prestellar} \citep[see][]{Andre2000-PPIV,WardTh_JCMT_GBS}. 
This section reviews, from both  observational and theoretical studies, the most recent updates on the properties of dense cores, 
which set up the initial conditions of subsequent protostellar evolution.

\subsection{Large Area Surveys in Star-forming Regions\label{sec:4.1}}

Cores are generally embedded within molecular clouds.
Large-area, high-resolution surveys of dust and gas are thus essential in improving our understanding of the 
internal state of star-forming regions down to the core scale.
While advanced continuum studies have provided an unprecedented view of the physical properties in 
star-forming regions from cloud to core, the dynamic information probed by spectroscopic observations is critical to complete the picture. 
Here we review the most updated results from survey-style studies performed in recent years.

\subsubsection{Dust Continuum Surveys}

Over the past decade, large surveys of the continuum emission from dust have provided numerous insights into the 
structure of star-forming clouds across a wide range of spatial scales. With the $\sim$15$\arcsec$ angular resolution 
at $200~\mu$m of the {\it Herschel} Space Observatory, the HGBS 
mapped, at 70--500~$\mu$m, large areas of star-forming {molecular clouds} within $\sim$500~pc 
of the Sun \citep[see review in][]{Andre2014-PPVI}.
As already described in Sect.~\ref{sec:fil_core}, these observations revealed that dense cores are found primarily 
within ubiquitous filamentary structures in molecular clouds, with most prestellar cores in particular located within 
transcritical or supercritical filaments \citep{Konyves_2015,Konyves_OrionB_2020,Benedettini+18,Bresnahan_2018_HerschelCrA, Ladjelate+20, Pezzuto_HGBS_2021}.

The JCMT SCUBA-2 Gould Belt Legacy Survey (GBS) \citep{WardTh_JCMT_GBS} mapped the densest regions 
of the Gould Belt clouds visible from the northern hemisphere, at 450 and 850~$\mu$m with higher (8--14\arcsec) resolution. 
While JCMT SCUBA-2 is less sensitive than {\it Herschel} to extended, lower column density material, 
JCMT GBS detects primarily the densest starless cores, more likely to be prestellar \citep{WardTh_2016_JCMTvsHerschel}. 
Core mass estimates tend to assume a specific power-law dependence of the dust emission with wavelength, 
often setting the dust emissivity, $\beta = 2$ or similar. 
Using the full range of continuum emission afforded by both {\it Herschel} and JCMT observations, \cite{Sadavoy_dust_2013} 
showed evidence for a lower value of $\beta$ indicative of grain growth in cores. Regions of low $\beta$ 
may be correlated with local temperature peaks in NGC 1333\index[obj]{NGC 1333} \citep{MChen_dust_2016}. Better understanding 
of where and when dust properties evolve will enable more precise measurements of core masses, as well as 
improving our understanding of the process of grain evolution from cloud to core.

Toward more distant regions, the Hi-GAL survey 
has provided comprehensive source 
catalogs toward the Galactic Plane \citep[][]{HiGal_2016, Elia_HiGAL_2021},
although without sufficient spatial resolution to resolve the cores. 
Nevertheless, IR-dark structures can in principle be considered as prestellar core candidates \citep[e.g.,][]{ASHES_19},
though most of the IR-dark studies are at clump scales.

\subsubsection{Molecular Line Surveys of Dense Gas}

In the past years, high-resolution and wide-coverage molecular line surveys of intermediate and higher density 
gas tracers have been performed in nearby star-forming regions, the Galactic plane, and toward the Galactic center 
to complement ongoing surveys of dust continuum emission.
These line surveys provide the necessary kinematic information required to fully assess the stability of 
cores and filaments, identify rotation and evaluate the evolution of angular momentum from large to small scales, 
and investigate the role of accretion in the mass evolution of filaments and cores.

\begin{figure*}[ht]
\centering 
\includegraphics[width=\textwidth]{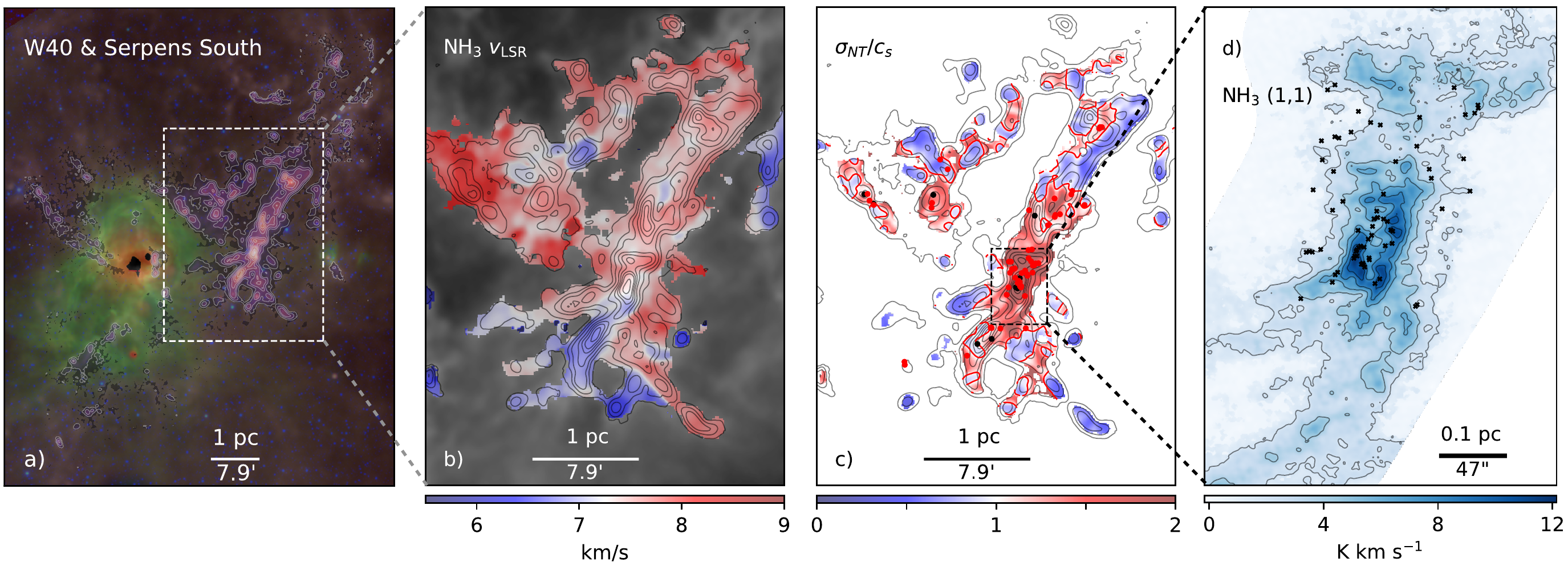}
\caption{Zooming in on the W40\index[obj]{W40} and Serpens South\index[obj]{Serpens South Molecular Cloud} star-forming regions: a) WISE 22~$\mu$m, 12~$\mu$m, 
and 4.6~$\mu$m (RGB), overlaid with $N(\htwo)$ (grayscale; \textit{Herschel} GBS) and \ammo integrated 
intensity contours (white; GAS) at 32\arcsec\ (0.07~pc) resolution. b) $N(\htwo)$ (grayscale; \textit{Herschel} GBS) 
and \ammo integrated intensity contours (gray; GAS), with \ammo-derived $v_\mathrm{LSR}$ (colorscale). 
c) Ratio of non-thermal velocity dispersion to sound speed in \ammo, adapted from \citet{Friesen2016-Serpens_South}. 
d) Dense gas surrounding the Serpens South young stellar cluster in \ammo (1,1) at 5\arcsec\ (0.01~pc) resolution 
({\it R.K. Friesen}, private communication), highlighting dense cores and compact substructures within filaments. 
Millimeter sources detected with ALMA are shown \citep{Plunkett_2018}. 
In all plots, a distance of 437~pc is assumed \citep{ortiz_leon_2018b}.
}
\label{fig-Serpens-NH3}
\end{figure*}

The kinematic transition from cloud to filament and core has often been traced by rotational line emission from CO and CO 
isotopologues, such as \ce{^{13}CO} and \CeightO, at millimeter to submillimeter wavelengths. 
Toward some Gould Belt clouds, JCMT GBS observations of the CO and isotopologues of the (3--2) 
rotational transitions ($\approx$14\arcsec\ resolution) were used to examine the interaction between outflows 
and star-forming gas \citep{White_JCMTGBSline_2015,Drabek-Maunder_JCMTCO_2016}, and informed virial analyses 
of dense cores identified in these regions \citep{Pattle2015-SCUBA2_Oph}. 
In order to resolve spatially individual cores along with their kinematics, surveys like the CARMA-NRO Orion\index[obj]{Orion A Molecular Cloud} 
survey \citep{kong_2018} use a combination of interferometer and single dish observations to discern structures 
on 8\arcsec\ (3300~au) scales in the Orion molecular cloud, while retaining sensitivity to large-scale structures. 
Toward more distant regions, single dish CO surveys such as the FOREST unbiased Galactic plane imaging survey 
with the Nobeyama 45-m telescope \citep[FUGIN;][]{Umemoto_FUGIN1_2017} are sensitive 
to molecular gas clumps on scales of $\sim$1~pc, and masses $\sim$100~\msun  at 10~kpc.

More recently, sensitive \ammo observations by the Green Bank Ammonia Survey \citep[GAS;][]{GAS_DR1} 
have enabled analysis of the cloud-core transition in a single dense gas tracer 
($n_\mathrm{eff} = 7.9 \times 10^2$~\cc to produce a 1~\kkms spectral line at $T = 15$~K; \citealt{Shirley_2015}). 
GAS takes advantage of the multiple-pixel K-band Focal Plane Array (KFPA) at the 100-m GBT 
to survey nearby star-forming regions within $\sim$500~pc at 32\arcsec\ angular resolution 
(0.02~pc to 0.07~pc spatial resolution), with a total areal coverage of nearly four square degrees across 
$\sim$12 clouds \citep[][]{GAS_DR1,Keown_GAS_2017,Redaelli_GAS_2017}. 
These high sensitivity observations show substantial evidence for multiple velocity components in the 
dense gas \citep{M_Chen+20, Choudhury_2020_supersonic, Choudhury2021-L1688_Extended}. 
While past observations revealed a sharp transition between the supersonic molecular cloud and 
the quiescent dense core \citep{Pineda2010-Coherence_B5}\index[obj]{Barnard 5}, wider-scale mapping has shown that larger 
areas of subsonic linewidth are also identified in dense gas within low-mass 
star forming regions \citep[][see Fig.~\ref{fig-Serpens-NH3}]{Friesen2016-Serpens_South}. 
Furthermore, kinematic transitions between the turbulent cloud and dense core may be more gradual, shown 
by fitting of multiple velocity components across core edges in L1688\index[obj]{L1688} 
\citep[][see also Sect.\,\ref{sec::CoreEvo} below]{Choudhury2021-L1688_Extended}.

Using Serpens South as an example, Fig.\,\ref{fig-Serpens-NH3} compares the spatial distribution 
of dust (traced by thermal continuum emission) and dense gas 
\citep[traced by \ammo;][]{Friesen2016-Serpens_South, GAS_DR1} from cloud to core scales, 
and demonstrates the correlation between YSOs\index{Young Stellar Object (YSO)} 
(traced by high-resolution millimeter continuum 
emission observed with ALMA) and dense cores of the hub-filament system. 
The gas kinetic information traced by \ammo also reveals the transition from super- to sub-sonic gas 
motions over extended scales in filaments and around the dense cores (panel c),
although as noted above, these transitions from supersonic turbulence to quiescent, 
velocity-coherent cores may not be as sharp as expected based on these maps.

Using the same tracer, the KFPA Examinations of Young STellar Object Natal Environments 
\citep[KEYSTONE;][]{Keown2019-KEYSTONE} mapped \ammo (and \ce{H2O} maser) emission 
toward more distant high mass star forming regions, 
while the Radio Ammonia Midplane Survey \citep[RAMPS;][]{hogge_RAMPS_2018} targeted portions of the Galactic plane.
The angular resolution of these single dish studies is less able to resolve individual cores at kpc distances, however.

The CARMA (Combined Array for Research in Millimeter-wave Astronomy) Large Area Star Formation 
Survey \citep[CLASSy;][]{CLASSy_Storm_2014} probe higher density material in emission from \ntwohp (1--0), with $n_\mathrm{eff} = 6.7 \times 10^3$~\cc. 
Utilizing the high angular resolution ($\approx$$7$$\arcsec$ at $\sim$$90$\,GHz) and sensitivity 
at a wide range of spatial scales of CARMA,
CLASSy mapped subregions in the Perseus\index[obj]{Perseus Molecular Cloud} 
and Serpens\index[obj]{Serpens Molecular Cloud} Molecular Clouds
to cover a range of star-forming activity levels \citep{CLASSy_Storm_2014,KLee_2014_CLASSy,Fernandez-Lopez2014-CLASSY,Storm_2016_CLASSy}. 
Further investigations toward five subregions covered by CLASSy are presented in \cite{Dhabal_2018} 
as CLASSy-II, which combined lines covered by CLASSy (\ntwohp, HCN, and \ce{HCO+}) with additional 
optically thin dense gas tracers \ce{H^{13}CO+} and \ce{H^{13}CN} to test whether the observed 
kinematics in CLASSy represents the bulk property of the material making up the filaments.
In general, both the CLASSy and CLASSy-II 
observations support the existence of finer structures within dense filaments and the presence 
of velocity gradients across them (see e.g., Sect.\,\ref{sec:accretion} and Fig.~\ref{velo_grad}).

Other high-resolution, wide-coverage maps provide a coherent picture of the gas dynamics in
both nearby \citep[e.g.,][]{Hacar2018-ALMA_Orion} and Galactic \citep[e.g.,][]{Sokolov2018-Coherence_IRDC} star-forming regions.
Combining several molecular lines, including \ce{^{13}CO} and \CeightO for larger-scale diffuse gas, and \ntwohp, \ce{HCO+}, SO, and CS, 
the TRAO (Taeduk Radio Astronomy Observatory) survey of Filaments, the Universal Nursery of Stars (FUNS) aims to provide complete pictures 
of the kinematics and chemistry in filaments and cores within 10 Gould Belt Clouds with $\approx 50$\arcsec\ angular resolution 
at $\sim$$100$\,GHz \citep{Chung_TRAO_2019,Chung_TRAO_2021}. 
In L1478 in the California\index[obj]{California Molecular Cloud} molecular cloud, 
cores traced by \ntwohp tend to lie in supercritical filaments as expected, but differ in 
how well their kinematic properties match that of their embedding filaments \citep{Chung_TRAO_2019}. Systematic shifts seen 
in core $v_\mathrm{LSR}$ as traced by \ntwohp and \CeightO in hub systems differ from previous results in the Perseus molecular 
cloud \citep{Kirk_07}. Kinematic studies such as these suggest core formation in filaments may follow multiple mechanisms, depending on the environment.

\subsection{Core-Environment Interaction}

\subsubsection{Defining Core Boundaries}
\label{sec:CoreIdn}\index{Dense Core!Boundary|(}

Different methods have been developed to identify and isolate 
the emission from dense molecular cores, and they are applied to continuum or spectral line data, involving both 2D images and 3D data cubes. 
Since many stars form in complex, clustered environments  \citep[see reviews by][]{Krumholz_cluster_2014,Reipurth_multi_2014},
disentangling emission from cores from that of their parent clouds, including complex large-scale background and filamentary structures, can be challenging. 
The assignment of emission to one source compared with another, and the method of assessment of background vs.~source flux, 
among other algorithmic decisions, can thus affect strongly the derived core properties, including radius and mass, although to 
date systematic comparisons of different methods are rare, partly because different methods are best matched to regions of varying complexity.

Previous methods introduced in \citet{Andre2014-PPVI} for extracting sources in {\it Herschel} continuum data included 
CuTEx \citep[Curvature Thresholding Extractor;][]{Molinari_2011_Cutex}, CSAR \citep[Cardiff Sourcefinding AlgoRithm;][]{JKirk_2013_HGBScore}, 
and \texttt{getsources} \citep[][]{mensh_2012_getsources}. 
CuTEx assesses curvature via second-order differentiation of the image brightness in, e.g., thermal continuum emission maps. 
This allows the disentanglement of closely-spaced, compact sources in clustered environments, but is less likely to detect unbound, 
more diffuse objects than \texttt{getsources} \citep{Bresnahan_2018_HerschelCrA}, where multi-wavelength images are spatially-filtered 
and background-subtracted to identify sources of interest. Deblending of apparently overlapping sources in the image plane is enabled 
by \texttt{getsources}, as well as by dendrogram algorithms like \texttt{astrodendro} \citep[][]{Rosolowsky_2008} in velocity space when 
applied to 3D position-position-velocity molecular line emission cubes. Other than in \texttt{getsources}, emission is assigned to individual 
sources at each pixel (or voxel, in \texttt{astrodendro}), following various thresholding or segmentation rules. 
More recently, \cite{Berry_FellWalker_2015} introduced \texttt{FellWalker}, which defines the peaks and boundaries of sources based on local gradients.

We note that the most appropriate method for identifying structures depends on the data and science goals.
Generally speaking, with {\it Herschel}'s multi-wavelength data, \texttt{getsources} is preferred 
\cite[e.g.,][]{Konyves_2015, Konyves_OrionB_2020, Marsh_2014_HerschelTaurusCores, Marsh_2016_HerschelL1495cores}, 
while \texttt{FellWalker} could provide better estimate on core boundaries when cores are well-separated, 
as in the case of JCMT GBS \citep{Kirk_JCMTGBS_2016,Mowat_JCMTLupus_2017}.\@ 
Though \texttt{getsources} only re\-turns the elliptical footprint of the core area instead of the actual boundary as provided by \texttt{FellWalker}, 
uncertainties in core boundary results in only a factor of a few error in mass estimation 
[for density profile following $\rho(r) \propto r^{-2}$, and thus mass $m(r) \propto r$] 
and thus \texttt{getsources} may be preferred for CMF/IMF studies because of its better performance on finding small cores in clustered environment.
In particular, \cite{Lane_JCMTGBS_2016} applied both \texttt{getsources} and \texttt{FellWalker} on their JCMT Orion~A data, and 
concluded \texttt{getsources} is more robust on highlighting small, compact cores when there is substantial background emission.
Also, as demonstrated in \cite{Goodman_2009} and \cite{CLASSy_Storm_2014}, clump-finding algorithms 
like \texttt{Cloudprops} \citep{Cloudprops_2006} and \texttt{clumpfind} \citep{Williams_clumpfind_1994} work best 
for small-\-scale segmentation in sparse fields that have resolved separations between objects, while dendrogram-type 
analysis is more appropriate to study features across a wide range of spatial scales.

With the complex filamentary structures in star-forming regions revealed by recent observations, background 
subtraction becomes an important topic in core definition, {particularly} since dense, 
potentially prestellar cores are strongly associated with filaments (see \S\,\ref{sec:fil_core}). 
Also, filamentary structures that overlap with noisy fluctuations could appear to be core-like,  
which further introduces complications for source extraction routines \citep{Menshchikov_getfilaments_2013}. 
Indeed, numerous algorithms have been developed and adopted to identify and characterize filamentary 
structures in both the diffuse ISM and dense molecular clouds in recent years, including 
the Rolling Hough Transform (RHT; \citealt{Clark2014}), Discrete Persistent Structures Extractor (\textsc{DisPerSE}; \citealt{Sousbie_DisPerSE_2011}), 
\texttt{getfilaments} \citep[as part of the \texttt{getsources} package;][]{Menshchikov_getfilaments_2013}, FilFinder \citep{Koch_FilFinder_2015}, 
template matching \citep[TM;][]{Juvela_TM_2016}, and \texttt{RadFil} \citep{RadFil2018}.
However, these routines generally work independently of the core identification algorithms,
and a joint approach like \texttt{getsf}, a recent improvement of \texttt{getsources} and \texttt{getfilaments} \citep{mensh_2021_getsf}, is desirable.

Because the strength of molecular line emission depends on density along with the total column density, maps of 
different molecular line transitions are biased toward varying ranges in density and column density.
Applying source extraction algorithms to line emission from the denser gas tracers described above therefore can simplify 
the identification and disentangling of core emission from the environment, {as there is less} contamination from diffuse gas along the line of sight.
Spectral line observations also allow for core identification in one additional dimension (i.e., the position-position-velocity, or PPV space). 
This functionality is available for PPV data cubes in both \texttt{clumpfind} \citep{Williams_clumpfind_1994} and 
\texttt{astrodendro} \citep{Rosolowsky_2008}, and each have been widely adopted \citep[e.g.,][]{Ikeda_2007,Friesen_2009,DC_B59_2012,Keown_GAS_2017,Nakamura_2017,Takemura_C18OCMF_2021}.
Alternatively, \cite{CLASSy_Storm_2014} developed a new non-binary dendrogram algorithm to allow for grouping more 
than two objects, thus providing a more statistically meaningful way to represent  hierarchical structure. This approach has 
been adopted in follow-up studies using the isolated hyperfine component of \ntwohp from the CLASSy survey \citep{KLee_2014_CLASSy, Storm_2016_CLASSy}.

While identifying structures in PPV space could be more desirable because of the additional dimension of information, 
structures in PPV space may not always be consistent with actual three-dimensional structures  
\citep[in position-position-position, or PPP space; e.g.,][]{BP_MacLow_2002,Burkhart_2013_structure,Beaumont_2013_MC_simulation,Clarke+18}. 
Furthermore, the hyperfine structures in the emission lines of traditional dense gas tracers like \ntwohp and \ammo 
can make structure identification in PPV space less sensitive. In some cases, structure analyses can be applied to single, 
isolated hyperfine components (e.g., in the \ntwohp 1--0 line). In other cases, the hyperfine line structure can be fit 
assuming initially a single velocity component, and a data cube reconstructed with a single Gaussian line (e.g., \ammo). 
In practice, many structure analyses of molecular line observations still identify cores in the integrated intensity maps \citep[e.g.][]{HChen-2019_Droplet}.

In simulations, the complete 3D information in principle makes defining the core boundaries more straightforward. 
Core-defining methods in theoretical studies are therefore more diverse, as most theoretical works either prescribed 
their own core-identification routines \citep[e.g.,][]{Gammie_2003}, or focused on the properties and/or evolution of 
the star-forming gas in general instead of highlighting cores.
In general, methods based on density contours are the most common, like the popular 
\texttt{clumpfind} \citep[e.g.,][]{Padoan_2007_clumpfind,Schmidt_2010_clumpfind,Bleuler_sink_2014,LiPS_2015,Pelkonen_CMF_2021}. 
In contrast, \texttt{astrodendro} was mostly used in simulations for characterizing the hierarchy of structures instead 
of defining cores \citep{Beaumont_2013_MC_simulation, Burkhart_2013_structure}, with a few exceptions \citep{Smullen_dendro_2020, Clarke_2020_filament}.
On the other hand, \cite{Dib_2007} developed a clump-finding algorithm based on a density threshold criterion 
and a friends-of-friends approach.
This is similar \mbox{to the HOP algorithm} \citep{HOP} adopted in \citet{Hennebelle_2018_FRIGG}, \cite{Ntormousi19_coreInfilament}, and \cite{Hennebelle_2019},
though \citet{Hennebelle_2019} pointed out that these ``cores'' may be too dense and clustered compared to observed cores.
Alternatively, the core-finding method GRID-core (Gravitational potential Identification of cores) \, uses \, the contours 
of the local gravitational potential to identify core boundaries \citep{GO11,CO14,CO15,CO18,GO15}, because the 
gravita\-tional potential is more directly linked to the fundamental physics during core evolution. 
A similar approach was adopted by \cite{Smith_2009}, who noted that using the gravitational potential instead of 
density yields smoother core boundaries.
Last but not least, recent work by
\cite{Kuznetsova_2018_IMF,Kuznetsova2019-Origin_J,Kuznetsova2020-Cores_MHD} considered ``sink patches'' 
to record gas properties in the immediate surrounding of the sink particles in the simulations, which in principle resemble star-forming cores.

\index{Dense Core!Boundary|)} 

\subsubsection{Stability and Evolution of Dense Cores from Observations}
\label{sec::CoreEvo} \index{Dense Core!Stability|(} 

The stability of starless and prestellar cores is often assessed via a virial analysis, where the second 
derivative of the moment of inertia $\ddot{I}$ is given by
\begin{equation}
    \frac{1}{2}\ddot{I} = 2\Omega_K + \Omega_G + \Omega_M + \Omega_P~,
\end{equation}
and $\Omega_K$, $\Omega_G$, $\Omega_M$, and $\Omega_P$ are the internal energy, gravitational 
potential energy, magnetic energy, and energy due to external pressure, respectively.\index{Magnetic field} 
The gravitational and external pressure terms are negative, while the internal energy and magnetic energy terms are positive.
A core is considered to be virially unstable and may collapse if $\ddot{I} < 0$.

In practice, most often a comparison is made between the internal and gravitational potential energy 
terms only, and the virial parameter $\alpha_{\rm vir}$ defined for a core of mass $M$ and radius $R$ such that
$ \alpha_{\rm vir} = {5 \sigma_v^2 R}/{a\, GM}$ \citep{Bertoldi_McKee_1992},
where $\sigma_v$ is the 1D velocity dispersion of the core, and can include thermal motions only, 
or thermal and non-thermal contributions if kinematic data are available;
and $a$ is a correction factor of order unity that accounts for the radial density profile 
\citep[$a=1$ for uniform density spheres and $a=1.22$ for a critical Bonnor-Ebert sphere (BES); see][]{Singh_2021_subvirial}.

Given the recent availability of large surveys of matched continuum and molecular line data toward many nearby 
clouds described in Sect.~\ref{sec:4.1}, studies have begun to address the full virial equation to investigate core stability 
\citep{Pattle2015-SCUBA2_Oph,Seo2015-Taurus_Core_Evolution,Pattle_2016,Kirk2017-OrionA_Pressure,HChen-2019_Droplet,Kerr2019-GAS_Pressure}. 
When including (presumed supportive) non-thermal motions in the full analysis, the internal kinetic energy term is 
not balanced by the gravitational term for many cores in nearby clouds, and they are instead bound by the 
external pressure \citep[e.g.,][]{Kirk2017-OrionA_Pressure,Kerr2019-GAS_Pressure}. 
In most studies, the external pressure is assumed to be a confining pressure derived from turbulent motions in the 
embedding cloud, but \citet{Gomez_2021} point out that it can also be interpreted as ram pressure from a collapsing, 
larger-scale envelope. 
We do not address the contributions from magnetic fields in detail here, although recent polarization surveys 
are starting to probe core-scale magnetic field structures across multiple regions. 
Instead, we refer the reader to recent reviews of the state-of-the-art theory and observations of magnetic fields\index{Magnetic field} 
in \cite{HennebelleInutsuka19}, \cite{pattleANDfissel2019}, \cite{PudritzANDRay2019-Review}, %
and \citet[][in this volume]{Pattle+2022}. 

\citet{HChen-2019_Droplet} identified a set of `coherent cores' in GAS \ammo  data toward the Ophiuchus and Taurus 
molecular clouds that were defined by both peaks in \htwo  column density as well as an observed transition in the 
size-linewidth relation \citep[the `transition to coherence';][]{Goodman_1993,Pineda2010-Coherence_B5}. 
These selection criteria identify a subset of core-like sources that appear confined by external pressure, 
rather than bound virially by their self-gravity. 
This subsonic velocity dispersion in high-density regions has also been reported in numerical simulations \citep{GO11}, 
who pointed out that such feature could be due to a combination of low post-shock velocities and the 
spatially limited scale highlighted by density.

Turbulence sometimes dissipates on much broader scales, however, as indicated by chains of velocity-coherent 
cores in the Taurus L1495/B213 filament \citep{Tafalla_Hacar_2015} 
and larger-scale deviations from the size-linewidth relation in filaments. 
Moreover, \cite{Pineda2021-Ions_Neutrals} found that within the velocity-coherent core Barnard 5\index[obj]{Barnard 5}, 
the ions (traced by \ntwohp) display a higher level of turbulence than the neutrals (traced by \ammo).
Assuming these two molecules trace similar volumes, this can be explained if the magnetic field\index{Magnetic field} 
within Barnard 5  is oscillating. Since the ions are more strongly coupled to the magnetic field, 
this would produce a larger velocity dispersion in \ntwohp.
Better understanding toward the decay and transport of turbulence in core-forming regions is still needed.

Several analyses, focused on high-mass (and more distant) star forming regions, reported very low 
values ($\ll1$) of the virial parameter \citep{Kauffmann+13,Urquhart2014_ATLASGAL_Virial,Traficante+18c,Traficante+18b, Keown2019-KEYSTONE}. 
This result would imply that these objects are not in equilibrium and that they would last for a fraction of the free-fall time. 
However, \citet{Singh_2019_virial} introduced a method to measure more directly the gravitational 
term from \htwo column density maps, negating the need to quantify a core radius. 
Applying this method to Gould Belt clouds mapped by both the HGBS and GAS, \cite{Singh_2021_subvirial} show 
that by including bulk motion in the kinematic term, larger cores and clumps have virial parameters closer to the stable value. 
This result shows that it is easy to underestimate the virial parameter, and provides a potential solution to the 
very low values of the virial parameter previously reported.

The observational result that many cores do not appear gravitationally bound may seem to be in conflict with models of core 
formation via gravitational instability of nearly isothermal filaments discussed in the previous section, but gravitational instability 
models proposed for self-gravitating prestellar cores are not inconsistent with the view that unbound cores form via 
turbulent cloud motions in the first place and a fraction of them grow (and eventually collapse) by gravitational instability.
Based on MHD\index{MHD} simulations, \citet{Offner2020-Core_Evolution} propose that the pressure-confined, coherent objects 
represent an early phase in core evolution, where dense structures form via turbulent cloud motions, 
turbulence dissipates within the dense structures, and some fraction become gravitationally unstable 
and eventually collapse \citep[see also][]{Seo2015-Taurus_Core_Evolution,Pattle_2016}. 
Core formation in simulations is discussed in more detail in the following section.
\index{Dense Core!Stability|)}

\subsubsection{Core Formation and Evolution in Simulations}
\index{Dense Core!Formation|(} 

It is well-established that dense cores are mostly associated with filaments 
\citep[see the review by][and \S\,\ref{sec:fil_core}]{Andre2014-PPVI}. This lends support to the core formation 
scenario that thermally supercritical filaments would fragment longitudinally into cores, or groups of cores 
(see Sect.~\ref{sec:filament} for details). 
This filament-fragment-to-core model has been studied extensively using both analytic 
approaches \citep{Inutsuka_Miyama_1992_filament,Inutsuka_Miyama_1997_filament, Inutsuka_2001_IMF, YNLee_2017} 
and numerical simulations \citep{Seifried_2015_filament, Gritschneder_2017_filament, Clarke_2017_filament,Clarke_2020_filament}.
Numerical simulations of accreting filaments showed that the fragmentation behavior can either grow 
from internal perturbation of an accreting filament \citep{Clarke_2016_filament} or originate from the 
clumpy nature of the accreting turbulent gas \citep{Clarke_2016_filament, Clarke_2017_filament}.
Indeed, the growth of structures could be enabled by nonlinear perturbations induced by turbulence at all scales. 
Several kinematic and structural studies of star-forming clouds also suggest that these systems 
may be undergoing global collapse \citep[e.g.,][]{Beuther_globalcollapse_2015,Csengeri_ATLASGAL_2017,Hacar_OMC1_2017,Barnes_2018,Jackson_2019,Trevino_Morales_MonR2_2019, Nony_Gcollapse_2021}.
\cite{VS_GHC_2019} elucidate these multiscale, non-homologous collapses with scale-dependent timescales 
as the scenario of global hierarchical collapse (GHC). In this picture, structures are continuously feeding their 
substructures while they accrete from their parent structures \citep[see also][]{Gomez_VS_14,NaranjoRomero_GHC_2015,VS_2017}.

While the geometry differs, the GHC scenario is consistent with\,recent\,simulation\,results\,that\,dense cores\, and\, 
fila\-ments develop simultaneously \citep[e.g.,][]{CO14,CO15, VanLoo_etal_14,GO15}, in contrast to 
the two-step, filament fragmentation scenario that has long been adopted in semi-analytic static models. 
In particular, the anisotropic contraction model described in \cite{CO14,CO15} enables rapid core formation 
even in strongly magnetized media and without ambipolar diffusion, consistent with the abundant observational 
examples of magnetically supercritical cores.\index{Magnetic field}

While numerical simulations of star formation have been significantly improved in the past decade, 
theoretical investigations on simulated dense cores are relatively sparse.
Since numerical simulations are limited by finite resolution, insertion of sink particles has become a common 
practice to replace the densest structures in simulations \citep[e.g.,][]{Hubber_2013_sink,Bleuler_sink_2014}. 
With star formation being the primary interest in most of the numerical works, detailed structures of the dense 
gas (which could be identified as cores) are not necessarily analyzed even when dense cores were identified 
during the course of the simulation in order to evaluate sink insertion and accretion 
\citep[e.g.,][]{Ballesteros-Paredes_2015_IMF, Bertelli_2016_IMF, Lee_2018_IMF_conditions, Bate_2019_IMF}.
In addition, spatial resolution at sub-pc scale is necessary for proper star-forming core identification and inspection, 
which is computationally expensive to achieve in large-scale, cluster-type simulations constructed at a few tens 
of pc \citep[e.g.,][]{Lee_2016_cluster, He_2019_IMF}. By far, few numerical simulations have covered such 
wide dynamic ranges (see e.g., \citealt{Padoan+16,Kuffmeier2018-episodic,GrudicKruijssen2021}; 
also see Sect.~\ref{subsec:stream_sim}).

Some cluster-type simulations have identified dense cores and analyzed their statistical properties 
\citep{Padoan_2001_cluster,GM_2007_core_simulation, Smith_2009, Hennebelle_2018_FRIGG, Ntormousi19_coreInfilament,Pelkonen_CMF_2021}, 
with the goal of providing a potential link between the CMF and the IMF. 
However, few existing works have tried to actually link the properties of the cores to those of the stars. 
\cite{Kuznetsova_2018_IMF} analyzed the gas ``patches'' from which sink particles accrete and explain 
the outcome of the stellar mass spectrum as a consequence of Bondi-Hoyle accretion from the reservoir. 
\cite{Colman_2020_IMF} identified tidally protected regions around newly-formed sink particles and found 
a good correlation with the final mass of the stars. Nonetheless, \cite{Pelkonen_CMF_2021} 
analyzed bound cores around sink particles from simulations described in \cite{Haugboelle2018} and 
suggested that the final mass of sink particles, though exhibiting a correlation with the reservoir mass, 
accretes not only from the bound reservoir and does not accrete the totality of the latter 
either \citep[e.g.,][]{Smith_2009}.
This is in line with the mass accretion scenario proposed in \cite{Padoan2020}, that the final stellar 
mass is dominated by pc-scale turbulent flows instead of the mass of the parent core. 
However, it is worth noting that \cite{Padoan2020} focused on massive star-forming cores
which, because of the significant feedback effects, shall have weaker correlation between core mass 
and stellar mass compared to the low-mass regime \citep[see e.g.,][]{Beuther_HighMassSF_2007}.
\index{Dense Core!Formation|)} 

\subsection{Dynamic Properties of Dense Cores}
\subsubsection{Infall}
\index{Dense Core!Infall|(} 

Those subset of cores that become gravitationally unstable should then collapse to form protostars. 
Core collapse models have been developed both analytically and numerically. 
Self-similar solutions of spherically symmetric starless core collapse predict the infall velocity evolution and mass accretion rates given initial density profiles. \cite{whitworth_1985} showed that broad families of self-similar solutions, including the Larson-Penston (LP) model \citep{Larson1969,Penston_1969}, and the singular isothermal sphere model \citep[SIS;][]{Shu1977}, can be described by two parameters that reflect the initial gravitational instability of the core, and the significance of the external bounding pressure. 
Models that are initially stable against collapse, with little influence due to external pressure, evolve quasi-statically toward instability. 
Contraction motions in an initially unstable core will be enhanced by external compression, leading to more rapid evolution and larger infall speeds. 
Since measured core density profiles tend to follow the common $\rho \propto r^{-2}$ profile at larger radii, 
testing theoretical collapse scenarios requires high spatial resolution to resolve and model 
the inner density and velocity profiles.

From observed molecular line emission profiles, it is difficult to get a direct indication of gravity-driven infall, because rotation 
and outflows may also produce comparable kinematic signatures, and the interpretation usually depends on comparisons to numerical models that involve various assumptions on gas properties (density, temperature, chemical abundance, etc.).
However, it has long been proposed that the opaque infalling gas in the foreground cloud could generate redshifted absorption against the continuum emission from the central protostar. 
The resulting {\it inverse P-Cygni profile}, with emission on the blueshifted side of the central velocity and absorption on the redshifted side, can therefore be used as an unambiguous indicator of inward motion of the foreground matter \citep{Leung_Brown_infall_1977}.

In general, starless and protostellar cores show more evidence for inward rather than outward motions \citep{Mardones1997_Infall,Lee_2011_infall}, with largely subsonic infall speeds where measured. 
In pointed, single-dish line surveys, between 10\% and 25\% of starless cores in nearby molecular clouds show blue-shifted line asymmetries indicative of infall or inflow \citep{Lee_1999_infall, Lee_2004_infall,Sohn_2007_infall,Tsitali_2015_infall,Campbell2016-Perseus_Infall}, although these studies do not all have common core selection criteria. 
In a broad survey of cores in Perseus\index[obj]{Perseus Molecular Cloud}, \cite{Campbell2016-Perseus_Infall} find that starless cores more massive than their Jeans mass are more likely to show infall line signatures. 
\citet{Keown_infall_2016} show that measured infall speeds vary with tracer and position across low mass cores 
L492\index[obj]{L492}, L694-2\index[obj]{L694-2}, and L1521F\index[obj]{L1521F}, as is expected for 
several contracting core scenarios, such that the true magnitude of the infall motions 
may not be accurately measured via single-pointing surveys.

The identification of inward and outward gas motions via the detection of blue- or red-shifted, self-absorbed optically thick emissions relative to an optically thin counterpart,
however, relies on high-enough spatial resolution to reduce the contamination from outflows and foreground large-scale clouds \citep{DiFrancesco_infall_2001}, and thus has thrived as ALMA became available
\citep[e.g.,][]{Pineda2012-Infall_IRAS16293, LeeCF_infall_2014, Evans_infall_2015, Mottram_2017_WILL, Su_infall_2019}.
With ALMA, we are now better able to probe the structure and kinematics of the innermost regions of dense cores, potentially identifying the infall velocity at the radius where the density profile transitions between distinct radial power-law slopes 
($\sim 10^2-10^4$\,au) to test collapse model predictions. 
Toward more evolved, protostellar cores, \cite{Maureira_L1451mm_2017} identify both rotation and infall 
in the First Hydrostatic Core (FHSC) candidate L1451-mm\index[obj]{L1451-mm} at 1000~au scales with ALMA, 
again finding subsonic infall speeds of $\sim$$0.17$~\kms \citep[see also][]{Tsitali_2013-ChaMMS1}. 

Note that radiative transfer modeling of the protostellar envelope with chemical abundance profiles of the corresponding molecules is still required to properly constrain the infall kinematics from the observed line profile \citep[e.g.,][]{YLYang_2020}.
Measurements of infall speeds through the inverse P-Cygni technique range from subsonic to supersonic, with values $\sim$$0.4$--$0.8$~\kms at $r$$\sim$100~au toward some sources\index[obj]{IRAS16293} \citep{Pineda2012-Infall_IRAS16293,Su_infall_2019}. 
Detailed measurements of infall speeds as a function of radius will help constrain collapse profiles. Toward B335\index[obj]{B335}, HCN and \ce{HCO^+} line profiles at 50~au scales are fitted well by an inside-out collapse model where $v_\mathrm{inf} \propto r^{-0.5}$ \citep{Evans_infall_2015}.

Moving toward more detailed infall analyses, \cite{Keto_2015_collapse} predict the emission line profiles of \ce{H2O} ($1_{10}$--$1_{01}$) and \CeightO (1--0) in the prestellar core L1544\index[obj]{L1544} for several core collapse models, and find only the quasi-static contraction of an initially unstable BES is in agreement with observations. In this model, the maximum infall velocity remains subsonic, and infall motions extend $\sim$1.6$\times 10^4$~au (0.08~pc) across the core. 
Similarly, \citet{Koumpia_2020} use radiative transfer modeling of several infall models to predict line profiles of dense gas tracers  \ntwohp and \ce{H2D+} across several starless cores. 
They find that while the SIS model is clearly a poor match to cores without central point sources,
a larger survey is needed to discern between other core contraction models.

\index{Dense Core!Infall|)} 

\subsubsection{Rotation}
\label{sec:coreRot}
\index{Dense Core!Rotation|(}\index{Angular Momentum|(}

Rotation is critical in the core collapse process leading to the creation of protostellar systems. 
A well-known problem is that
molecular cloud cores have far greater angular momentum (by 6--7 orders of magnitude) than is measured in individual stars \citep{McKeeOstriker07}. 
Indeed, if angular momentum conservation holds during core collapse, 
large disks develop very fast 
(before the Class 0 stage) in hydrodynamic simulations\index{HD}, in contradiction with 
observations \citep[e.g.,][]{Maury_2010,Segura-Cox2018-Disk_Size,Maury2019-Calypso}.
Magnetic braking has been considered as the main mechanism for cores to lose angular momentum, 
but it can be too efficient, such that no disk could form during core collapse and protostellar formation.
Non-ideal MHD\index{MHD} effects could provide a way to allow for the decoupling between core material and magnetic field\index{Magnetic field}
\citep[see the review by][]{Li_PPVI}.
The angular momentum of dense cores, $L$, is therefore an important quantity in shaping the outcome of subsequent evolution.

\begin{figure*}[htb]
\centering
\includegraphics[width=\textwidth]{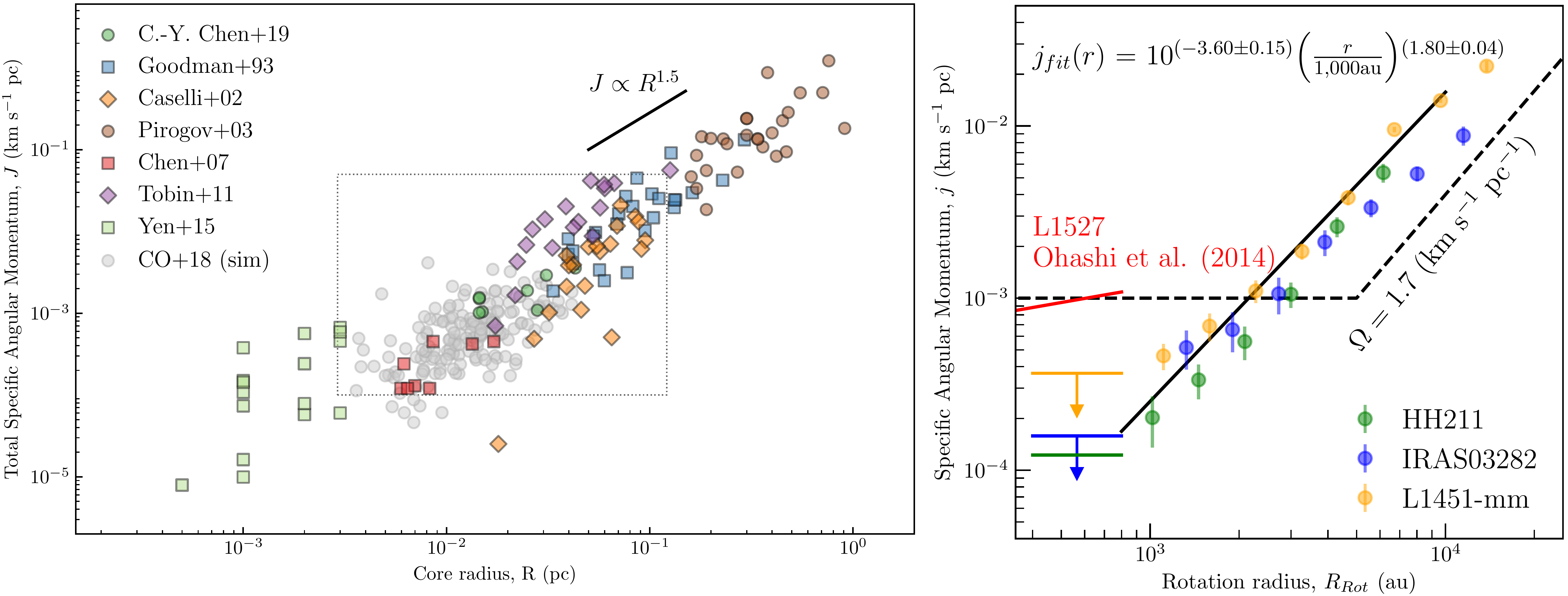}
\caption{\small
{\it Left:} Total specific angular momentum$-$size ($J-R$) correlation for a sample of dense cores \citep[adapted from][]{CChen_2019}. 
The dashed box marks the region shown in the right panel. 
{\it Right:} Radial profile of the specific angular momentum, $j(r)$, for three YSOs \citep{Pineda2019-j}.\index{Young Stellar Object (YSO)}
The solid black line shows the best-fit power-law relation to the data between 800 and 10,000 au. 
The dash curve shows the previously proposed specific angular momentum profile \citep{Belloche_2013}. }
\label{fig:j-r}
\end{figure*}

The size-dependence of the total specific angular momentum, {$J \equiv L/M$,} at sub-pc scales ($\sim$0.01$-$1~pc) within cores and clumps is long-known and well-established from early observations \citep{Goodman_1993,Caselli_2002,Pirogov_2003}. 
Recent interferometric observations \citep{XChen_JR_2007, Tobin2011, Yen_JR_2015} and numerical simulations \citep{CO18} confirmed and extended this trend down to scales $\sim$0.001~pc.
These studies revealed a power-law correlation between $J$ and the core size, $R$, such that $J \propto R^\alpha$, with $\alpha \approx 1.5$ (see the left panel of Fig.~\ref{fig:j-r}; also see reviews by \citealt{Belloche_2013} and \citealt{Li_PPVI}).

The $J-R$ correlation 
over three orders of magnitude in spatial scales is puzzling, and may suggest that gas motion in cores originates at scales much larger than the core size \citep{Walch2010-Turbulent_Cores,CO18}. 
This is inconsistent with the classical theory, which envisioned 
cloud cores to form after losing magnetic and turbulent support from the cloud material \citep[][]{Shu1987}. 
In fact, in addition to the $J-R$ correlation, the ratio between rotational energy and gravitational energy \citep[$\beta_E$; see the definition in][]{Goodman_1993} is typically $\sim$0.02 and is relatively independent of core/clump size. 
\cite{CO18} estimated the rotational-to-kinetic energy ratio in simulated cores, and found that {the ratio between rotational energy and the total gas kinetic energy,} $E_{\rm rot}/E_K\sim 0.1$ on average, is also independent of core/clump size. 
These results show that rotation is not dominant dynamically within dense cores. 

Calculations of $J$, as in Fig.~\ref{fig:j-r} (left), generally assume a constant density for the core, which may contribute to the scatter in the observed data points if the included cores have varying radial density profiles.
In addition, linear fitting of the observed velocity gradient across individual cores is commonly adopted to estimate the core's angular momentum, which is based on the assumption of rigid-body rotation.
However, recent high-spectral resolution ($\Delta v \approx 0.025$~\kms) observations by \cite{CChen_2019} found that 
while the fitted linear velocity gradients across their dense core targets nicely agree with the $J \propto R^{1.5}$ correlation (see the colored symbols in the left panel of Fig.~\ref{fig:j-r}), 
the detailed gas structures in the PV space are not consistent with rigid-body rotation \citep[see also][]{Belloche2002-IRAM04191}. 
Magnetic braking is especially important at $r$$<$1000\,au, and therefore 
these observations give support to turbulence (or some cloud-core interaction) 
as the main driver of the core's angular momentum, which we discuss in more detail later (Sect.~\ref{sec:coreJ}).
\index{Angular Momentum|)} 

\index{Specific Angular Momentum|(} 
Instead of using a single-value estimate of total angular momentum for the entire core,
observing the spatial and temporal evolution of angular momentum within one core is more challenging {but} has become feasible recently. 
Early work by \cite{Yen2010, Yen2011} combined multiple observations with different molecular lines to probe the radial profile of the specific angular momentum within individual cores, $j(r) = r \times \delta v$, in the B335\index[obj]{B335} Bok globule. 
In the case of rigid-body rotation and uniform density the total specific angular momentum corresponds to the radial specific angular momentum at that scale, $J = j(R)$, while for other cases there is a normalization correction \citep[see][]{Pineda2019-j}.
While the data points were sparse, the transition from rotation ($j\propto r^\alpha$) to infall ($j\approx$ constant) was clear, which was later confirmed with numerical modeling \citep{Kurono2013} and extended to $\sim$10~au scale by additional ALMA observations \citep{Yen15_B335disk}.
\cite{Ohashi_2014} also reported that the $j(r)$ profile of the inner core is almost flat for L1527 IRS\index[obj]{L1527 IRS} down to 54 au.

While interferometers like ALMA and SMA provide the necessary angular resolution at 
high frequencies ($\gtrsim 100$~GHz) for studying gas kinematics at 
$10-10^3$~au scales \citep[e.g.,][]{Yen_jprofile_2013, Yen2017, Tobin_BHR7_2018}, 
observations {toward the innermost envelopes}
($\sim$5000~au or $0.01$~pc) are difficult because of the dynamic range {needed over broad} spatial scales.
\cite{Pineda2019-j} re-analyzed  \ammo data from VLA, and 
resolved the complete radial profile of the specific angular momentum $j(r)$ in three young objects down to several hundreds of au.
The inferred angular momentum profile follows a relation $j \propto r^{1.8}$, indicating that the core is close to rigid-body rotation ($j\propto r^2$) while the imprint from the initial turbulence is still present. 
More recently, \cite{Gaudel2020-j} combined IRAM PdBI and 
30-meter observations with two molecular lines to probe dense gas kinematics in 12 Class~0 protostellar envelopes between $\sim$50--5000~au. 
They found a similar trend that $j \propto r^{1.6}$ above 1600~au and $j\approx$ constant between 50--1600~au.

We note that angular momentum conservation during the collapse is usually invoked 
as a simple explanation 
for the regime with flat $j$ distribution, while \cite{Takahashi2016-Origin_j} proposed a model 
that considers the widening of the region during the collapse to account for a regime
of constant angular momentum regardless of its initial distribution. 
On the other hand, measurements of velocity gradient at 1000 au scale around young objects do present a wide range of angular momentum, showing possible varieties of disk-forming conditions \citep[e.g.,][]{Yen_JR_2015}. 
Nevertheless, the observed angular momentum could have large uncertainties due to projection effects, as suggested by \cite{Zhang_obsJ_2018} via synthetic observations.
\index{Dense Core!Rotation|)}\index{Specific Angular Momentum|)}

\subsubsection{The Origin of Angular Momentum in Cores}
\label{sec:coreJ}\index{Angular Momentum|(} 

There are various proposals for the origin of angular momentum at the core scale, which are not mutually exclusive. 
We roughly categorize them into (1) anisotropic accretion, 
(2) large-scale rotation in the parent structure, and (3) local turbulence.

\cite{Kuznetsova2019-Origin_J,Kuznetsova2020-Cores_MHD} followed the angular momenta within the immediate surrounding of forming protostars in both HD\index{HD} and MHD\index{MHD} simulations, 
and concluded that episodic accretion from filaments onto cores
induced by multi-directional flows around the cores dominates the gas dynamics at core scales. 
This suggests the star-forming environment is highly heterogeneous, which has been noted in previous simulations by \cite{Kuffmeier2017-GMCtodisk,Kuffmeier2018-episodic}. 
Considering that cores form from filament fragmentation, \citet{Misugi_FilaFrag_2019,Misugi_FilaFrag_2023} proposed a model 
that conserves the angular momentum originated from larger scales during the anisotropic contraction 
that forms the filaments and the further axial fragmentation that leads to the formation of cores. 
They pointed out that the observed specific angular momentum within cores could be reproduced by simulated cores formed from 
filament fragmentation\index{Filament!Fragmentation} if the velocity structure within the filament follows a one-dimensional Kolmogorov power spectrum 
or with an anisotropic model with more power in the transverse direction.

Observations have shown stellar spin alignment in some open clusters \citep{Corsaro_2017_star_j, Kovacs_2018_star_j_obs}, 
and coherent global rotation was shown for one of those clusters \citep{Kamann_2019_star_j_obs}. 
Nevertheless, numerical simulations suggest that the stellar spin alignment happens only when the 
global 
rotation of the star-forming clump is significant with respect to the turbulence \citep{Corsaro_2017_star_j}. 
Alternatively, alignment could also occur within sub-clusters if the rotation at larger scales is less important \citep{Rey-Raposo_2018_star_j_simu}.
Furthermore, if core-embedding filaments have non-zero angular momenta as suggested by a recent observation \citep{Hsieh_2021_angmom}, 
it is possible that cores inherit rotation from their parent filaments. 
This behavior is consistent with the observations of the massive infrared dark cloud G28.37+0.07\index[obj]{G28.37+0.07}, 
where outflows (as a proxy of rotation axis) 
are aligned mostly perpendicular to the parent filaments \citep{Kong2019-Aligned_Outflows}.

However, angular momenta within cores do not seem to always align with the large-scale rotation.\index{Magnetic field}
For example, the rotating cores in the Orion A cloud do not seem to be correlated to the rotation 
of the large-scale filament \citep{Tatematsu_2016}.
In general, protostellar outflows (which are theoretically perpendicular to the disk plane) are not 
aligned with their parent structures \citep{Stephens_2017} and ambient magnetic field directions \citep{Hull_TADPOL_13}.
The same rotation-magnetic field misalignment is also found in simulated cores by \cite{CO18}, 
who concluded that cores must acquire rotational motions from local turbulence so that 
their rotational axes are independent of the magnetic field directions. 
This is consistent with the results discussed in \cite{Kuznetsova2019-Origin_J}, 
who showed that the angular momenta of individual cores are not strongly affected by the rotation of the parent cloud. 
\cite{Hennebelle_2018_FRIGG} also pointed out that core rotation may be primarily inherited from the initial turbulence within the core itself.

As originally pointed out by \cite{Burkert2000-Turbulent_j}, 
the observed rotation-like features of a linear velocity gradient may arise from sampling of turbulence at a range of scales.
{Since,} $J = L/M \sim R \cdot v_{\rm rot}$, the correlation $J \propto R^{1.5}$ simply suggests that $v_{\rm rot}\propto R^{0.5}$ \citep{CO18}.
Such a relation is similar to the so-called Larson relation of the turbulent interstellar medium, according to which 
turbulent velocities (traced by spectral linewidth along the line of sight) increase roughly as $\ell^{1/2}$ where $\ell$ is the size of the system.
Taken together, these results may thus suggest that the rotational velocity in cores is inherited from the overall turbulent cascade.

We note that it is possible that angular momentum at core scale does not transport all the way to the disk-forming scale, and the rotation of the protostellar disk is independent of that in the core/envelope \citep[see recent observations by][]{CChen_2019}.
This would require a separate mechanism for disks to gain angular momenta. 
Alternatively, magnetic braking could be highly dependent on individual core properties. 
Outward transport of angular momentum from the inner part could differ significantly between individual cores/envelopes. 
A highly-magnetized core with identical initial angular momentum likely forms a significantly smaller disk than a weakly-magnetized core \citep[see e.g.,][]{Li_PPVI}.

\cite{Gaudel2020-j} found difference
of rotation direction of the outer core with respect to that of the inner core, which suggests that the rotation of the disk itself might not be inherited from the turbulence of the environment directly. 
\cite{Verliat_2020} proposed a model that forms a disk without any initial rotation with respect to the geometric center of a prestellar core, or even without any turbulence. 
Since the angular momentum is defined with respect to a certain point, it can be easily generated during the core collapse as long as there exists some asymmetry in density or velocity that deviates the collapse from the original 
center of mass. 
The rotation thus generated is not related to the larger scale kinematics, but to density perturbations. 

\index{Angular Momentum|)}

\subsubsection{Core Fragmentation}\index{Dense Core!Fragmentation|(}
One of the unknowns in our understanding of star formation is the determination of how dense cores fragment during the star formation process 
\citep[see][in this volume]{Offner+2022}.
This is important since it can change the mapping between the CMF and the IMF \citep[e.g.][]{Goodwin2008-CMF_IMF}, and it can also affect the properties of multiple systems \citep[e.g.][]{Offner2010-Turbulent_Fragmentation,Walch2012-Core_Fragmentation}.
Several observations of samples of \emph{starless} cores have attempted to determine the degree of fragmentation by carrying out relatively shallow dust continuum observations with interferometers \citep{Schnee2010-Starless_CARMA, Dunham2016-ALMA_Cha, Kirk2017-ALMA_Oph, Tokuda2020-ACA_Taurus_Survey, zha21, Sahu2021-Orion_Cores_ALMA}, 
yielding a small number of continuum detections.
Despite this low number of detections, these results are mostly compatible with the predictions (using synthetic observations) from numerical simulations of turbulent fragmentation \citep[e.g.,][]{Offner2012-Predictions_Observations, Kirk2017-ALMA_Oph}. 

The multi-scale fragmentation study by \citet{Pokhrel_Perseus_2018} indicates that the number of observed fragments is generally lower than 
expected from thermal Jeans fragmentation. This suggests that the 
mass scale of fragment formation is somewhat larger than the thermal Jeans mass, 
as is the case in magnetized clumps/cores with mildly supercritical 
mass-to-magnetic-flux ratios \citep{Das2021}.

ALMA observations of the L1521F\index[obj]{L1521F} protostellar core \citep{Tokuda2014-ALMA_L1521F} revealed arc-like features and high-density substructures in different  molecular transitions, suggesting dynamic gas interaction. 
Other efforts have focused on deeper observations of dense gas in the Barnard 5\index[obj]{Barnard 5} region \citep{Pineda2015-Fragmentation_B5, Schmiedeke2021-Filament_Properties, Pineda2021-Ions_Neutrals,ChenMike2022-B5}, 
which revealed the presence of fragmenting narrow filaments in the process of forming a wide separation quadruple system. 
The substructures found in this region are at scales much smaller than the Jeans scale, and therefore these observations reveal directly the turbulent fragmentation process within a core.

Recent deep ALMA observations of the prestellar core L1544\index[obj]{L1544} have 
revealed fragmentation in the inner 1500 au radius of a truly starless core \citep{Caselli2019-ALMA_L1544}. 
These fragments (also called `kernels') are detected at 
densities higher than $10^6$ \cc, where the dense core density profile is smooth and flat. 
These observations are 
in agreement with non-ideal MHD\index{MHD} simulations of a contracting dense core, for which the synthetic interferometric observations are a good match of the observations.
This result highlights the need for deeper interferometric observations of additional prestellar cores to test fragmentation scenarios.
\index{Dense Core|)}\index{Dense Core!Fragmentation|)}

\section{\uppercase{From Cores to Accreting Disks:\\Implication from Kinematics}\label{sec:disk}}
The classical picture of star formation focuses on the material in an isolated parental dense core 
that undergoes gravitational collapse \citep{Larson1969,Shu1977,Terebey1984}. 
In addition, all the material used to form stars and planets must pass through the dense core.
In this scenario, star- and disk-formation can be studied in numerical simulations of isolated/closed boxes, allowing the implementation of a range of physical processes 
 \citep[e.g.,][]{Zhao2018,MachidaBasu2019,Marchand2020}.

However, as discussed in the previous sections (e.g., Sect.~\ref{sec:fil_core}), 
\emph{Herschel} observations have consolidated the result that molecular clouds 
are highly sub-structured.
While there are isolated cores (e.g., Bok globules), it is well
established that most dense cores are harbored in filaments 
\citep[see also][]{Andre2014-PPVI}.
As such, most cores do not form and will not evolve in isolation, and the interplay between dense cores and molecular clouds must be understood to improve our understanding of star and planet formation.

In this section we discuss how the dense core material is delivered to the disk forming scales. 
We present the observational and numerical evidence that disks cannot be considered 
as isolated entities detached from the dense gas at larger scales. 
The large scale environment is important to fully capture the disk- and star-formation process.

%
%
\subsection{Non-axisymmetric Accretion onto Disks}
\index{Streamer|(}\index{Young Stellar Object (YSO)|(}
The evidence for velocity coherent narrow structures funneling material, `streamers', around young stellar objects (YSOs) at different evolutionary stages is growing. 
The kinematic structure of streamers 
combines a smooth velocity gradient (driven by gravity) 
and some amount of rotation. 
Therefore, modelling is required to truly confirm their nature.
Here we present and discuss different examples of streamers at different evolutionary stages.
We divide the discussion based on the observationally-defined Classes of YSOs \citep{Lada1987-classes,Andre2000-PPIV,Dunham2014-PPVI}, 
from deeply embedded, envelope-dominated protostars with high ratios of submillimeter to bolometric luminosity (Class~0), 
to more evolved protostars in which the circumstellar envelope no longer dominates and the central star$+$disk system 
is visible in the near-infrared (Class I),
to pre-main sequence stars with protoplanetary disks and negligible circumstellar envelopes (Class II).

\subsubsection{Highly Embedded Class~0 Protostars}
Non-axisymmetric structures around Class 0 objects were observed with 
{\it Spitzer}, and they were interpreted as the result of the collapse of 
non-equilibrium structures \citep{Tobin2010-Spitzer}.
Initial observations suggesting the presence of accretion streamers in  young Class 0 objects (Serpens SMM1 and Ser-emb-8)\index[obj]{Serpens SMM1, Ser-emb-8} were presented by \cite{legouellec2019-streamers}.
These ALMA dust continuum polarization observations at 870 $\mu$m, with a resolution better than 150 au,
revealed the presence of two 
magnetically-aligned filamentary features
reaching down to the disk scales. 
The authors suggest that at least one of these features 
might indeed be related to accretion streamers. 
Unfortunately, there are no complementary molecular line observations that 
could confirm the nature of the accretion stream.

\begin{figure}[ht]
    \centering
    \includegraphics[width=\columnwidth]{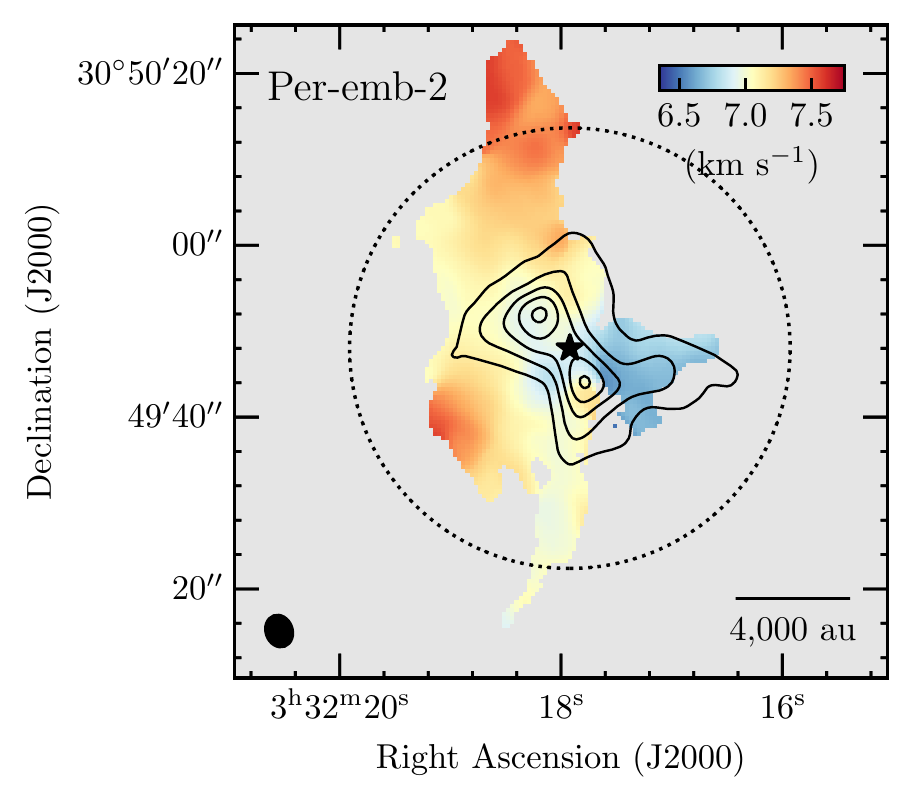}
    \caption{\small
    A large scale streamer and the dense core around Per-emb-2\index[obj]{Per-emb-2}. 
    The centroid velocity from \ce{HC3N} (10--9) is in the background, original data from \cite{pineda2020-streamer}.
    The dense core as traced in \ntwohp (1--0) integrated intensity is shown in contours.
    The beam size and a scale bar are shown at the bottom left and right corners, respectively.}
    \label{fig:Per-emb-2}
\end{figure}

Recently, NOEMA observations of a Class 0 source in Perseus (Per-emb-2)\index[obj]{Per-emb-2} revealed the presence of a large scale streamer \citep{pineda2020-streamer}.
These observations detected the streamer in molecular line emission, but not in the 
(less sensitive) dust continuum emission. 
The derived velocity map in the streamer is smooth and 
shows a streamer that begins at $\approx$10,000 au from the central YSO (beyond the classical dense core seen in \ammo and \ntwohp).
It displays a velocity gradient well matched by a 
model of free-falling material\index{Free-fall} and rotation.
Figure~\ref{fig:Per-emb-2} shows the velocity  derived using \ce{HC3N}, while the dense core traced with \ntwohp is shown in contours.
The estimated average infall rate from the streamer onto the disk forming scales is $10^{-6}$ \msunyr, 
which is comparable to the current {protostellar} accretion rate of $7\times 10^{-7}$ \msunyr \citep{frimann2017-burst_masses,hsieh2019-perseus_survey}.
This suggests that the streamer could modify protostellar accretion by funnelling extra material to the central region.

A similar analysis was performed for Lupus 3-MMS\index[obj]{Lupus 3-MMS} \citep{Thieme2022-Lupus_Streamer}, which 
highlights the presence of multiple accretion streams seen in \CeightO with ALMA. 
A total of four different structures are well modelled with infall models (including rotation), 
with a total infall rate on the streamers of  $0.5-1.1 \times 10^{-6}$ \msunyr. 
This is the first case of multiple streamers modeled in a single source.

Recently, observations of molecular line emission revealed more streamers 
toward B335\index[obj]{B335} \citep{Cabedo2022-B335}, Per-emb-8\index[obj]{Per-emb-8} ({\it Segura-Cox et al.}, private communication) 
and IRAS16293A\index[obj]{IRAS16293A} \citep{Murillo2021-IRAS16293}.
In particular, it is surprising that the detection of the streamer in the well-studied IRAS16293A had remained unidentified, 
highlighting the relevance of re-evaluating previous observations in view 
of these new features.

\subsubsection{Class I YSOs}

\begin{figure}[ht]
    \centering
    \includegraphics[width=\columnwidth]{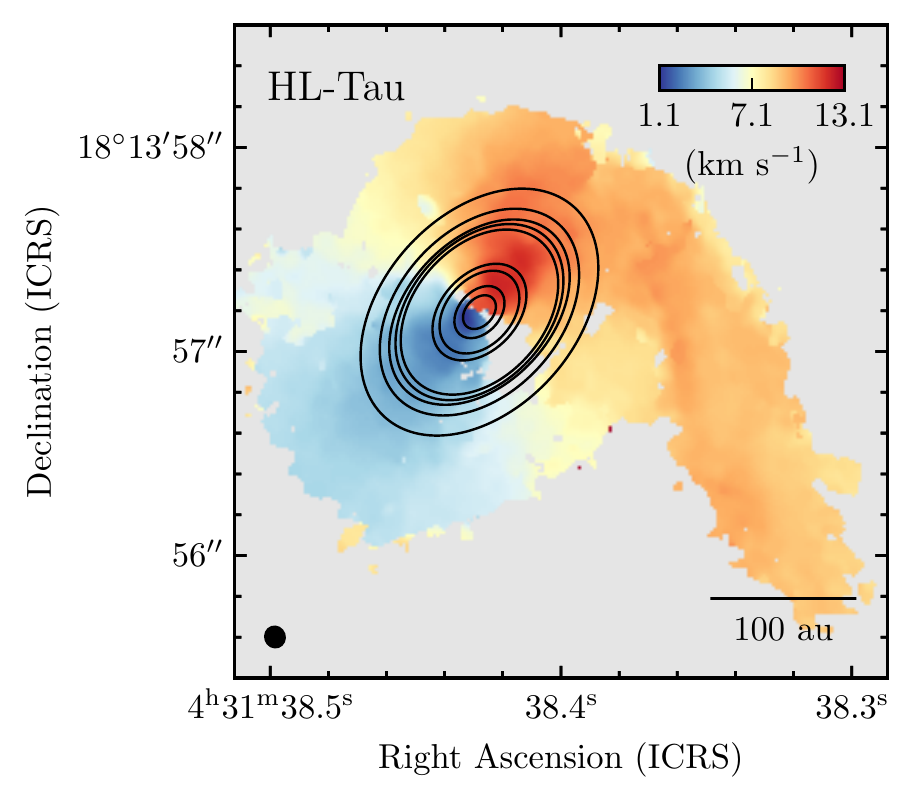}
    \caption{\small
    The streamer and disk around HL Tau\index[obj]{HL Tau}. 
    The centroid velocity from \ce{HCO+} (3--2) is in the background, original data from \cite{yen_2019-hltau_spiral}.
    Dust continuum rings and gaps in the disk, as identified in \cite{ALMA_2015-LBC_HLTau}, are shown in contours.
    The beam size and a scale bar are shown at the bottom left and right corners, respectively.}
    \label{fig:HLTau}
\end{figure}

The surprising observations of rings and gaps in the disk around HL Tau\index[obj]{HL Tau} with ALMA \citep{ALMA_2015-LBC_HLTau} suggested that planet formation might already be ongoing in these early stages. 
Therefore, several observations have focused on studying the disk properties and their relation with the surrounding envelope.
ALMA observations of HL Tau revealed an accretion stream of $\approx$200 au in length \citep{yen_2019-hltau_spiral}, that brings material down to the disk-forming scales in an asymmetrical fashion. Figure~\ref{fig:HLTau} shows the extent of the gas disk kinematics and the streamer, as traced with \ce{HCO+} (3--2), while the rings and gaps in the disk are marked by the contours. 
Using a parametric modelling of the kinematics, \cite{yen_2019-hltau_spiral} show that the gas kinematics of the streamer are a combination of infall and rotation. 
Recently, \cite{garufi2021_streamer} also detected the streamer in CS (5--4) with ALMA, 
which when combined with \ce{HCO+} allows for a good fit of the emission using a streamline model as in the Class 0 case \citep{pineda2020-streamer}.
Moreover, \cite{garufi2021_streamer} present evidence for shocks\index{Shock}, 
which are usually traced in SO and \ce{SO2} in disks \citep{Sakai2014-L1527_SO}, 
that would mark the interaction region between disk and streamer. 
This example shows that the streamer delivers material to the disk,  which 
is then transported through the disk without perturbing the dust structures in the midplane.

In a similar way, the youngest Class I disk with rings \citep[IRS 63;][]{seguracox2020-irs63_rings}\index[obj]{IRS 63} also presents a clear streamer detected in \ce{HCO+} ({\it D. Segura-Cox \etal}, private communication) 
This streamer is also well described with a streamline model, and the derived streamer infall rate is comparable with the accretion rate\index{Disk:Accretion}.
This is a second example of rings identified in 
dust emission, which are unperturbed by the streamer infall, and with clear evidence 
of shocks\index{Shock} thanks to the detection of bright SO emission where streamer and disk meet. 
Further analysis of the gas disk could shed some more light on how the streamer would modify the gas disk.

Finally, NOEMA observations revealed the presence of a streamer in \ce{H2CO} and \CeightO toward 
the Class I object Per-emb-50\index[obj]{Per-emb-50} \citep{valdiviamena2021-streamer}.
This streamer has an extent $>$2,000\,au and 
is well described with a streamline model, confirming that this material is infalling to the central disk scales. 
The derived streamer infall rate is comparable to the accretion rate derived from NIR spectroscopy \citep{fiorellino2021-NGC1333}, 
which once again points to the relevant role of streamers in the accretion process of young stars.
Also, \cite{valdiviamena2021-streamer} quantify
the infall rate along the streamer, showing 
that the variations are within a factor of $\approx3$ from the average value and systematically above the current protostellar accretion rate.

\subsubsection{Class II Objects}

\begin{figure}[ht]
    \centering
    \includegraphics[width=\columnwidth]{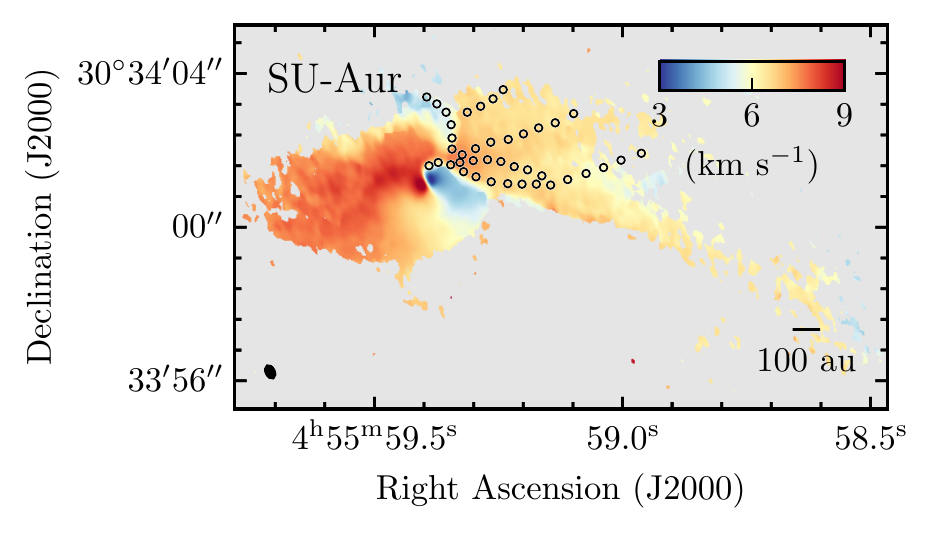}
    \caption{\small
    Several streamers toward SU Aur\index[obj]{SU Aur} identified in CO and scattered light images. 
    The centroid velocity from CO (3--2) is in the background, reprocessed data from \cite{Ginski2021-Streamer,akiyama2019-suaur_streamer}. 
    The disk rotation is clearly seen in the central region, while several streamers are seen, both kinematically and in scattered light images.
    The circles show the streamers identified in the SPHERE scattered light images \citep{Ginski2021-Streamer}. 
    The long streamer extending south-west matches very well the streamer previously seen in HST Coronagraphic observations \citep{Grady2001-HST_Streamer}.
    }
    \label{fig:stream-ClassII}
\end{figure}

Recently, streamers have also been identified toward Class~II objects. 
Despite the low level of envelope surrounding them, sensitive ALMA observations 
have revealed the presence of gas streamers.
In the case of SU Aur\index[obj]{SU Aur}, a long tail ($\approx$3000 au at a distance of 158 pc) is detected in \hbox{CO (2--1)} with ALMA \citep{akiyama2019-suaur_streamer} with a coherent centroid velocity map. 
This suggests that the structure is associated with a single infall event, although the original discovery paper
suggested that  a collision with a (sub)stellar intruder or a gaseous blob {was} probably the most plausible explanation.
Complementary SPHERE observations of SU Aur ruled out a recent close encounter or the ejection of a dust clump for their origin  \citep{Ginski2021-Streamer}, 
suggesting that material is falling onto the central system (see Fig.~\ref{fig:stream-ClassII})\index{Disk:Accretion}.
Moreover, these observations revealed that more than one streamer is approaching the disk. 
This source represents one of the clearest examples of late accretion onto a disk.
Apart from SU Aur, there are at least two other stars in the L1517 cloud that show similar patterns of external streamers: 
AB Aur\index[obj]{AB Aur} \citep{Grady1999} and GM Aur\index[obj]{GM Aur} \citep{Huang2021}.

Similarly, new ALMA observations of DG Tau\index[obj]{DG Tau} revealed the presence of two streamers reaching the disk \citep{garufi2021_streamer}, and one of these streamers was previously identified as an accretion flow in an outflow/wind study \citep{Guedel2018-DGTau}. 
The streamers are detected in the CO (2--1) and CS (5--4) molecular lines and 
extend from $\approx$300 au in projected distance down to the disk edge.
The molecular line observations allow for a determination of a smooth velocity gradient in the streamers, which are well modelled with streamlines. 
The presence of these streamers shows the emission landing on the disk, 
and the landing spot on the disk is identified with typical shock tracers (SO and \ce{SO2})\index{Shock}. 
This again supports the idea that streamers can deliver material down to the disk itself.

High resolution ALMA observations of [BHB2007] 1\index[obj]{[BHB2007] 1} \citep{alves2020_bhb2007-1} show a nearly edge-on gapped disk in 
dust continuum emission.
The complementary molecular line observations revealed two streamers identified in the  CO (2--1) transition from 
$\approx$2,000\,au down to the disk edge.
A follow-up analysis of VLA and NACO observations confirmed the presence of a substellar object \citep{Zurlo2021-BHB2007-1_BD}.
This presents further evidence that the presence of a streamer does not hinder the formation of companions.
\index{Streamer|)}

\subsection{Numerical Simulations}
Considering that streamers have been observed for YSOs of Class 0, I, and II, it raises the question whether all streamers emerge through the same mechanism or whether there are differences depending on the evolutionary stage of the objects.  
As there is little information about the origin of streamers available yet, one of the key questions is whether streamers actually correspond to 
streams of gas or whether their origin is of a different kind.  
In this subsection, we provide an overview of currently suggested mechanisms that can explain the presence of such streamers.

\begin{figure*}[ht]
    \centering
    \includegraphics[width=0.99\textwidth]{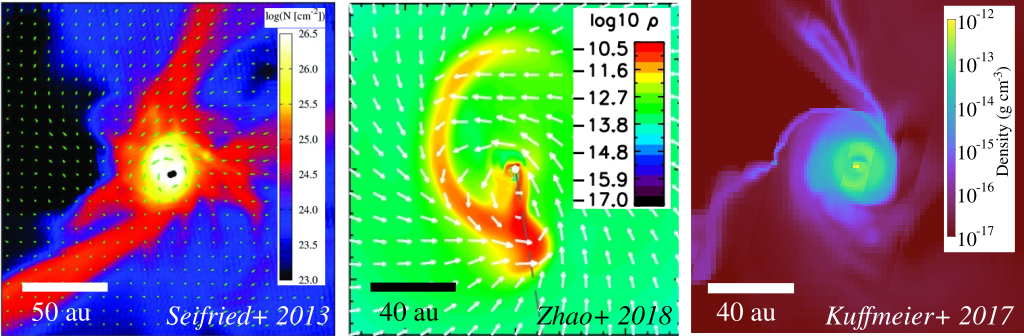}
    \caption{\small
    Streamers\index{Streamer} around disks in different MHD\index{MHD} simulations.
    These simulations include different setups, but similar streamers emerge as a result of asymmetric infall. 
    The left panel shows results from simulations first presented in \cite{Seifried2013}, the middle results presented in \cite{Zhao2018}, and the right panel results first presented in \cite{Kuffmeier2017-GMCtodisk}.   
    }
    \label{fig:streamers_disk}
\end{figure*}

\subsubsection{Accretion Channels from Core to Disk}
Numerical simulations of Bonnor-Ebert spheres using smoothed particle hydrodynamics (SPH) showed that the perturbations generated 
by the turbulence velocity would end up in irregular streams feeding the central disk \citep{Walch2010-Turbulent_Cores}.
Similarly, numerical simulations including turbulence and magnetic fields\index{Magnetic field} \citep{Seifried2013,SeifriedBanerjeePudritzKlessen2015} 
showed that turbulence causes accretion onto the disk along distinct channels during the deeply embedded stage, 
in contrast to the previous picture that considered a coherent rotational structure during the collapse phase. 
Several MHD\index{MHD} simulations from different groups also confirm the presence of channeled accretion along magnetized accretion streamers 
\cite[e.g.,][]{Joos2013,Li14_turbulenceDisk,Masson2016,Kuffmeier2017-GMCtodisk,Matsumoto_2017,Zhao2018,Lam19_diskformation,Hennebelle2020}.    
While different types of streamers have been seen in models already in the past $\sim$ ten years (Fig.~\ref{fig:streamers_disk}), 
relatively little work has been done in analyzing the properties and effects of accretion streamers. 
Carrying out parameter studies using isolated core collapse models, 
the focus has been more on studying how effects such as (non-)ideal MHD \citep[e.g.,][]{Tomida_nonidealMHD_2015,Masson2016,Wurster2016} 
including the effect of the ionization degree \citep{Wurster2018ionization,KuffmeierZhaoCaselli2020}, 
turbulence \citep[e.g][]{Seifried2012,Joos2013,Gray2018,WursterLewis2020}, orientation of the magnetic field\index{Magnetic field} 
with respect to the total angular momentum vector \citep[e.g.,][]{Joos2012,Krumholz2013,Tsukamoto2018misalignment,Hirano2020}, 
or the grain size distribution \citep{Zhao2016nosmalldust-disk,Marchand2020} affect the size of circumstellar disks.
Therefore, modelers and theorists have not paid much attention to streamers as they
presumably violate the picture of symmetrical collapse and as they are difficult to approximate in simpler 2D or 1D models.

Core collapse models that start with an initially perturbed velocity distribution of the gas in the spherical core  
frequently show the presence of accretion streamers that feed the young disk with fresh material from the dense core 
\cite[e.g.,][]{Walch2010-Turbulent_Cores,Seifried2013,Seifried_2015_filament,Joos2013,Hennebelle2020,Zhao2018}.
Carrying out a parameter study including different levels of initial turbulence, \cite{Mignon-Risse2021-massive_core} found that the streamers were wider for higher levels of initial turbulence, 
possibly caused by turbulent reconnection \citep[e.g.,][]{LazarianVishniac1999,SantosLima2013}.

\subsubsection{Streamers from Interstellar Scales to Protostellar Scales\label{subsec:stream_sim}}

\begin{figure*}[ht]
    \centering
    \includegraphics[width=0.99\textwidth]{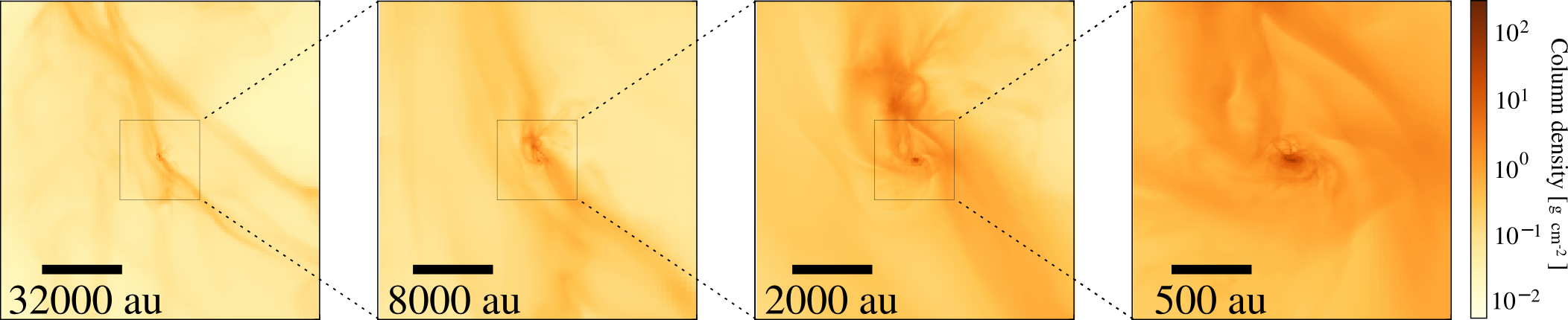}
    \caption{\small
    Zoom-in onto a forming protostar that is embedded in a turbulent birth environment of a molecular cloud. 
    It shows the presence of a bridge connecting the central protostar with the parental molecular cloud. 
    The snapshot is taken from multi-scale MHD\index{MHD} models by \cite{Kuffmeier2019-Bridge}.  
    }
    \label{fig:zoom-in}
\end{figure*}

While collapse models have become increasingly sophisticated in terms of including more physics in the simulations, 
they have fundamental limitations in their setups as highly idealized initial and boundary conditions remain. 
Since stars form inside the large-scale environment defined by MCs, 
and because some streamers are seen reaching out beyond the dense core \citep{pineda2020-streamer}, a 
more realistic framework for star formation would include the larger-scale environments of the MC in which stars are embedded.

Although several  specifically designed parameter studies investigated infall onto an existing disk  \citep{Vorobyov_extinfall_2015,Bae_infallondisk_2015,Lesur2015,LeeCharnozHennebelle2021,Kuznetsova2022}, 
only a few groups studied disk formation while covering the dynamical range from (giant) MC scales down to the disk in one self-consistent model.
\cite{Bate2018} and \cite{Lebreuilly2021} independently carried out population synthesis studies of the formation of individual disks in stellar clusters that emerged from the collapse of a massive clump of 100--1000 M$_{\odot}$ in mass and about 0.1 to 1 pc in length. \cite{Bate2018} carried out SPH models without magnetic fields, \cite{Lebreuilly2021} models with non-ideal MHD\index{MHD} using adaptive mesh refinement (AMR).  
While these models show the presence of accretion streamers and provide 
first statistical constraints on disk formation, they cannot take into account
the dynamics on MC scales that are responsible for inflow (see Sect.~\ref{sec:accretion}).

\begin{figure}[ht]
    \centering
    \includegraphics[width=\columnwidth]{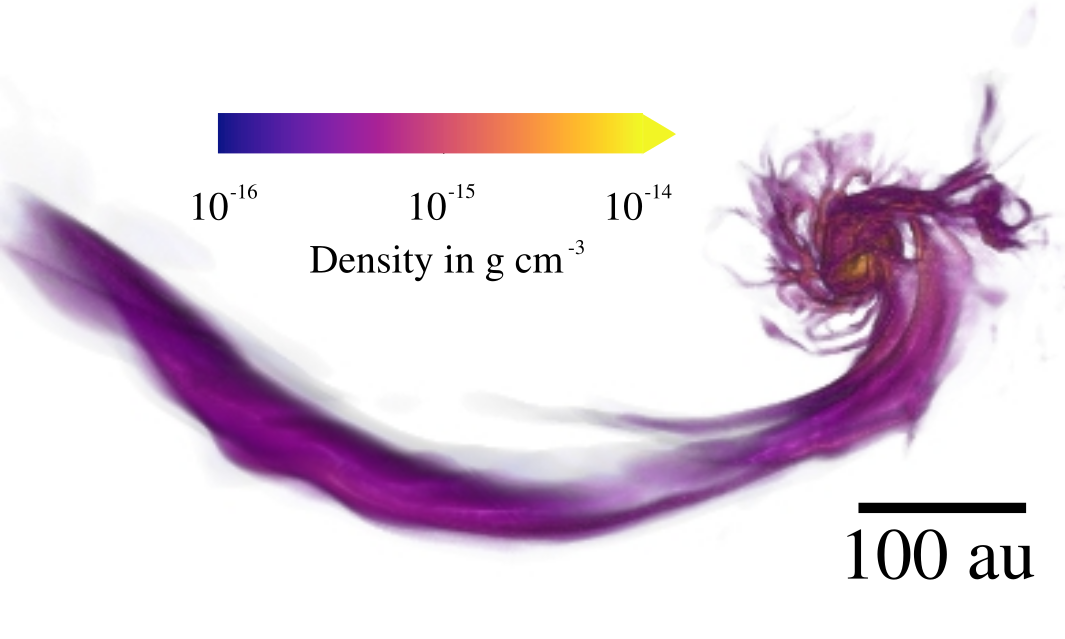}
    \caption{\small
    A streamer feeding an embedded disk \citep[visualization made by using data from][]{Kuffmeier2019-Bridge}.
    }
    \label{fig:streamer_and_blob}
\end{figure}

To cover the range of scales from the MC down to the disk \citep[see Fig.~\ref{fig:streamer_and_blob} and][see also sections \ref{sec:accretion} or \ref{sec::CoreEvo}]{Padoan2020,Pelkonen_CMF_2021}, \cite{Kuffmeier2016} started a sequence of papers started their models from a MC of $\sim$10$^5$ M$_{\odot}$ in mass and ($40$ pc)$^3$ in volume and analyzed the star-disk formation process in 3D MHD\index{MHD} zoom-in simulations with a resolution down to 0.06 au \citep[][]{Kuffmeier2018-episodic}, which marks the largest spatial coverage so far (see Fig.~\ref{fig:zoom-in} for a zoom-in onto a deeply embedded protostar forming in a filament; \citealt{Kuffmeier2019-Bridge}).
Their models \citep{Kuffmeier2017-GMCtodisk} confirm that accretion within the core scale happens via accretion channels \citep{Seifried2013}. They also predict the possibility of replenishing the disk (and thereby the mass reservoir for planet formation) with material that was originally not gravitationally bound to the collapsing core \citep[as observed for Per-emb-2][]{pineda2020-streamer}. 
Regardless of the (significant) differences in the %
physics and spatial coverage of these models \citep{Kuffmeier2017-GMCtodisk,Bate2018,Lebreuilly2021},
they agree on the key result that disk formation is a heterogeneous, environmentally dependent process. 

\index{Young Stellar Object (YSO)|)}

\subsubsection{Late Interaction with the Protostellar Environment} 
Stellar encounters can lead to the formation of spiral- or tail-like structures that may resemble the aforementioned accretion streamers. 
It is important to highlight that there are other mechanisms that can cause the formation of streamer-like structures. 

\subsubsubsection{Gas Capture}
Filamentary arms around YSOs might arise after the initial collapse phase if the star experiences an encounter with interstellar gas \citep{Bate2018}. 
The scenario was tested more in parameter studies adopting a setup of cloudlet-infall/capture with the codes PLUTO \citep{Dullemond2019} and AREPO \citep{KuffmeierGoicovicDullemond2020, Kuffmeier2021}. 
Sweeping up gas via Bondi-Hoyle accretion can lead to the formation of filamentary arms that can appear as accretion streamers. Interestingly, a (late) encounter can even induce the formation of a new, second-generation disk \citep{KuffmeierGoicovicDullemond2020} that is likely misaligned with the primordial disk from the initial protostellar collapse \citep{Bate2018,Kuffmeier2021}. 
Infall may therefore not only rejuvenate primordial disks, but also induce new misaligned disks that can explain shadows in scattered light observations of 
some systems \citep[e.g.,][]{Avenhaus2014,Marino2015,Benisty2017,Benisty2018,Casassus2018,Ginski2021-Streamer}.

\subsubsubsection{Stellar Fly-By}
\index{Fly-By|(}
We point out that a spiral or tail-like structure around a disk may not necessarily be associated with infall. 
It can also be a consequence of the interaction of a star-disk system with an external perturber, such as a binary component or a stellar fly-by. 
The possibility of an external star that perturbs the disk has been studied and discussed already for a few decades \citep[e.g.,][]{Clarke1993,Pfalzner2003}. 
Recent hydrodynamical\index{HD} models show that a star that encounters an existing star-disk system triggers a spiral structure \citep{Vorobyov2017-intruder,Cuello2020-flyby2}. 
When such a structure is observed from the `right' angle, 
it resembles the aforementioned accretion streamers 
proposed for Z CMa\index[obj]{Z CMa} \citep{Dong2022}. 
Apart from that, perturbations by stellar companions in the disk might be responsible for tearing up the disk, and hence induce misalignment of inner and outer disks \citep[e.g.,][]{Nealon2020-HD100453} 
similar to disk tearing in accretion disks around black holes \citep[e.g.,][]{Nixon2012}. 
However, it is difficult for an external perturber alone to tear up protostellar/-planetary disks because of the low viscosity in these disks \citep{Gonzalez2020-HD100453,NealonCuelloAlexander2020,Smallwood2021}.
An external large-scale streamer around a misaligned outer disk might therefore hint at second-generation disk formation 
\cite[see also Sect. 4.3 on misalignment in][in this volume]{Pinte2022}. 
\index{Fly-By|)}

\section{\uppercase{Conclusions and {Future} Challenges}}

\begin{figure*}[ht]
    \centering
    \includegraphics[width=0.8\textwidth]{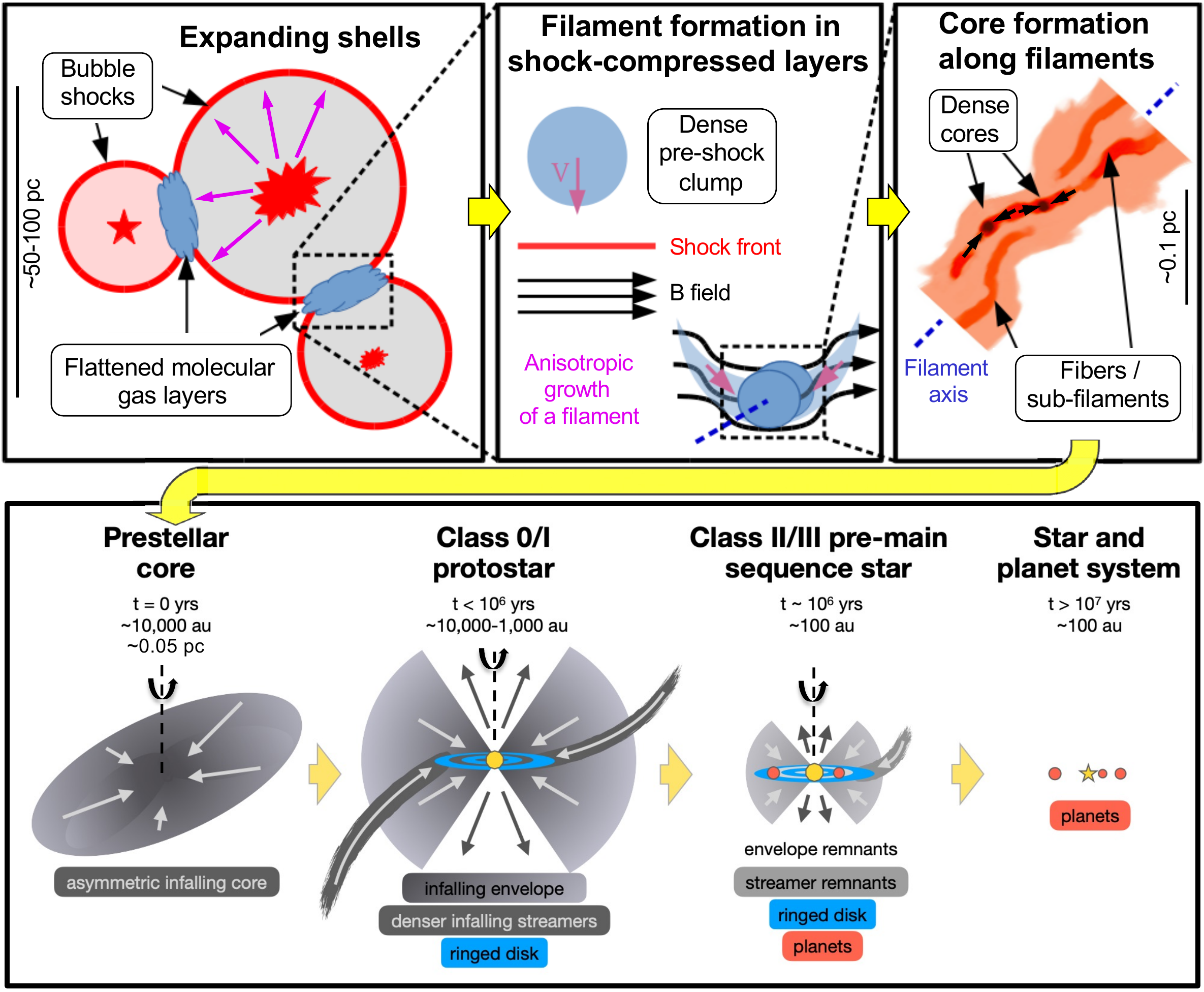}

    \caption{\small
    Sketch of the star-formation process on various scales, emphasizing the anisotropic growth of dense structures 
    in the cold ISM leading to the formation of stars and planetary systems. 
    It highlights the role of large-scale expanding bubbles in compressing interstellar matter in flattened gas layers and producing molecular filaments, 
    which themselves form rotating dense cores through axial gravo-turbulent fragmentation (top row, see Sects.~\ref{sec:bubble} and \ref{sec:filament}).
    The sketch also emphasizes the role of angular momentum and non-axisymmetric streamers at core/disk scales 
    (bottom row based from {\it D. Segura-Cox}, private communication); 
    the classical dense core and these streamers coexist at different evolutionary stages of protoplanetary disk formation (see Sects.~\ref{sec:core} and \ref{sec:disk}).
    The orientation of the streamers is not necessarily aligned  
    with the rotation axis of the dense core, but it does follow a 
    trajectory consistent with free-fall and rotation.
    }
    \label{fig:stream-sketch}
\end{figure*}

{Here, we summarize the coherent picture 
of the overall star formation process 
arising from the current observations and models.} 
Thanks to a new array of observations, we are now able to better connect the anisotropic flow of material through different levels of structures in the hierarchical ISM. 
In particular, the high spatial dynamic range of submillimeter dust continuum images from {\it Herschel} and {\it Planck} has been essential to connecting 
column density structures observed across a wide range of scales from giant \hone  and \hion bubbles to filaments and protostellar cores.
Large molecular line surveys have been crucial in providing kinematic tests of
filament formation and accretion models (see Sects.~\ref{sec:bubble} and \ref{sec:filament}), 
and tracing gas motions and angular momentum from cloud to core and disk scales (see Sects.~\ref{sec:core} and \ref{sec:disk}). 
In parallel, analytic models and numerical simulations have provided detailed predictions that can be tested directly against observational results. 
Over this wide range of scales, bubbles, filaments, cores, and disks 
appear to be fundamental, interconnected levels of ISM structures, differing in their 3D geometry and dominated by different physics 
(see sketch in 
Fig.\,\ref{fig:stream-sketch}). 
Specifically, compressive turbulent flows driven by the quasi-spherical expansion of large bubbles 
dominate on $\sim$1--100 pc scales, generating shock-compressed layers\index{Shock!Shock-compressed layers} of magnetized molecular gas. 
Dense, self-gravitating filaments form within these shock-compressed layers on $\sim$pc scales as a result 
of the combined action of magnetic fields\index{Magnetic field} and large-scale flows.  
The gravo-turbulent fragmentation of self-gravitating filaments leads to the formation of rotating prestellar cores 
on $\simlt$0.1 pc scales. 
Eventually, 
the gravitational collapse of rotating, non-axisymmetric dense cores 
produces protostellar systems with ringed disks and infalling streamers on  $\simlt$1000~au scales.\index{Streamer} 

In contrast to other scenarios, 
our proposed picture (cf. Fig.\,\ref{fig:stream-sketch})  
emphasizes the role of geometrical effects at different scales in the process. 
It also provides new clues toward understanding the origin of 
1) the {core/star} formation inefficiency, 2) the stellar initial mass function (IMF), and 3) the angular momentum of protostellar systems. 

Observations suggest a connection between the expanding bubbles 
in the ISM (in \hion and \hone) and the 
formation of dense filamentary structures, 
though detailed numerical investigations are needed to provide more direct comparisons to the observed bubble-filament connection.
Additional observational and theoretical studies are also required 
to understand the possible evolutionary link between atomic and molecular filaments 
and the role of \hion regions (i.e., formation and/or feedback) on the surrounding filamentary ISM.

While multiple mechanisms for filament formation have been presented, the current data are consistent with 
filaments being formed from converging flows within magnetized shock-compressed layers\index{Shock!Shock-compressed layers} 
(type-O/C mechanisms in Fig.~\ref{fig_class}) with further mass growth due to gravity. 
The critical feature of filament formation is thus one of compressive flows.
Several scenarios have been invoked to explain the origin of velocity-coherent fibers within filaments. 
These fibers are commonly, but not universally, observed, and observational tests are needed that 
can distinguish between, e.g., fragment-and-gather and fray-and-fragment models.

The filament line mass function (FLMF{-- Fig.~\ref{flmf_CMF}a}) appears as an important 
property of the population of molecular filaments. 
The critical line mass for nearly isothermal filaments sets a natural 
transition for the core formation efficiency within filamentary molecular clouds,  
between a regime with negligible formation of prestellar cores at low densities in 
subcritical filaments and a regime with relatively efficient core 
formation at high densities in thermally supercritical filaments 
(Fig.~\ref{aquila_cfe}). 
The shape of the prestellar core mass function, and by extension that of the stellar 
IMF, may be partly inherited from the FLMF {(see Sect.~\ref{sec:FLMF})}. 
Observed core spacings do not generally 
follow the simple periodic predictions of 
gravitational fragmentation models within near-equilibrium cylindrical filaments.
Filamentary fragmentation likely unfolds over multiple scales where either gravitational fragmentation 
of supercritical filaments, or gravity-induced turbulence dominate.

Prestellar cores are primarily found within thermally transcritical or supercritical filaments. 
One of the key properties of cores is the amount of rotation present, which is directly related to the origin of the angular momentum. 
The total angular momentum seen in cores is consistent with being dominated 
by the turbulent motions injected at the largest scales 
and may be partly inherited from the 
formation and fragmentation process of the parent filaments (e.g. Sect~\ref{sec:coreJ} and Fig.~\ref{fig:j-r}). 
While resolved observations of cores reveal
clear differential rotation, the scale on which specific angular momentum is constant (and gravity dominates) 
is just being resolved at smaller radii than previously suggested ($<$1,000 au).
Therefore, magnetic braking  must be quite important to explain the small disk radii observed around Class 0 objects.

\index{Streamer|(}
With the spatial resolution in nearby regions made possible with interferometers, we can now measure infall and rotation 
within cores and follow the gas from core to disk. 
Observations on these scales have revealed a new component: streamers. 
The detection of streamers has modified our view of the 
mass delivery at core/disk scales.
The sketch at the  bottom of Fig.~\ref{fig:stream-sketch} 
starts with a prestellar core, which has begun gravitational collapse.
The next stage shows the classical picture of the dense core feeding the disk 
with the addition of possible multiple streamers. 
Later on, once most of the surrounding envelope is gone, some late accretion events would be driven by streamers.
Since streamers deliver material non-axisymmetrically to disks, they may be in part responsible for luminous outbursts, and help solve the YSO\index{Young Stellar Object (YSO)} luminosity problem. 
Streamers could play an important role {in} the formation and evolution of disks, since in addition to the sudden mass delivery at small scales, these streamers likely bring a  different amount (or orientation) of angular momentum to the disk-forming scales, which has a direct relation to the maximum disk size. 
Streamers might bring grains which have not undergone as much grain-growth as in the classical inner envelope, 
changing the fraction of small grains at disk scales, which also has a direct role in the disk formation process. 
Similarly, streamers are particularly abundant in chemically fresh (carbon rich) species in the gas, which would 
directly affect the {chemistry and gas phase} composition of material available during planet formation.
\index{Streamer|)}

\paragraph{Future Challenges:}
Determining the role of magnetic fields\index{Magnetic field} in the formation and evolution of filaments remains one of the main challenges for future observations and 
models. 
Further insight will be achieved thanks to ongoing and planned polarization surveys at far-infrared and submillimeter wavelengths with SOFIA-HAWC$+$, 
SCUBA2-POL2, and NIKA2-POL, and in the future with a space-based high-dynamic-range polarimeter such as Millimetron or the Origins Space Telescope.
Combined with gas kinematics surveys on the same spatial scales, 
the new polarization data will provide critical tests of the models of filament formation within magnetized shock-compressed layers\index{Shock!Shock-compressed layers}.

Systematic dust polarization studies at sub-arcmin resolution
will determine whether magnetic fields\index{Magnetic field} remain perpendicular to the long axis of star-forming filaments or typically switch to a parallel configuration in their dense $\sim$0.1\,pc interior. 
This will have profound implications for our understanding of filament stability and fragmentation into prestellar cores.
The scales involved in filament fragmentation are difficult to assess statistically given the small numbers of cores found in any
individual filament. 
The importance of gravo-turbulence  
in filament evolution and fragmentation is highlighted by dynamic models with and without B-fields, and future observational analyses 
with millimeter and submillimeter interferometers such as ALMA and NOEMA 
will continue to test these models.

Characterizing critical stages in protostellar evolution, such as the break in the specific angular momentum profile 
of individual protostellar cores (e.g., Fig.~\ref{fig:j-r}b), is a crucial challenge for future interferometric studies 
of gas kinematics at core scale.
Given the complexity in chemistry, evolutionary stages, and kinematics, the interpretation of these observations 
will critically depend on the comparison with synthetic line observations from simulated dense cores.

Current observational evidence and theoretical models indicate
that streamers are dominated by free-falling motions,
however, it is still unknown how often and for how long streamers affect the disk scales. 
Moreover, there is a strong need for further observations to better quantify the roles 
of streamers, while also determining if there are environmental factors. 
In addition, since there are many different paths in which numerical simulations are able to generate streamer-looking features 
and non-axisymmetric accretion flows, 
it is crucial to post-process different numerical simulations to establish which of the possible scenarios 
for their origin provides a better match to the observations.
One of the biggest challenges for modelers in the upcoming years will 
be to provide better constraints on the kinematics of streamers. 
Multi-scale, multi-physics simulations covering a large physical range 
should provide important constraints on the origin of chemical differences 
between streamers and dense cores, and help to diagnose observable 
tracers of streamers via synthetic observations.

\bigskip

\noindent\textbf{Acknowledgments}
JEP acknowledges the support by the Max Planck Society. AZ acknowledges the support of the Institut Universitaire de France (IUF). 
PhA acknowledges support  from ``Ile de France'' regional
funding (DIM-ACAV$+$) and from the French national programs 
on stellar and ISM physics (PNPS and PCMI). 
SDC is supported by the Ministry of Science and Technology (MoST) in Taiwan through grant MoST 108-2112-M-001-004-MY2.
YNL acknowledges funding from the Ministry of Science and Technology, Taiwan (109-2636-M-003-001 and 110-2124-M-002-012), the grant for Yushan Young Scholar from the Ministry of Education, Taiwan. 
JDS acknowledges funding from the European Research Council under the Horizon 2020 Framework Program via the ERC Consolidator Grant CSF-648505. 
MK acknowledges funding from the European Union’s Framework Programme for Research and Innovation Horizon 2020 (2014-2020) under the Marie Sk{\l}odowska-Curie Grant Agreement No. 897524.
This work was performed under the auspices of the U.S.~Department of Energy (DOE) by Lawrence Livermore National Laboratory under Contract DE-AC52-07NA27344 (C.-Y.C). 

\bibliographystyle{pp7}
\bibliography{refs_full} 
\end{document}